\documentclass[usenatbib]{mn2e}
\usepackage{multirow}
\usepackage{rotating}
\usepackage{colortbl}
\usepackage{color}
\usepackage[fleqn]{amsmath}
\usepackage{amssymb}
\usepackage{amsfonts}
\usepackage{verbatim}
\usepackage{scalefnt}
\usepackage{lscape}
\usepackage{booktabs}

\usepackage[bookmarks,bookmarksnumbered,colorlinks=true, citecolor=blue, linkcolor=cyan]{hyperref}

\newcommand{\apjl}{ApJ}% Astrophysical Journal, Letters
\newcommand{\apjs}{ApJS}% Astrophysical Journal, Supplement
\newcommand{\ao}{Appl.~Opt.}% Applied Optics
\newcommand{\aap}{A\&A}% Astronomy and Astrophysics
\newcommand{\beq}{\begin{equation}}
\newcommand{\eeq}{\end{equation}}

\title[On the population of AGN]{Supermassive black holes in cosmological simulations II: the AGN population and predictions for upcoming X-ray missions}
\author[Habouzit et al.]{M\'{e}lanie Habouzit$^{1,2}$\thanks{E-mail: habouzit@mpia.de}, Rachel S. Somerville$^{3,4}$, Yuan Li$^{5}$, Shy Genel$^{3,6}$, James Aird$^{7,8}$,
\newauthor
Daniel Angl\'es-Alc\'azar$^{9,3}$, 
Romeel Dav\'e$^{7}$, Iskren Y. Georgiev$^{1}$, Stuart McAlpine$^{10}$, 
\newauthor
Yetli Rosas-Guevara$^{11}$, Yohan Dubois$^{12}$, Dylan Nelson$^{2}$, Eduardo Banados$^{1}$,
\newauthor
Lars Hernquist$^{13}$, S\'ebastien Peirani$^{14,12}$, Mark Vogelsberger$^{15}$\\
\\
 $^1$ Max-Planck-Institut f\"ur Astronomie, K\"onigstuhl 17, D-69117 Heidelberg, Germany\\
 $^2$ Zentrum für Astronomie der Universit\"at Heidelberg,
 ITA, Albert-Ueberle-Str. 2, D-69120 Heidelberg, Germany\\
 $^3$ Center for Computational Astrophysics, Flatiron Institute, New York, NY 10010, USA\\
 $^{4}$ Department of Physics and Astronomy, Rutgers University, 136 Frelinghuysen Rd, Piscataway, NY 08854, USA\\
 $^5$ Department of Physics, University of North Texas, Denton, TX 76203, USA\\
 $^6$ Columbia Astrophysics Laboratory, Columbia University, 550 West 120th Street, New York, NY 10027, USA\\
 $^{7}$ Institute for Astronomy, Royal Observatory, University of Edinburgh, Edinburgh EH9 3HJ, UK\\
 $^{8}$ School of Physics \& Astronomy, University of Leicester, University Road, Leicester LE1 7RJ, UK\\
 $^{9}$ Department of Physics, University of Connecticut, 196 Auditorium Road, U-3046, Storrs, CT 06269-3046, USA\\
 $^{10}$ Department of Physics, University of Helsinki, Gustaf H\"{a}llstr\"{o}min katu 2a P.O. Box 64, FI-00014 University of Helsinki, Finland\\
 $^{11}$ Donostia International Physics Centre (DIPC), Paseo Manuel de Lardizabal 4, 20018 Donostia-San Sebastian, Spain\\
 $^{12}$ Institut d'Astrophysique de Paris, Sorbonne Universit\'es, CNRS, UMR 7095, 98 bis bd Arago, 75014 Paris, France\\
 $^{13}$ Harvard-Smithsonian Center for Astrophysics, 60 Garden Street, Cambridge, MA 02138, USA\\
 $^{14}$ Université Côte d'Azur, Observatoire de la Côte d'Azur, Laboratoire Lagrange, Bd de l'Observatoire, CS 34229, Nice, France\\
 $^{15}$ Department of Physics, Kavli Institute for Astrophysics and Space Research, MIT, Cambridge, MA 02139, USA
}

\date{2021}                  

\begin{document}
\maketitle

\begin{abstract}

In large-scale hydrodynamical cosmological simulations, the fate of massive galaxies is mainly dictated by the modeling of feedback from active galactic nuclei (AGN). The amount of energy released by AGN feedback is proportional to the mass that has been accreted onto the BHs, but the exact sub-grid modeling of AGN feedback differs in all simulations. 
Whilst modern simulations reliably produce populations of quiescent massive galaxies at $z\leqslant 2$, it is also crucial to assess the similarities and differences of the responsible AGN populations.
Here, we compare the AGN population of the Illustris, TNG100, TNG300, Horizon-AGN, EAGLE, and SIMBA simulations.
The AGN luminosity function (LF) varies significantly between simulations.  
Although in agreement with current observational constraints at $z=0$, at higher redshift the agreement of the LFs deteriorates with most simulations producing too many AGN of $L_{\rm x, 2-10 keV}\sim 10^{43-44}\, \rm erg/s$.
AGN feedback in some simulations prevents the existence of any bright AGN with $L_{\rm x, 2-10 keV}\geqslant 10^{45}\rm erg/s$ (although this is sensitive to AGN variability), and leads to smaller fractions of AGN in massive galaxies than in the observations at $z\leqslant 2$. 
We find that all the simulations fail at producing a number density of AGN in good agreement with observational constraints for both luminous ($L_{\rm x, 2-10 keV}\sim 10^{43-45}\, \rm erg/s$) and fainter ($L_{\rm x, 2-10 keV}\sim 10^{42-43}\, \rm erg/s$) AGN, and at both low and high redshift. 
These differences can aid us in improving future BH and galaxy subgrid modeling in simulations.
Upcoming X-ray missions (e.g., Athena, AXIS, and LynX) will bring faint AGN to light and new powerful constraints.
After accounting for AGN obscuration, we find that the predicted number density of detectable AGN in future surveys spans at least one order of magnitude across the simulations, at any redshift.

\end{abstract}

\section{Introduction}
In the local Universe, we observe supermassive black holes (BHs) with masses in the range $M_{\rm BH}=4\times 10^{4}-10^{10}\, \rm M_{\odot}$ in galaxies of different types (star-forming, quiescent galaxies), and from dwarf to large elliptical galaxies \citep{2019arXiv191109678G}. BHs are ubiquitous in our Universe, and are believed to play a crucial role in the evolution of galaxies through their energetic feedback \citep[][and references therein]{2012RAA....12..917S,2015ARA&A..53...51S}. Evidence for the co-evolution between BHs and their host galaxies can be found in empirical relationships between BH mass and e.g, galaxy total stellar mass, bulge mass, velocity dispersion \citep[e.g.,][]{Magorrian1998,Haring2004,Gultekin2009}.
Beyond the local Universe, we have no choice but to observe only a fraction of the BH population: the active and accreting BHs, i.e., the Active Galactic Nuclei (AGN), which are the focus of this paper.

Hydrodynamical cosmological simulations, such as Illustris, TNG100, Horizon-AGN, EAGLE, and SIMBA \citep{2013MNRAS.436.3031V,2014MNRAS.445..175G,2014MNRAS.444.1518V,2015MNRAS.452..575S,2018MNRAS.475..624N, 2014MNRAS.444.1453D,2016MNRAS.463.3948D,2016MNRAS.460.2979V,2015MNRAS.446..521S,2015MNRAS.450.1937C,2016A&C....15...72M,2019MNRAS.486.2827D,2019MNRAS.487.5764T,2020NatRP...2...42V}, are a great tool to study the properties of the AGN population and its connection to the full BH population. 
BHs are modeled as collisionless sink particles, each of them being able to accrete surrounding gas, merge with other BH sink particles (often immediately after galaxy mergers), and to release energy into the neighboring gas cell/particle elements. The latter process is called AGN feedback, and is thought to be able to shape the massive end of the galaxy mass function \citep[e.g.,][]{2012RAA....12..917S}.
In these simulations, we can follow the accretion rates onto the BHs, and therefore, 
assess the luminosity of the BHs by assuming that a given fraction of the accreted mass is converted to light and radiated away. The radiative efficiency typically ranges from 10 to $20\%$ in the simulations, and is often used to calibrate the efficiency of AGN feedback and reproduce the empirical $M_{\rm BH}-M_{\star}$ scaling relations. Large-scale simulations with $\sim 100\, \rm cMpc$ side length unfortunately do not have sufficient resolution to resolve the small scales needed to physically capture the physics of the AGN accretion disk \citep[][and references therein]{2017MNRAS.467.3475N,2020arXiv200812303A}. Nevertheless, it is possible to estimate an accretion rate following the Bondi-Hoyle accretion model \citep{Bondi1944}, which describes the spherical stationary inflow of a perfect, non-viscous, non self-gravitating gas onto a BH. In practice, we assume in most of these simulations that the accretion rate is proportional to $M_{\rm BH}^{2}$ and is related to the properties of the surrounding gas. One exception is the SIMBA simulation which employs a gravitational torque accretion model \citep{2011MNRAS.415.1027H,2017MNRAS.464.2840A}, in which the accretion rate is almost independent of BH mass. Large-scale simulations produce a large number of galaxies with stellar mass in the range $M_{\star}=10^{9}-10^{13}\, \rm M_{\odot}$. They allow us to understand the population of AGN in diverse environments and in a broad galaxy mass range.
However, simulations also carry a lot of uncertainties through their sub-grid modeling. Looking in detail at the active BHs can provide us with additional channels to constrain the sub-grid physics of the simulations. It is important to notice that large-scale cosmological simulations were not calibrated to reproduce any of the AGN properties, which are thus true predictions from the simulations.

Observationally, the AGN luminosity function constrains a combination of BH quantities: the BH mass distribution and the accretion rate, or the Eddington ratio distributions. This provides information on the growth of BHs through cosmic times.  
Constraints on the luminosity function and on the number density of AGN have shown over the years that the population of AGN strongly evolves with time. The number of AGN reaches a peak at $z\sim1-2$, and declines at lower redshift. The peak of activity depends on the luminosity of the AGN, with more luminous AGN of $L_{\rm x, \,2-10 \, keV}\sim 10^{44-45}\,\rm erg/s$ having most of their activity at $z\sim 2$ and a sharp decline afterwards. Fainter AGN with $L_{\rm x, \,2-10 \, keV}< 10^{44}\,\rm erg/s$ peak at $z\sim1$ but present a smoother decline later on compared to brighter AGN \citep{2014ApJ...786..104U,2015ApJ...802...89B,2015MNRAS.451.1892A}.
While they provide crucial information, the various observational constraints on the luminosity function and the number density of AGN show some differences.
At $z\leqslant 3$, the observational discrepancies on the luminosity functions remain small and a good agreement between the results of e.g., \citet{2015ApJ...804..104M,2015ApJ...802...89B,2017arXiv170901926K} is found.
Differences increase at higher redshift ($z\geqslant 4$). For example, the hard X-ray luminosity function of \citet{2015MNRAS.453.1946G} has a much lower normalization at the faint end ($L_{\rm x, \,2-10 \, keV}\leqslant 10^{44}\,\rm erg/s$) than the functions derived by \citet{2010MNRAS.401.2531A,2014ApJ...786..104U,2014MNRAS.445.3557V,2015ApJ...802...89B,2016MNRAS.463..348V}. 
For the bright end ($L_{\rm x, \,2-10 \, keV}\geqslant 10^{44}\,\rm erg/s$), \citet{2015A&A...578A..83G} find a much lower normalization than \citet{2015ApJ...802...89B}, while having consistent results for the faint end.
The AGN population is complex and observationally there are still differences on the shape of the faint and bright ends of the luminosity function (particularly at $z\geqslant 3$), and in their overall normalization.

A crucial aspect of AGN is how many of them are significantly obscured. Obscuration arises from the gas and dust both near the BHs and further away in the host galaxies
\citep[][for the relative contributions of the small scale vs galaxy scale gas/dust content]{2017MNRAS.465.4348B,2017NatAs...1..679R}. Most of the obscuration is likely occurring on small scales that can not be resolved by large-scale cosmological simulations.
The fraction of heavily obscured AGN, i.e., the Compton-thick AGN embedded in hydrogen column densities of $N_{\rm H}\geqslant 10^{24}\, \rm cm^{-2}$, is almost entirely derived from X-ray surveys \citep{2015A&ARv..23....1B}. 
Optical AGN surveys are biased against even moderately obscured AGN. Mid-infrared emission is, a priori, not biased against obscuration but the emission from the galaxy component can be significant. 
At low redshift many Compton-thick AGN have been observed but at $z\geqslant 3$ their detection becomes challenging with current X-ray telescopes. Thus far only a few $z\geqslant 3$ Compton-thick AGN have been detected, with the most distant one at $z=4.76$ \citep{2014MNRAS.445.3557V,2016MNRAS.463..348V,2016ApJ...827..150M,2011ApJ...730L..28G}.
These AGN could potentially represent $30-50\%$ or more of the AGN population \citep{2007A&A...463...79G,2007PThPS.169..286G,2014ApJ...786..104U,2014MNRAS.437.3550M}. Obscuration is a key unknown of the AGN population.

Improving the knowledge of the fraction of obscured AGN will require the use of new X-ray instruments with higher sensitivity, but also the ability to explore larger areas on the sky to gain statistics. 
The upcoming Athena X-ray mission \citep{2013arXiv1306.2307N} and AXIS \citep{2018SPIE10699E..29M} and LynX \citep{2018arXiv180909642T} concept X-ray missions, will increase by at least one order of magnitude the current X-ray flux sensitivity, and aim at observing the Universe up to high redshifts to reveal fainter and fainter AGN. These missions will follow the large number of successful X-ray surveys that have been employed in the field over the last decades (e.g., eRASS, XMM-XXL, Stripe-82X, XMM-Atlas, X-Bootes, DEEP2-F1, XMM-COSMOS, C-COSMOS, X-UDS, J1030, COSMOS Legacy, SSA 22, AEGIS-XS CDFS, CDFN), showing that X-ray selection is powerful to understand BH growth in the distant Universe \citep[][for a review]{2015A&ARv..23....1B}.

In the first paper of this series \citep{2020arXiv200610094H}, we examined the BH population of the Illustris, TNG100, TNG300, Horizon-AGN, EAGLE, and SIMBA simulations  \citep{2013MNRAS.436.3031V,2014MNRAS.445..175G,2014MNRAS.444.1518V,2015MNRAS.452..575S,2018MNRAS.475..624N, 2014MNRAS.444.1453D,2016MNRAS.463.3948D,2016MNRAS.460.2979V,2015MNRAS.446..521S,2015MNRAS.450.1937C,2016A&C....15...72M,2019MNRAS.486.2827D,2019MNRAS.487.5764T,2020arXiv201011225T,2017MNRAS.464.2840A}. While all being calibrated with an empirical scaling relation, the shape and normalization of the $M_{\rm BH}-M_{\star}$ mean relation and its evolution vary from one simulation to another \citep{2020arXiv200610094H}, because these aspects are driven by the sub-grid physics of both the BH and the galaxy models (e.g., seeding, supernova (SN) and AGN feedback, BH accretion modeling). 
Given this, and the difficulty of measuring $M_{\rm BH}$ and the galaxy properties in a wide range of galaxies, even in the local Universe, the $M_{\rm BH}-M_{\star}$ does not appear as the most ideal way of constraining further the BH population in cosmological simulations nowadays. Therefore, in this second paper we explore the AGN population produced by the six large-scale cosmological simulations.

We aim at providing the reader with the fundamental quantities that characterize the demographics of active BHs in cosmological simulations. We assess how different models can affect the AGN population, and show that Illustris, TNG, Horizon-AGN, EAGLE and SIMBA all produce different populations of AGN. These populations are in good agreement with some observational constraints, but can also show significant differences with some others. We will show that in general it appears very challenging for a given simulation to produce a population of AGN in good agreement with observations at both high and low redshift, and for both faint and bright AGN.
We also deliver predictions on the AGN population that the Athena, AXIS, and LynX missions will be able to see. To confront our results with future observations, we apply empirically-motivated models for the fraction of obscured AGN to our initial catalogs of simulated AGN. We show that accessing the faint regime of the AGN population could help discriminate different cosmological simulation models. 
While an interesting goal of these space missions is to improve our knowledge beyond $z=6$, here we restrict our analysis to $z\leqslant 6$, a redshift range for which cosmological simulations can already be compared and constrained with current observations.

We first investigate what population of BHs power the AGN in Section~\ref{sec:whatBHs}.
In Section~\ref{sec:edd_ratio}, we present the distributions of the Eddington ratios of the BH populations. We compute the AGN luminosity functions in Section~\ref{sec:lum_fct}, and the AGN number density in Section~\ref{sec:number_density}.
In the following sections, we investigate which galaxies the AGN live in. In particular, we derive the probability of galaxies to host an AGN (i.e., the galaxy occupation fraction) in Section~\ref{sec:agn_fraction} and compare it with constraints in massive galaxies. 
Finally, in Section~\ref{sec:predictions} we synthesize the AGN population that will be detectable by the upcoming Athena mission, and the AXIS and LynX concept missions, and explore how we could use these new constraints to improve the BH/galaxy sub-grid models in simulations.

\section{Methodology: Cosmological simulations, AGN luminosity and obscuration}
\subsection{Cosmological simulations}
We use the six Illustris, TNG100, TNG300, Horizon-AGN, EAGLE, and SIMBA large-scale cosmological hydrodynamical simulations. These simulations model the time evolution of the dark matter and baryonic matter content in an expanding space-time. Due to the large dynamical range needed to follow the non-linear evolution of galaxies, the simulations all employ sub-grid modeling for e.g., star formation, stellar and SN feedback, BH formation, evolution and feedback. While all the same in spirit, sub-grid models vary from simulation to simulation as explained in the first paper of this series \citep{2020arXiv200610094H}. 
Detailed descriptions of the simulations and their BH modeling can be found in \citet{2014MNRAS.445..175G,2014MNRAS.444.1518V} for Illustris, \citet{2017arXiv170302970P,2018MNRAS.479.4056W} for TNG, \citet{2016MNRAS.463.3948D,2016MNRAS.460.2979V} for Horizon-AGN, \citet{2015MNRAS.446..521S,2015MNRAS.454.1038R,2016MNRAS.462..190R,2018MNRAS.481.3118M,2017MNRAS.468.3395M} for EAGLE, and \citet{2019MNRAS.486.2827D,2019MNRAS.487.5764T,2020arXiv201011225T,2017MNRAS.464.2840A} for SIMBA.
These simulations were calibrated to reproduce one of the empirical scaling relation between BH mass and galaxy properties identified in the local Universe. No calibration on the properties of the active population of BHs were used.

BH particles are seeded either in massive halos of $\geqslant 10^{10}\, \rm M_{\odot}$, or in galaxies of $M_{\star}\geqslant 10^{9.5}\, \rm M_{\odot}$, or based on the local gas properties \citep{2016MNRAS.463.3948D}. Initial BH masses range in $M_{\rm BH}\sim10^{4}-10^{6}\, \rm M_{\odot}$.
BHs can growth by BH-BH mergers and gas accretion. Most of the simulations model BH gas accretion with the Bondi-Hoyle-Lyttleton model, or some variations of its formalism, e.g, including a magnetic field component \citep[TNG,][]{2017arXiv170302970P}, or a viscous disk component \citep[EAGLE][]{2015MNRAS.446..521S,2015MNRAS.454.1038R}. The SIMBA simulation employs a two mode gas accretion model \citep[the two modes can be simultaneous,][]{2019MNRAS.486.2827D,2017MNRAS.464.2840A}: gravitational torque-limited accretion model for the cold gas component ($\rm T<10^{5}\,\rm K$) and the Bondi-Hoyle-Lyttleton model for the hot gas component ($\rm T>10^{5}\,\rm K$).
Finally, BHs release energy proportionally to their accretion rate. AGN feedback is modeled in one or two modes, and the released energy can be e.g., thermal and/or kinetic. Illustris employs a two mode feedback, both with release of thermal energy \citep{2015MNRAS.452..575S}, and a transition for $f_{\rm Edd}=0.05$. TNG uses a two mode feedback: thermal in the high-accretion mode, and kinetic in the low-accretion mode \citep{2017MNRAS.465.3291W}. The transition between modes takes place at $f_{\rm Edd}=\min(2\times10^{-3}\times \left( M_{\rm BH}/10^{8}\rm \, M_{\odot}\right)^{2}, 0.1)$. Horizon-AGN uses a thermal mode for high-accretion BHs and a kinetic mode for low-accretion BHs, with a transition at $f_{\rm Edd}=0.01$ \citep{2016MNRAS.463.3948D}. EAGLE employs a single thermal mode \citep{2015MNRAS.446..521S}. Finally, SIMBA uses two different kinetic modes with a transition at $f_{\rm Edd}=0.2$ (with a maximum jet speed reached for $f_{\rm Edd}=0.02$). A complete description of these models can be found in \citet{2020arXiv200610094H}.

In this paper, we only consider AGN in galaxies that are well resolved in all the simulations, i.e., galaxies with total stellar mass of $M_{\star}\geqslant 10^{9}\, \rm M_{\odot}$. 

\subsection{Computation of AGN luminosity}
We compute the luminosity of the BHs following the model of \citet{Churazov2005}, i.e. explicitly distinguishing radiatively efficient and radiatively inefficient AGN. The bolometric luminosity of radiatively efficient BHs, i.e. with an Eddington ratio of $f_{\rm Edd}=\dot{M}_{\rm BH}/\dot{M}_{\rm Edd}>0.1$, is defined as:
\begin{eqnarray}
L_{\rm bol}=\frac{\epsilon_{\rm r}}{1-\epsilon_{\rm r}} \dot{M}_{\rm BH} c^{2}.
\label{eqn:method1}
\end{eqnarray}
Most of the studies based on large-scale cosmological simulations have computed the luminosity of AGN assuming that all the AGN were radiatively efficient, i.e. using Eq.\ref{eqn:method1}.

BHs with smaller Eddington ratios, $f_{\rm Edd}\leqslant 0.1$, are considered to be radiatively inefficient and their bolometric luminosities are computed as:
\begin{eqnarray}
L_{\rm bol}=0.1 L_{\rm Edd} (10 f_{\rm Edd})^{2}=(10 f_{\rm Edd}) \epsilon_{\rm r} \dot{M}_{\rm BH} c^{2}.
\end{eqnarray}
The hard X-ray luminosities are then computed by applying the bolometric correction (BC) of \citet{Hop_bol_2007}:
\begin{eqnarray}
\log_{10} L_{\rm 2-10\, keV,\odot}=\log_{10} L_{\rm bol,\odot} - \log_{10} \rm BC,
\end{eqnarray}
with
\begin{eqnarray}
{\rm BC} = 10.83 \left(\frac{L_{\rm bol,\odot}}{10^{10}\, \rm L_{\odot}} \right)^{0.28} + 6.08 \left(\frac{L_{\rm bol,\odot}}{10^{10}\, \rm L_{\odot}} \right)^{-0.020}.
\end{eqnarray}
Recently however, \citet{2020A&A...636A..73D} showed that the hard X-ray correction could be slightly lower than the \citet{Hop_bol_2007} correction in the range $\log_{10}\, L_{\rm bol}=10^{10.5}-10^{12.5}\, \rm L_{\odot}$. Using the correction of \citet{2020A&A...636A..73D} changes the hard X-ray luminosity function of the simulations, which is slightly shifted towards more luminous AGN, but does not affect the conclusions of this paper. We discuss this in Section 4.1.

We use the radiative efficiency parameter that has been used to derive the accretion rate self-consistently in the simulations. Therefore, we use $\epsilon_{\rm r}=0.2$ for Illustris, TNG100, and TNG300, and  $\epsilon_{\rm r}=0.1$ for Horizon-AGN, EAGLE and SIMBA. The choice of the efficiency parameter will affect the normalization of the functions that we study here, and we discuss this aspect when needed in the different sections below.
We point out that in theory the radiative efficiency depends on BH spin. All the simulations studied here employ a single fixed value of $\epsilon_{\rm r}$; a more physical approach would be to draw values of $\epsilon_{\rm r}$ from a distribution that reflects the distribution of BH spins, and this could impact the properties of BHs and AGN. We discussed this in \citet{2020arXiv200610094H}.

\subsection{AGN obscuration}
In this paper, we compare the AGN population produced by the different simulations to several observational constraints. These constraints already include corrections for AGN obscuration. For that reason we do not add any further correction for obscuration in the first sections of the paper. Thus, 
Fig.~\ref{fig:lum_bhmass}-\ref{fig:agn_variability_03}, and Fig.~\ref{fig:agn_occ_all}, Fig.~\ref{fig:agn_occ_massivegal}
do not include a correction for obscuration.
However, in Section~\ref{sec:predictions}
we predict the number of AGN that we could detect with the future X-ray upcoming or concept missions Athena, AXIS, and LynX. To do so, we correct the simulated populations of AGN with empirically-motivated models for obscuration as described below.

The gas and dust content of a galaxy and/or of the surroundings of its AGN can be the source of obscuration. Photons emanating from an AGN can be absorbed along the line-of-sight to the observer, and consequently the apparent luminosity of the AGN can be lower than its intrinsic luminosity. The hard (2-10 keV) band is less susceptible to obscuration, which means that Compton-thin AGN with hydrogen column densities of $10^{22} \leqslant N_{\rm H}/\rm cm^{-2} \leqslant 10^{24}$ are not significantly impacted.
However, some AGN could be heavily obscured (i.e., Compton-thick AGN) with column densities of $N_{\rm H}\geqslant 10^{24}\, \rm cm^{-2}$ and be completely missed even by hard X-ray surveys. There is evidence showing that the Compton-thick AGN fraction could be constant with both redshift and luminosity \citep{2015ApJ...802...89B}. 
There is also recent work indicating that Compton-thick torii could be present in all AGN, independent of the Eddington ratio \citep[see Fig. 4 of][]{Ricci:2017abh,2017MNRAS.465.4348B}.
Thus far there is still no consensus on the amount of obscured AGN in the Universe, and how the fraction of obscured AGN could evolve with the AGN luminosity and/or redshift.

\begin{figure}
\centering
\includegraphics[scale=0.463]{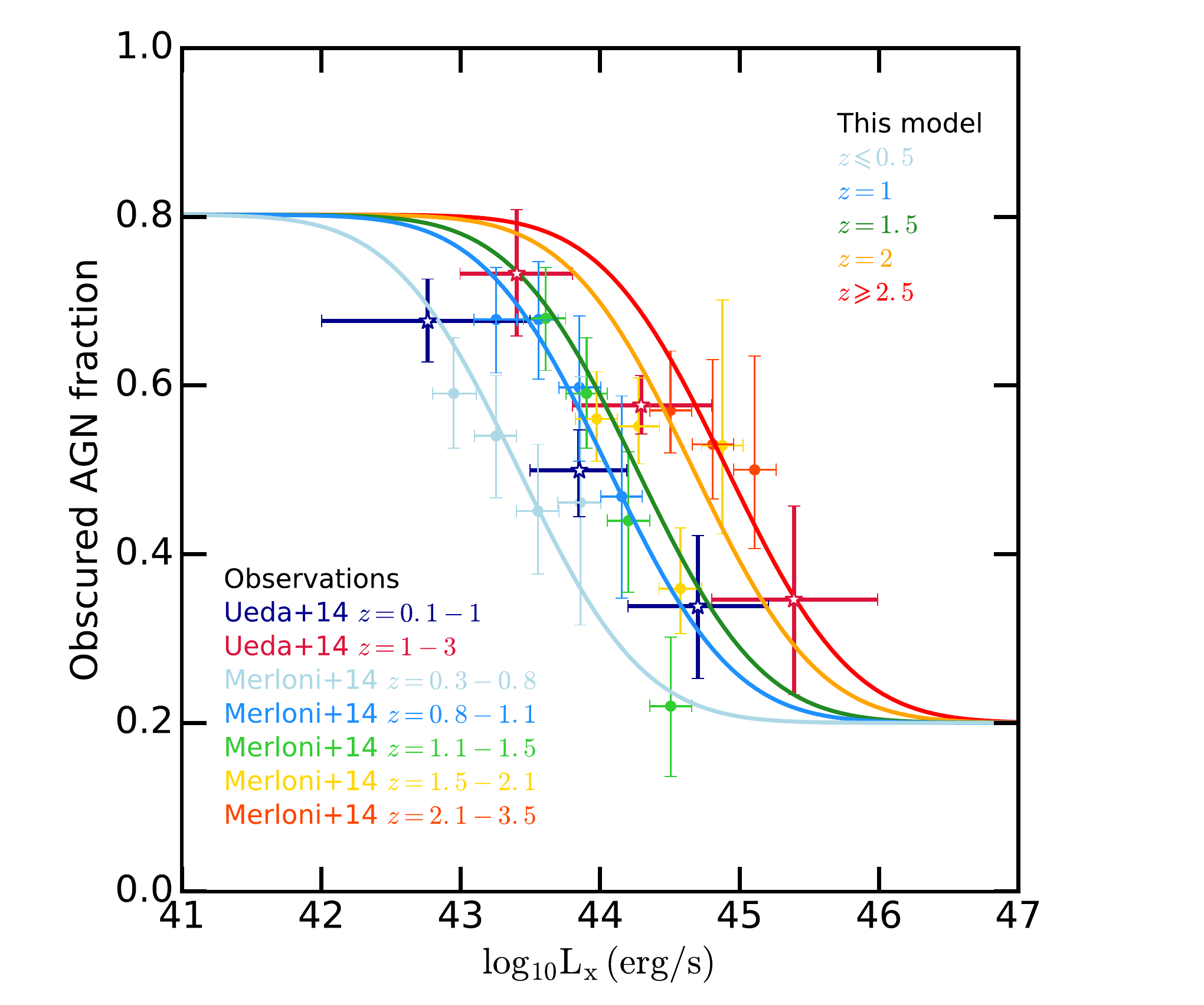}
\caption{We use two different models for the fraction of obscured AGN. The first model assumes that $40\%$ of the AGN are obscured at any redshift, and independently of their luminosities.
The figure shows our second model, which assumes an anti-correlation between the fraction of obscured AGN and their hard X-ray luminosities, and more obscured AGN at higher redshifts following the empirical results of \citet{2014ApJ...786..104U,2014MNRAS.437.3550M}.
}
\label{fig:obscuration_model}
\end{figure}

\begin{figure*}
\centering
\hspace*{-0.14cm}
\includegraphics[scale=0.477]{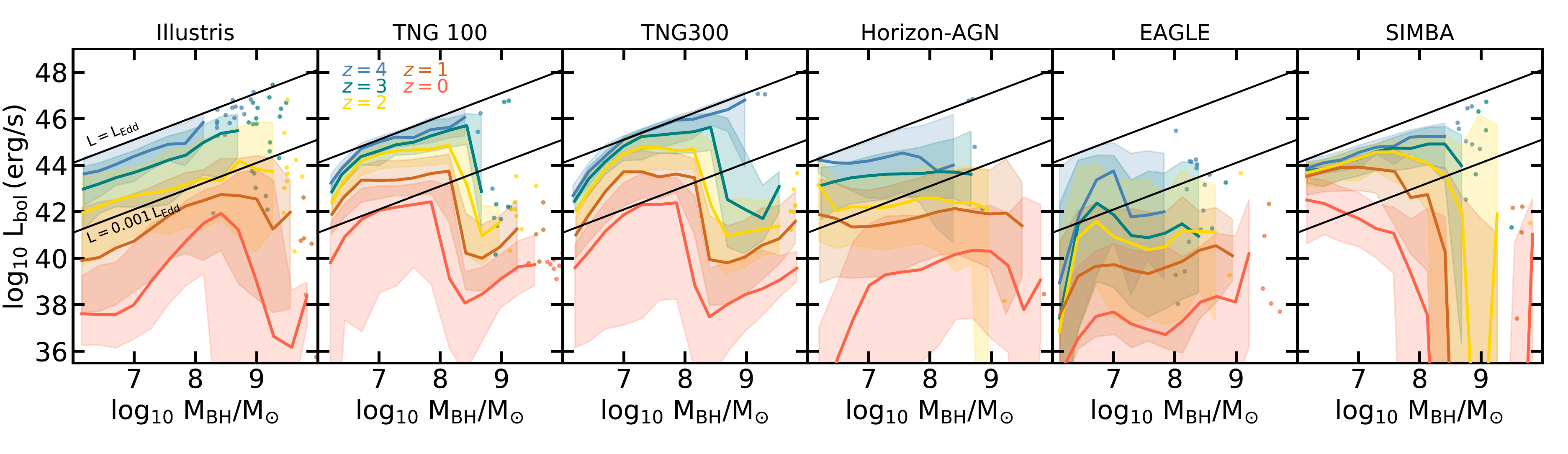}
\hspace*{-0.15cm}
\includegraphics[scale=0.477]{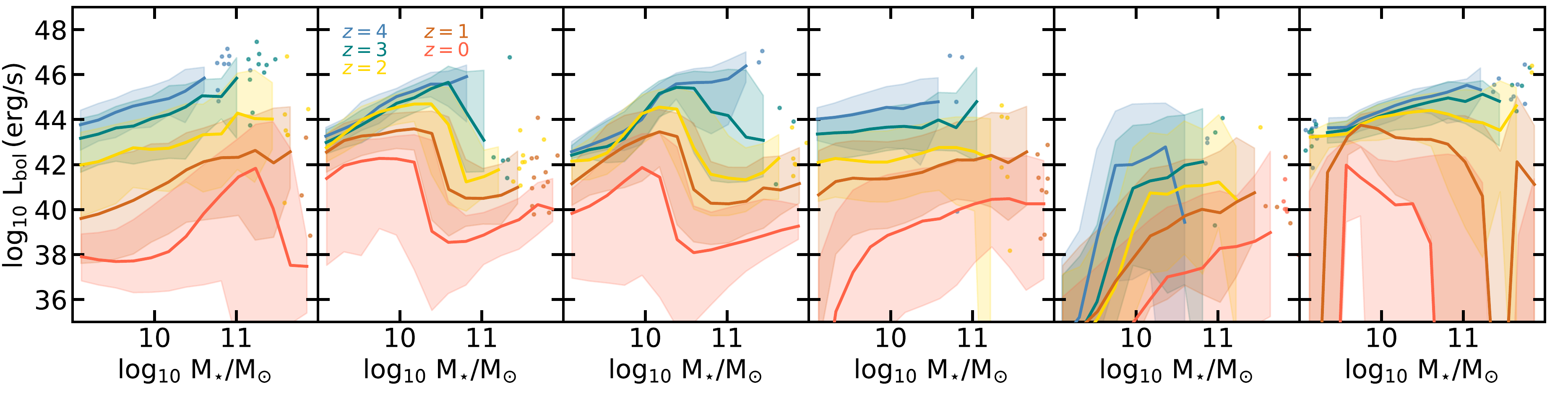}
\caption{{\it Top panels:} Median bolometric luminosity as a function of BH masses for all the simulations, at different redshifts. Dots indicate single BHs when bins contain less than ten BHs, and shaded areas the 15th-85th percentiles of the distributions. The Eddington and $0.1\%$ Eddington luminosity are shown as black solid lines to guide the eye. Most of the simulated BH populations have, on average, luminosities lying between these two references at high redshift. At fixed BH mass, the median bolometric luminosity of the BHs decreases with time, for all the simulations. We detail the specifics of each simulation in the text. {\it Bottom panels:} Median bolometric luminosity as a function of the stellar mass of BH host galaxies.} 
\label{fig:lum_bhmass}
\end{figure*}

\begin{figure*}
\centering
\hspace*{-0.45cm}
\includegraphics[scale=0.475]{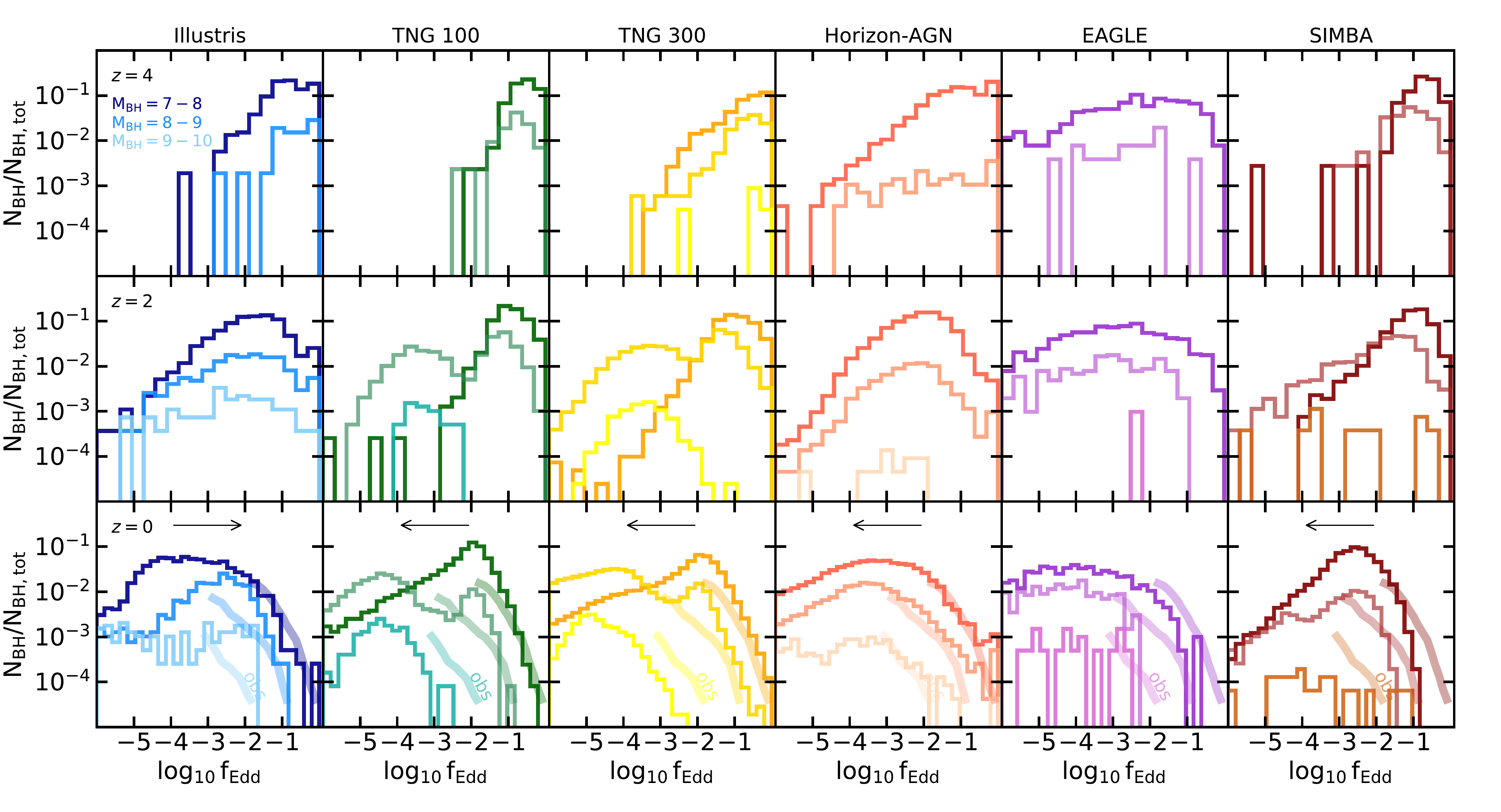}
\caption{Eddington ratio distributions (the y-axes use logarithmic scales). Each bin is normalized to the total number of BHs in the three mass bins. While the peaks of the TNG100, TNG300, Horizon-AGN and SIMBA $f_{\rm Edd}$ distributions move towards lower $f_{\rm Edd}$ for more massive BHs (as indicated by the black arrows), this is not the case for Illustris and EAGLE.
Observational constraints from SDSS at $z=0$ \citep{2004ApJ...613..109H} are shown as thick shaded lines for the $z=0$ panels. There is a good agreement with the simulations for the low-mass BH bins of $M_{\rm BH}=10^{7-8}\, \rm M_{\odot}$ and $M_{\rm BH}=10^{8-9}\, \rm M_{\odot}$. For the most massive BHs of $M_{\rm BH}=10^{9-10}\, \rm M_{\odot}$, we find that the simulations either overestimate the Eddington ratios of these BHs (Illustris), or underestimates them (TNG), at $z=0$. We identify a strong signature (bimodal $f_{\rm Edd}$ distribution) of the transition between the two modes of the AGN feedback modeling in the TNG and SIMBA simulations.}
\label{fig:fedd_ratio}
\end{figure*}

In order to account for obscured AGN, we employ and test two different models: 
\begin{itemize}
    \item First model: we simply assume that a fixed fraction of the AGN is obscured ($40\%$).
    \item Second model: we follow the observational constraints of \citet{2014ApJ...786..104U,2014MNRAS.437.3550M}\footnote{The observational constraints of \citet{2014ApJ...786..104U,2014MNRAS.437.3550M} initially represent Compton-thin AGN, but the presence of Compton-thick AGN in these observations cannot be ruled out.} and build a redshift- and AGN hard X-ray luminosity-dependent fraction of obscured AGN. Our model, shown in Fig.~\ref{fig:obscuration_model}, assumes that there is an anti-correlation between the fraction of obscured AGN and their X-ray luminosities, and that they are more numerous at higher redshift. Our model is defined as:
    \begin{eqnarray}
    \begin{split}
    f_{\rm obsc. \, AGN}(L_{\rm x}, z)=&0.2+(0.5/1.66)\times\\
    &{\rm erfc}[\log_{10} (L_{\rm x}/{\rm (erg/s)})-\alpha(z)],
    \end{split}
    \end{eqnarray} 
    with erfc the complementary function ($\rm erfc(x)=1-erf(x)$), and $\alpha=44.909, 44.666, 44.246,44.060,43.414$ for $z\geqslant 2.5, z=2, z=1.5, z=1, z\leqslant 1$, respectively.
\end{itemize}
These models modify the luminosity and/or the number of AGN at a given luminosity and redshift. The differences of these models are investigated in Section~\ref{sec:predictions}.
In our models we do not explicitly distinguish between Compton-thin and Compton-thick AGN, but rather assume that the models represent the fraction of all obscured AGN. 
For both obscuration models, we either completely remove the obscured AGN from our samples (hereafter called the {\it removed} model), or we assume that their apparent hard X-ray luminosity is one order of magnitude smaller than their intrinsic luminosity ({\it fainter} model).

\section{Results: Eddington ratios}
\subsection{What BHs power the AGN in different simulations?}
\label{sec:whatBHs}
In Fig.~\ref{fig:lum_bhmass} (top panels), we show the median relation between the bolometric luminosity of the BHs (we do not restrict to AGN but rather include all BHs) and their masses, for different redshifts.   
At fixed BH mass, the median $L_{\rm bol}$ decreases with time (blue to red lines) for all the simulations. 
The median $L_{\rm bol}$ is generally the lowest in EAGLE, which produces BHs with lower average accretion rates than the other simulations.
At $z\geqslant 2$, most of the simulations (except EAGLE) have a median bolometric luminosity at $0.1\%-100\%$ of the Eddington luminosity (i.e., lying between the two black lines in Fig.~\ref{fig:lum_bhmass}). At lower redshifts, and even more so for massive BHs, the median $L_{\rm bol}$ drops below $0.1\%$ of the Eddington luminosity. 
The redshift evolution of the simulations is predominantly due to the decrease of the average amount of gas available in galaxies with time, i.e. to cosmic starvation. However, the time evolution and the variations between simulations are due to the specific BH and galaxy subgrid physics of the simulations.
The sharp decrease found in some simulations for massive BHs is due to AGN feedback ($M_{\rm BH}\geqslant 10^{8}\, \rm M_{\odot}$). This is also noticeable in the $L_{\rm bol}-M_{\star}$ plane (Fig.~\ref{fig:lum_bhmass}, bottom panels). Given the tight correlation between BH mass and galaxy mass in all the simulations \citep{2020arXiv200610094H}, the median $L_{\rm bol}$ scales in the same way with $M_{\rm  BH}$ and $M_{\star}$.

Regarding the redshift evolution, we note that Illustris, Horizon-AGN, and EAGLE, have a stronger evolution than TNG and SIMBA for $M_{\rm BH}\leqslant 10^{8}\, \rm M_{\odot}$.
For example, the median in Illustris spans the range $\log_{10} L_{\rm bol}/\rm (erg/s)=38-45$ in the redshift range $z=5-0$, and only $\log_{10} L_{\rm bol}/\rm (erg/s)=42-44$ in TNG. 
At $z\leqslant 2$, low-mass BHs of $M_{\rm BH}\leqslant 10^{8}\, \rm M_{\odot}$ in Illustris, Horizon-AGN and EAGLE lose their ability to accrete gas efficiently, while the same-mass BHs have about on order of magnitude higher median $L_{\rm bol}$ at higher redshift. 
The TNG and the SIMBA simulations produce a population of BHs with $M_{\rm BH}\leqslant 10^{8} \, \rm M_{\odot}$ able to accrete gas efficiently even at low redshifts, compared to the other simulations.
Quenching of BHs in satellite galaxies can be responsible for the decrease of the low-mass BH luminosity \citep{2020arXiv200800005D}.

The median luminosity decreases for massive BHs of $M_{\rm BH}\geqslant 10^{8}\, \rm M_{\odot}$ in all of the simulations (with the exception of the EAGLE simulation), and illustrates the impact of their own feedback\footnote{To fully understand and quantify the self-regulation of AGN, running cosmological simulations with and without AGN feedback is important, but computationally expensive. This has been done for the Horizon-AGN and Horizon-noAGN \citep{2017MNRAS.472.2153P}.}.  
In the TNG simulations, we clearly see the impact of the strong low accretion rate state AGN feedback: AGN luminosities are strongly reduced at all redshifts. 
In TNG, the transition between the high accretion rate AGN feedback mode (injection of thermal energy) and the low accretion rate feedback mode (kinetic mode) takes place at $f_{\rm Edd}=\min [2\times 10^{-3} (M_{\rm BH}/10^{8}\, \rm M_{\odot})^{2},0.1]$ \citep{2017MNRAS.465.3291W,2017arXiv170302970P}.
Most of the BHs with $M_{\rm BH}\gtrsim 10^{8}\, \rm M_{\odot}$ (corresponding to galaxies with stellar masses of a few times $10^{10}\, \rm M_{\odot}$) have low accretion rates, and thus transition to the more efficient kinetic mode. This mode is responsible for regulating BH and star formation activity in the TNG galaxies \citep{2018MNRAS.479.4056W,2019MNRAS.484.4413H,2019arXiv190602747T,2019arXiv191000017L}.
We also see the effect of quenching in Illustris (which uses a different modeling of AGN feedback), but only at low redshift ($z\leqslant 1$). There is also a strong indirect self-regulation of AGN in SIMBA at $z\leqslant 2$, but starting at different BH masses for different redshifts. In SIMBA, AGN feedback heats the CGM of galaxies, which leads to their quenching. This curtails the dominant growth mode of BH torque-limited accretion, and results in an indirect self-regulation of the BHs.  
Indeed, BH growth in SIMBA is quenched for BHs of $M_{\rm BH}\geqslant 10^{7.5}\, \rm M_{\odot}$ at $z=0$, but only more massive BHs get quenched at higher redshift, on average. This is likely due to the AGN feedback modeling in SIMBA, and particularly the low-accretion jet mode that is responsible for galaxy quenching and shutting down BH growth.
In this AGN feedback mode, the velocity of AGN-driven winds increases for lower $f_{\rm Edd}$, and only reach maximum velocity for $f_{\rm Edd}\leqslant 0.02$.
Since Eddington ratios decrease with time even for relatively low-mass BHs of $M_{\rm BH}\leqslant 10^{8}\, \rm M_{\odot}$ (see Fig.~\ref{fig:fedd_ratio}), the feedback becomes more impactful at lower BH mass with time.
In other words, the $f_{\rm Edd}=0.02$ threshold for maximum jet velocity is reached at lower BH mass at low redshift: at $z=0$ any BH with $M_{\rm BH}\geqslant 10^{7.5} \, \rm M_{\odot}$ can transition to the jet mode due to the low Eddington ratios, however at higher redshifts only BHs of $M_{\rm BH}\geqslant 10^{8.5} \, \rm M_{\odot}$ ($z=3$) can start transitioning to the jet mode because Eddington ratios are on average high.

There is a sharp decrease in Illustris for $M_{\rm BH}\geqslant 10^{8}\, \rm M_{\odot}$, but only at low redshift.
We do not identify a sharp decrease of $L_{\rm bol}$ for the massive BHs in Horizon-AGN (except at $z=0$). In Horizon-AGN, the most massive BHs at fixed stellar mass tend to either power faint AGN or are inactive BHs. As a result, when binned in $M_{\rm BH}$ or $M_{\star}$ the median bolometric luminosity appears almost completely flat.

In EAGLE, the impact of AGN feedback is effective in galaxies with BHs of $M_{\rm BH}\geqslant 10^{7}\, \rm M_{\odot}$ \citep[see Fig. 3 of][]{2020arXiv200610094H}, but the effect is masked by the strong SN feedback  regulating the median bolometric luminosity for the low-mass BHs. Indeed, the $L_{\rm bol}$ luminosity is reduced for both the low-mass BHs stunted by SN feedback ($M_{\rm BH}\leqslant 10^{6.5}\, \rm M_{\odot}$) and the BHs self-regulated by their feedback ($M_{\rm BH}\geqslant 10^{7}\, \rm M_{\odot}$). Only BHs of $M_{\rm BH}=10^{6.5-7}\, \rm M_{\odot}$, between the two regulation phases, power slightly brighter AGN in EAGLE.

From Fig.~\ref{fig:lum_bhmass}, the self-regulation of the BHs and also the quenching of the galaxies appears to be different in different simulations. While it seems to be most efficient in the TNG and SIMBA simulations with a sharp decrease of $L_{\rm bol}$ for massive BHs \citep[see also][]{2019MNRAS.485.4817D,2020arXiv200800004D,2019MNRAS.486.2827D}, other simulations like EAGLE also have a strong quenching but masked by the average low $L_{\rm bol}$ for all BH masses. 
We emphasize here that the features and redshift evolution identified in this section also likely depend on gas availability and fueling, in addition to the specific coupling of accretion and feedback in each simulation.

The imprint of the different sub-grid models of the simulations can already be seen in the median AGN luminosity as a function of BH mass. The AGN populations predicted by different simulations are powered by different BHs.

Observational samples of AGN with BH mass estimates from the continuum and emission lines (not dynamical mass measurements) lie in the range $L_{\rm bol}=(0.001-1) \, L_{\rm Edd}$ (i.e., the two black lines in Fig.~\ref{fig:lum_bhmass}). The sample of \citet{2019MNRAS.487.3404B} in the low-redshift Universe does not show a sharp decrease of $L_{\rm bol}$ such as the one found in some simulations for massive BHs (or similarly massive galaxies), but does show a strong $L_{\rm bol}-M_{\rm BH}$ correlation.
Analyses of the relation between the AGN luminosity and the stellar mass of the host galaxies have been carried out in hard X-ray (2-10 keV) at higher redshift ($z=1-2$). When only selecting star-forming galaxies, a strong linear correlation was found \citep{Mullaney2012,2017arXiv170501132A} in general good agreement with the simulations for the median X-ray luminosity in galaxies with $M_{\star}\leqslant 10^{10.5}\, \rm M_{\odot}$. 
\citet{2017arXiv170501132A} also identify a flattening at the low-mass end of the $L_{\rm x}-M_{\star}$ relation, which could hint the effect of SN feedback in these galaxies.

In massive galaxies, we compare the simulations qualitatively to the analysis of the full galaxy population of \citet{2017MNRAS.469.3232G}. While the normalization of the AGN luminosity is consistent with the values of \citet{Mullaney2012,2017arXiv170501132A}\footnote{These works rely on star-forming samples. Similar stellar mass and SFR samples are needed to compare these constraints to simulations. We investigate this in the next paper of our series.} and a linear relation is found at $z\geqslant 2$, \citet{2017MNRAS.469.3232G} identify a flattening (or slight decrease) of the relation for galaxies with $M_{\star}\geqslant 10^{10.5}\, \rm M_{\odot}$ for $z<2$. This could be evidence of the impact of AGN feedback in massive galaxies, as seen in some simulations even if the flattening/decrease is not as strong as in the simulations.
The flattening of the relation at the massive end is indeed seen in massive galaxies with reduced sSFR at $z=1,2$ \citep[][for the relation between the galaxy total X-ray luminosity and galaxy stellar mass]{2018ApJ...865...43F}.
However there is no consensus yet, as \citet{2020A&A...642A..65C} recently find a linear increasing relation with stellar mass for quiescent massive galaxies up to $z=3$ (shallower relation than for star-forming galaxies).
This highlights the potential discrepancies with simulations showing a strong decrease due to AGN feedback in massive galaxies.
The comparisons above are qualitative as we do not apply here the same detection limits and selection biases as these observational studies.

\subsection{Eddington ratio distributions}
\label{sec:edd_ratio}

The evolution of accretion rates onto BHs of different masses is key to understanding not only how the BH population grows with time statistically but also how BHs can self-regulate through AGN feedback.
The accretion onto BHs is connected to two quantities. The first one is the radiative efficiency $\epsilon_{\rm r}$ which quantifies the fraction of accreted mass radiated away, and therefore links the growth of BHs to their bolometric luminosity. The second parameter is the Eddington ratio $f_{\rm Edd}=\dot{M}_{\rm BH}/\dot{M}_{\rm Edd}=L_{\rm bol}/L_{\rm Edd}$, linking the bolometric luminosity of the BHs to the Eddington luminosity $L_{\rm Edd}=4\pi c G m_{\rm H} M_{\rm BH}/\sigma_{\rm T}$.

We show in Fig.~\ref{fig:fedd_ratio} the distribution of the BH Eddington ratios binned in three BH mass bins: $\log_{10} M_{\rm BH}/\rm M_{\odot}=7-8, 8-9, 9-10$. All distributions are normalized to the total number of BHs in the simulations (and not by the number of BHs in a given BH mass bin), in order to compare with the observational constraints of \citet{2004ApJ...613..109H}. 
Varying our choice of the bin size slightly alters the normalization of the distributions, but not our conclusions below.  
We find that the simulations present several different important features, which we detail in the following. 
We find that all the simulations studied here peak at different Eddington ratios at fixed BH mass bins and redshift: e.g. for $\log_{10} M_{\rm BH}/\rm M_{\odot}=7-8$ at $z=0$,
$\log_{10}\, f_{\rm Edd, peak} \sim -4, -2, -2, -3, -4, -2.5$ for Illustris, TNG100, TNG300, Horizon-AGN, EAGLE, and SIMBA, respectively.

\subsubsection{Time evolution in a given BH mass bin}
For all the simulations, the $f_{\rm Edd}$ distributions within the fixed BH mass bins of $\log_{10} M_{\rm BH}/\rm M_{\odot}=7-8, 8-9, 9-10$ move to lower Eddington ratios $f_{\rm Edd}$ with time. From $z=4$ to $z=0$, the peaks of the distributions shift from by between one order of magnitude in $f_{\rm Edd}$ up to several orders of magnitude, depending on the simulation. In general, the ability of the simulated BHs, at a given $M_{\rm BH}$, to accrete gas diminishes with time.

\begin{figure*}
\centering
\hspace*{-0.45cm}
\includegraphics[scale=0.475]{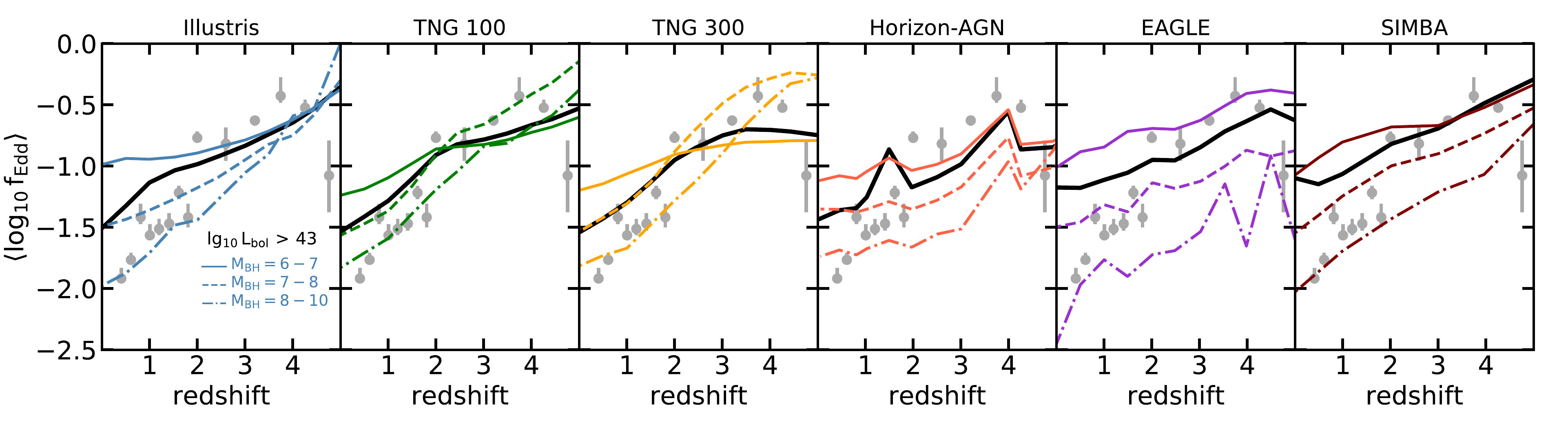}
\caption{Black lines show the mean of the logarithm of the Eddington ratio distributions as a function of redshift. For each simulation, we show the contributions of different mass BH populations, i.e., $\log_{10}\, M_{\rm BH}/\rm M_{\odot}=6-7, 7-8, 8-10$ (colored solid, dashed and dashed dotted lines). The mean of the Eddington ratios of all the simulations decreases with time in the redshift range $z=3-0$, as suggested by the observational constraints shown here. Grey symbols are observational constraints from \protect\citet{2012ApJ...746..169S}. In both simulations and observations, only AGN with $L_{\rm bol}\geqslant 10^{43}\, \rm erg/s$ are considered. To some level, the results of this figure can be connected to Fig.~\ref{fig:fedd_ratio}, but not entirely as we only consider AGN here.}
\label{fig:fedd_ratio_mean}
\end{figure*}

\subsubsection{Evolution across BH mass bins}
The ability of a population of BHs to accrete also depends on their masses, and we focus on $z=0$ in the following to describe our results.
For most of the simulations (TNG, Horizon-AGN, SIMBA), we find that the mean of the Eddington ratio distributions moves towards lower ratios for more massive BHs: more massive BHs globally accrete proportionally at lower rates than their less massive counterparts. We add arrows on Fig.~\ref{fig:fedd_ratio} at $z=0$ to illustrate the effect. 
In Illustris, we find the opposite trend: the distribution for BHs in the range $M_{\rm BH}=10^{8-9}\, \rm M_{\odot}$ peaks at a higher Eddington ratio than the distribution for $M_{\rm BH}=10^{7-8}\, \rm M_{\odot}$. In EAGLE, while the Eddington ratio distributions of all BH mass bins extend to low Eddington ratios, there is no clear evolution of the mean of the distributions with the BH mass bins. Massive BHs in EAGLE do not on average accrete less gas; this can be seen in Fig.~\ref{fig:lum_bhmass} with e.g., more massive BHs being more luminous at $z=0$.

The TNG and SIMBA simulations produce a bimodal Eddington distribution for the intermediate BH mass bin ($M_{\rm BH}=10^{8-9}\, \rm M_{\odot}$) for $z\leqslant 2$ (only in the $z=0$ panel for SIMBA). This bimodality reflects the transition between two modes of AGN feedback in these simulations. BHs transition from the high-accretion mode at high redshift (i.e., when BHs have high $f_{\rm Edd}$ ratios) to the low-accretion mode at lower redshift (i.e., when BHs have lower $f_{\rm Edd}$ ratios).

In TNG, the peak at $\log_{10} f_{\rm Edd}\sim-2$ corresponds to BHs in the high-accretion thermal mode of the feedback. When reaching the characteristic mass of $\log_{10} M_{\rm BH}/\rm M_{\odot}\sim 8$, many of these BHs transition to the kinetic low-accretion feedback. This mode being more efficient by design \citep{2018MNRAS.479.4056W,2019MNRAS.484.4413H}, the BHs accrete at lower rates (lower $f_{\rm Edd}$), leading to the appearance of the second peak at $\log_{10} f_{\rm Edd}\sim -4$ at $z\leqslant 2$. This second peak is more prominent at $z=0$ than $z=2$ since with time more and more BHs in the mass bin $M_{\rm BH}=10^{8-9}\, \rm M_{\odot}$ transition to the efficient AGN feedback mode.
In SIMBA, the velocity of the AGN winds scale with BH mass for the high accretion rate mode \citep[$f_{\rm Edd}>0.2$,][]{2019MNRAS.486.2827D}. 
For lower accretion rates, the velocity is further increased by a factor which is inversely proportional to the Eddington ratio, so that the feedback is stronger for lower Eddington ratios. By design, only BHs with $M_{\rm BH}\geqslant 10^{7.5}\, \rm M_{\odot}$ can enter the low accretion rate AGN feedback regime. The peak of the distribution at $\log_{10} f_{\rm Edd}\sim -4$ for BHs of $M_{\rm BH}=10^{8-9}\, \rm M_{\odot}$ represents the BHs that have already transitioned to this strong mode of the AGN feedback modeling in SIMBA.

The transition between AGN feedback modes in Horizon-AGN and Illustris does not produce such strong signatures in the Eddington ratio distributions for our intermediate $M_{\rm BH}$ bin. However, these two simulations tend to have two-peaked distributions for more massive BHs ($M_{\rm BH}=10^{9-10}\, \rm M_{\odot}$) at $z=0$, which is not found for the other simulations. Indeed, in the $z=0$ panel of Illustris, the BHs with $M_{\rm BH}=10^{9-10}\, \rm M_{\odot}$ peak at $\log_{10} f_{\rm Edd}\sim -5$ and $\log_{10} f_{\rm Edd}\sim -2$. Same in Horizon-AGN with peaks at $\log_{10} f_{\rm Edd}\sim -6$ and $\log_{10} f_{\rm Edd}\sim -3.5$. The Eddington ratio distributions of the other simulations for these massive BHs on average peak at $\log_{10} f_{\rm Edd}\sim -4.5$.

\subsubsection{Comparison to observational constraints}
We compare the distributions to observations from SDSS at $z\sim 0$ \citep[][thick shaded lines in the bottom panels of Fig.~\ref{fig:fedd_ratio}]{2004ApJ...613..109H}.
For the lowest BH mass bin ($\log_{10} M_{\rm BH}/\rm M_{\odot}=7-8$), the simulations are globally in agreement with the SDSS observations. However, since the observations only probe the high Eddington ratio tail of the $f_{\rm Edd}$ distribution (i.e., $\log_{10} f_{\rm Edd}\geqslant -2$), we cannot make any strong statement regarding the modeling of the simulations here.
The Illustris and Horizon-AGN simulations are in good agreement with \citet{2004ApJ...613..109H} in the range $\log_{10} f_{\rm Edd}\geqslant -2$. We note that the TNG simulations may produce too many AGN with $\log_{10} f_{\rm Edd}\sim -2$ with $\log_{10} M_{\rm BH}/\rm M_{\odot}=7-8$, while the EAGLE and SIMBA simulations may not form enough AGN with $\log_{10} f_{\rm Edd}=-2,-1$.

The Eddington ratio distributions of more massive BHs in the range $\log_{10} M_{\rm BH}/\rm M_{\odot}=8-9$ start deviating from the constraints of \citet{2004ApJ...613..109H}. The Illustris, TNG and SIMBA distributions peak at higher $f_{\rm Edd}$ than found in SDSS. The agreement is better for Horizon-AGN, EAGLE, and SIMBA for the peak of the $f_{\rm Edd}$ distribution. However, the EAGLE, and SIMBA simulations still seem not to produce enough efficient accretors with $\log_{10} f_{\rm Edd}\geqslant -2$ for these more massive BHs of $\log_{10} M_{\rm BH}/\rm M_{\odot}=8-9$.

Finally, for the most massive BHs in the range $\log_{10} M_{\rm BH}/\rm M_{\odot}=9-10$ the simulations are not successful in reproducing Eddington ratio distributions that agree with observational constraints.
The Illustris BH population peaks at a much higher $f_{\rm Edd}$ than the constraints, meaning that a large population of the massive Illustris AGN are accreting too efficiently.
The TNG BH population peaks at a lower $f_{\rm Edd}$ (because of the efficient kinetic AGN feedback) than the observational constraints of \citet{2004ApJ...613..109H}. The EAGLE and SIMBA simulations hardly produce BHs in this mass range ($\log_{10} M_{\rm BH}/\rm M_{\odot}=9-10$), which leads to very poor statistics for the Eddington ratio distribution. 
Horizon-AGN produces a better global agreement in this BH mass bin. However, the simulation seems to over-predict the number of the most efficient accretors compared to the SDSS constraints (for all the BH mass bins).
Illustris also presents this feature for the BH mass bin $\log_{10} M_{\rm BH}/\rm M_{\odot}=7-8$. This regime of $\log_{10}\, f_{\rm Edd}\sim 0$ needs to be taken with a grain of salt because it suffers from poor statistics in the simulations.

\subsection{Time evolution of the mean Eddington ratios of AGN}
We quantify the time evolution of the mean Eddington ratio for the relatively luminous AGN that are constrained by observations; i.e., $\log_{10}\, L_{\rm bol}/\rm (erg/s)>43$.
We show the mean Eddington ratios of the simulations as a function of redshift in Fig.~\ref{fig:fedd_ratio_mean} with a black solid line, selecting only the AGN with $\log_{10}\, L_{\rm bol}/\rm (erg/s)>43$. To some level the time evolution of the mean Eddington ratio can be seen in the previous Fig.~\ref{fig:fedd_ratio}, but not completely as here we only look at luminous AGN to be able to compare to observational constraints.
The mean value $\langle \log_{10}\, f_{\rm Edd}\rangle$ moves towards lower Eddington ratios with decreasing redshift for all the simulations. This is in qualitatively good agreement with the observational constraints of \citet{2012ApJ...746..169S} in the redshift range $z=0-4$ (grey symbols in Fig.~\ref{fig:fedd_ratio_mean}), obtained from the analysis of $\sim 60000$ SDSS DR7 AGN ($0.3\leqslant z\leqslant 5$). 
This is also in agreement with the observational constraints presented in \citet{2006ApJ...648..128K,2010ApJ...719.1315K,2010A&A...516A..87S,2013ApJ...764...45K}, and also in semi-analytical models and other simulations \citep[e.g.,][and references therein]{2014MNRAS.442.2304H}.
In the constraints of \citet{2012ApJ...746..169S}, there is a turnover at higher redshifts $z>4$ (although with large uncertainties).
In our analysis, we identify this turnover in several simulations: the mean $f_{\rm Edd}$ values decrease for $z>4$ in Horizon-AGN, and EAGLE, but not in Illustris, TNG100, and SIMBA.

The simulations do not provide an exact quantitative agreement with observations. For example, all the simulations seem to have higher mean Eddington ratios at $z<1$ than the observations, meaning that the AGN are on average accreting more than in the observations. 
We also note that SIMBA and EAGLE over-estimates the mean Eddington ratios for $z<2$, while providing a good agreement at higher redshifts with the measurements of \citet{2013ApJ...764...45K}. The higher mean Eddington ratios in SIMBA are due to both the seeding of the simulation and the accretion model. In SIMBA low-mass BH seeds of $M_{\rm BH}=1.43 \times 10^{4}\, \rm M_{\odot}$ are placed in relatively high-mass galaxies (compared to other simulations) of $M_{\star}\geqslant 10^{9.5}\, \rm M_{\odot}$. Just after seeding, BHs are undermassive with respect to the local $M_{\rm BH}-M_{\star}$ scaling relation, i.e. undermassive for their galaxies. Since $f_{\rm Edd}$ scales with $\propto \dot{M}_{\rm BH}/M_{\rm BH}$, this results in higher Eddington ratios $f_{\rm Edd}$ for these BHs than if they would have been on the scaling relation. The torque accretion model is also almost independent of BH mass, with $\dot{M}_{\rm BH}\propto M_{\rm BH}^{1/6}$ \citep{2013ApJ...770....5A,2015ApJ...800..127A}, so that young BHs catching up to get on the scaling relation can have a broad range of accretion rates (which is not the case for the Bondi accretion model scaling as $M_{\rm BH}^{2}$).
Fianlly, the spikes identified in the mean $f_{\rm Edd}$ of Horizon-AGN are likely due to the creation of new refinement levels in the simulation grid.

\subsubsection{Evolution across BH mass bins} 
In Fig.~\ref{fig:fedd_ratio_mean}, we also show the contributions of different BH mass bins\footnote{Only the mean Eddington ratios are presented in \protect\citet{2012ApJ...746..169S}, and not the contribution of different BH mass bins.} to the mean Eddington ratios $\langle \log_{10} \, f_{\rm Edd}\rangle$: the contributions of BHs with $M_{\rm BH}=10^{6-7}\, \rm M_{\odot}$ are shown with colored solid lines, those of $M_{\rm BH}=10^{7-8}\, \rm M_{\odot}$ BHs with dashed lines, and those of $M_{\rm BH}=10^{8-10}\, \rm M_{\odot}$ BHs with dashed dotted lines.
Lower-mass BHs always have higher mean Eddington ratios, for all the simulations except TNG at $z>2$. In the TNG simulations, the sample composed of more massive BHs of $M_{\rm BH}\geqslant 10^{7}\, \rm M_{\odot}$ have higher Eddington ratios at $z\geqslant 2$ than the BHs of $M_{\rm BH}= 10^{6-7}\, \rm M_{\odot}$. As shown in our previous paper \citep[e.g., Fig. 5 of][with the time evolution of the median $M_{\rm BH}-M_{\star}$ relation]{2020arXiv200610094H}, the stronger SN feedback of the TNG simulations implies that the initial growth of the TNG BHs is delayed (particularly for $z\geqslant 2$), compared to the Illustris BHs for example. These BH seeds are not able to accrete, and therefore have, on average, lower Eddington ratios than more massive BHs.  
While the $\langle \log_{10} \, f_{\rm Edd}\rangle$ means are relatively high in Illustris and TNG for massive BHs of $M_{\rm BH}=10^{7-8}\, \rm M_{\odot}$ and $M_{\rm BH}=10^{8-10}\, \rm M_{\odot}$ at high redshift $z>2$, this is not the case for Horizon-AGN, EAGLE, and SIMBA.
The EAGLE simulation shows an interesting behavior: while the efficient AGN powered by low-mass BHs of $M_{\rm BH}=10^{6-7}\, \rm M_{\odot}$ have very high $\langle \log_{10} \, f_{\rm Edd}\rangle$ compared to most of the other simulations, the ones powered by more massive BHs of $M_{\rm BH}=10^{8-10}\, \rm M_{\odot}$ have the lowest mean Eddington ratios through cosmic time, below the other simulations.

We have demonstrated here that while the redshift evolution of the mean $\langle \log_{10} \, f_{\rm Edd}\rangle$ of the simulated AGN with $\log_{10}\, L_{\rm bol}\rm/(erg/s)>43$ is similar for all the simulations, the evolution for different BH mass bins varies from simulation to simulation owing to variations in sub-grid modeling.

\begin{figure*}
\centering
\hspace*{-0.4cm}
\includegraphics[scale=0.5]{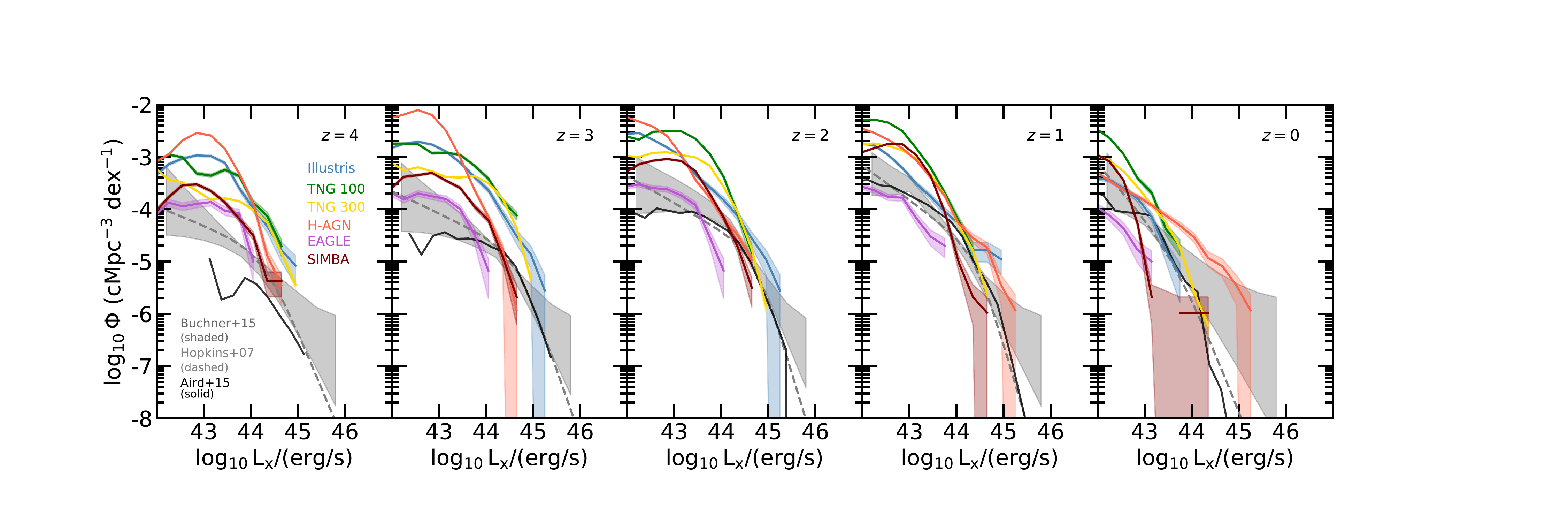}
\hspace*{-0.4cm}
\includegraphics[scale=0.5]{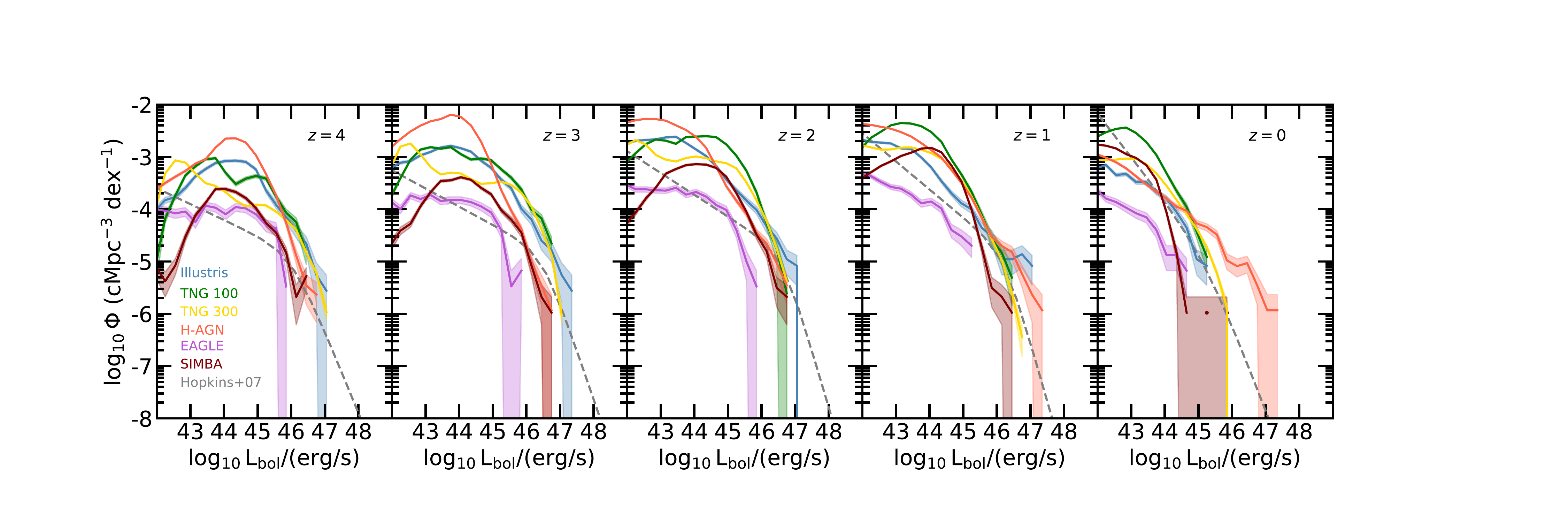}
\caption{{\it Top panels}: Hard X-ray (2-10 keV) AGN luminosity function for the simulations (shaded areas represent Poisson error bars). No AGN obscuration model is applied to the simulations. Observational constraints are shown in grey \citep{Hop_bol_2007,2015ApJ...802...89B,2015MNRAS.451.1892A}. The shapes of the LFs are in good agreement with the constraints. An excess of AGN with luminosities of $\log_{10}\, L_{\rm x}/{\rm (erg/s)}\leqslant 44$ is found in all the simulations at $z\geqslant 2$. This excess vanishes to some extent at lower redshifts. 
Most of the simulations (except Horizon-AGN) produce very few bright AGN with $\log_{10}\, L_{\rm x}/{\rm (erg/s)}\geqslant 44$, in disagreement with observational constraints. 
EAGLE produced the lowest number of AGN of any luminosity. As a result, the EAGLE faint-end is in better agreement with the observations at $z\geqslant 1$ with respect to the other simulations, but is not sufficient to reach the observational constraints at $z<1$.
We use the radiative efficiency $\epsilon_{\rm r}=0.2$ for Illustris and TNG, and $\epsilon_{\rm r}=0.1$ for Horizon-AGN, EAGLE, and SIMBA. 
{\it Bottom panels:}
Bolometric AGN luminosity functions, and observational constraints in grey \citep{Hop_bol_2007}.}
\label{fig:LF}
\end{figure*}

\section{Results: Number density of AGN}
\label{sec:lum_fct}
\subsection{Bolometric and hard X-ray (2-10 keV) AGN luminosity functions}
The AGN luminosity function is one of the fundamental quantities that characterize the demographics of active BHs. It represents the AGN comoving space density as a function of their luminosity.
We show the hard X-ray (2-10 keV) and bolometric luminosity functions in Fig.~\ref{fig:LF}. 
As described in Section 2.6 of \citet{2020arXiv200610094H}, none of the simulations studied here have been calibrated with the AGN luminosity function, thus making them true predictions of the simulations. 
To derive the hard X-ray AGN luminosities we used the bolometric correction of \citet{Hop_bol_2007}. We also tried the new correction of \citet{2020A&A...636A..73D}, which slightly shifts the AGN luminosity functions of the simulation towards more luminous AGN, as shown in Fig.~\ref{fig:LF2} (bottom panels), without affecting the conclusions that we draw below.

In this section, we compare the AGN luminosity functions from the simulations to the observed X-ray luminosity functions of \citet{2015ApJ...802...89B,2015MNRAS.451.1892A}. We also add the analysis of \citet{Hop_bol_2007}; i.e. we translate their bolometric luminosity function into hard X-ray constraints (in the same way as for the simulated AGN in Section 2.2).
Since the empirical luminosity functions include corrections for Compton-thick and Compton-thin AGN, we do not add any corrections for AGN obscuration to the simulation data in this section.

For clarity, we show only the three measurements discussed above in Fig.~\ref{fig:LF}, but many more constraints have been derived for low and high redshifts \citep[][and references therein]{Aird,2010MNRAS.401.2531A,2012MNRAS.425..623L,2014ApJ...786..104U,2014MNRAS.445.3557V,2015MNRAS.453.1946G,2015MNRAS.451.1892A,2015ApJ...804..104M,2015A&A...578A..83G,2016MNRAS.463..348V,2017arXiv170901926K,2019ApJ...871..240A}. At $z=0$, all the observations agree at a good level with the luminosity function of \citet{2015ApJ...802...89B}. A slightly lower normalization was recently found in \citet{2019ApJ...871..240A} at $z=0.1$ for $L_{\rm x}>10^{44.5}\, \rm erg/s$ (with a good agreement at higher redshift).
The bright end of the luminosity function could have a slightly lower normalization for these luminosities for $z\leqslant 3$, as found by studies based on larger surveys than \citet{2015ApJ...802...89B}.
The discrepancies start emerging at higher redshifts, especially at $z\geqslant4$. 
For example, the luminosity functions of \citet{2010MNRAS.401.2531A,2014ApJ...786..104U,2014MNRAS.445.3557V,2016MNRAS.463..348V} have lower normalization than the one from \citet{2015ApJ...802...89B}. The lower normalization compared to \citet{2015ApJ...802...89B} is even more pronounced at the faint end of the luminosity function from \citet{2015MNRAS.453.1946G} ($L_{\rm x}<10^{44}\, \rm erg/s$), or for the bright end of \citet{2015A&A...578A..83G} ($L_{\rm x}>10^{44}\, \rm erg/s$). 
The constraints of e.g., \citet{2014MNRAS.445.3557V} and \citet{2015MNRAS.453.1946G} could be more reliable at $z\geqslant 4$ as they are based on soft X-ray selection. 
In summary, there are still some differences among the measurements.\\

In the following, we analyze the results for the hard X-ray AGN luminosity functions (Fig.~\ref{fig:LF}), yet we find similar results for the bolometric luminosity function. 

\subsubsection{Luminosity functions at $z=0$}
We find a generally good agreement between the hard X-ray AGN luminosity functions from
the simulations and the observations at $z=0$ (right panel in Fig.~\ref{fig:LF}). We note that TGN100 produces an excess of faint AGN with $L_{\rm x}\leqslant 10^{43.5}\, \rm erg/s$ compared to the measurements, whilst EAGLE underestimates the number of these AGN. SIMBA also produces a lower number of AGN in the range $L_{\rm x}\sim 10^{43}-10^{44}\, \rm erg/s$ at $z=0$.
Regarding the bright end of the hard X-ray luminosity function, we find that Horizon-AGN is the simulation producing the brightest AGN ($L_{\rm x}\geqslant 10^{44}\, \rm erg/s$), in agreement with the observations of \citet{2015ApJ...802...89B} (but too many AGN at these luminosities compared to the constraints of \citet{2015MNRAS.451.1892A}). 
EAGLE has a harder time producing these powerful AGN, at any redshift.

\subsubsection{Luminosity functions at higher redshift}
The agreement with the observations becomes weaker toward higher redshifts ($z\geqslant 1$). Most of the simulations have a peak in their luminosity function in the range $L_{\rm x}=10^{42}-10^{43.5}\, \rm erg/s$ (depending on the simulation, and redshift).
Most of the simulations (except EAGLE) overpredict the number of AGN with $L_{\rm x}\leqslant 10^{44}\, \rm erg/s$, by up to one order of magnitude compared to the constraints of \citet{2015ApJ...802...89B,2015MNRAS.451.1892A,Hop_bol_2007}. However, these simulations remain in good agreement for brighter AGN.

EAGLE shows the opposite trend, better matching the faint end of the luminosity function given its lower normalization, but EAGLE does not produce enough bright AGN with $L_{\rm bol}\geqslant 10^{45}\rm \, erg/s$ compared to the constraints of \citet{Hop_bol_2007}.

\begin{figure*}
\hspace*{-0.5cm}
\includegraphics[scale=1]{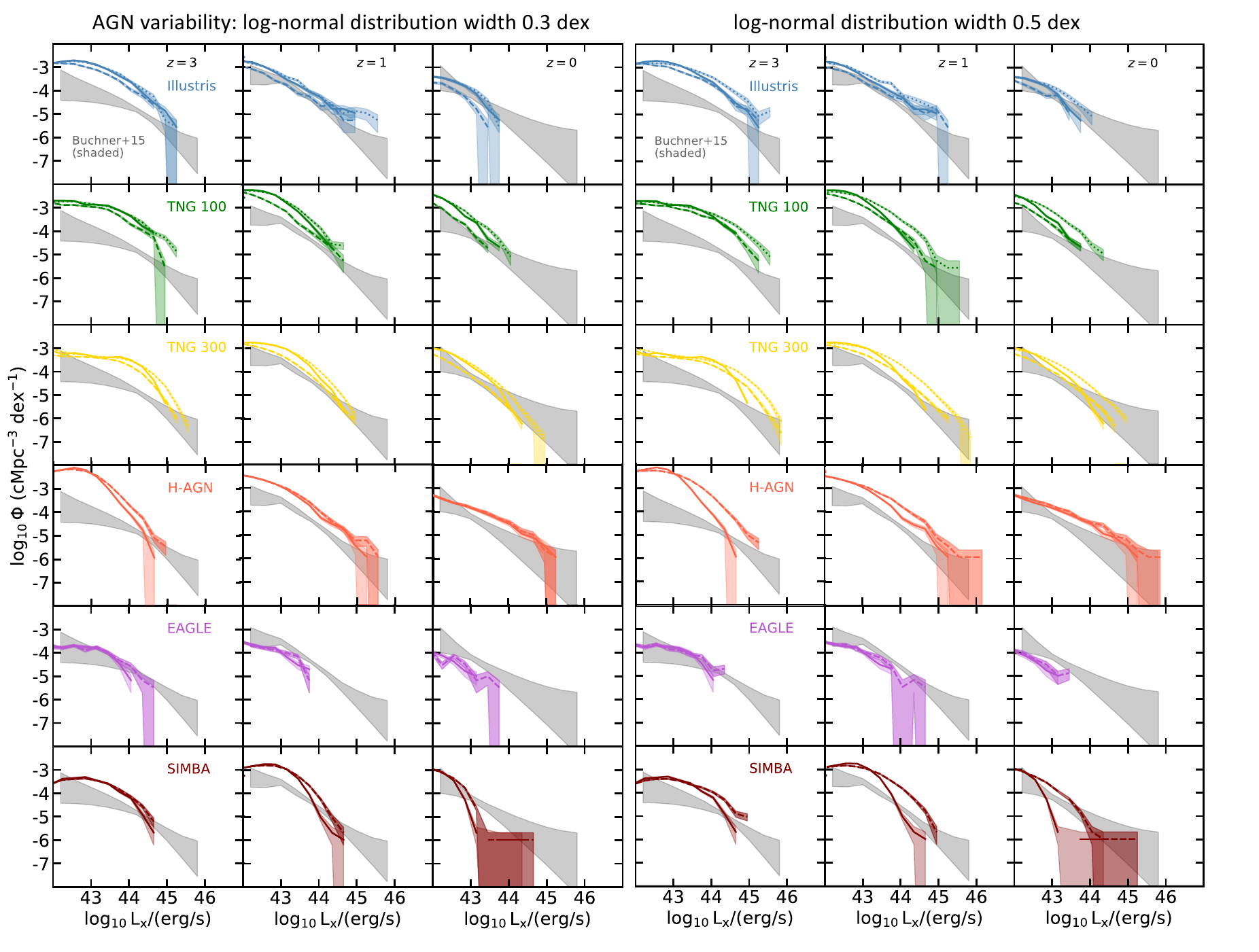}
\caption{Hard X-ray (2-10 keV) AGN luminosity function with a correction for short timescale AGN variability.
We re-compute the luminosity of each simulated AGN by randomly selecting a new $L_{\rm x}$ from a log-normal distribution centered on the initial $L_{\rm x}$ and with a width of 0.3 dex (left panels) and 0.5 dex (right panels). Only one realization of the luminosity function corrected for AGN variability is shown. Solid lines show the simulation AGN luminosity function (without the variability correction) with $\epsilon_{\rm r}=0.2$ for Illustris and TNGs, and $\epsilon_{\rm r}=0.1$ for Horizon-AGN, EAGLE, and SIMBA. The dashed lines represent the luminosity function with $\epsilon=0.1$ for all the simulations and our correction for AGN variability. This allows a direct comparison with the luminosity function of the other simulations, all having the same $\epsilon_{\rm r}=0.1$.
In the panels for Illustris, TNG100, TNG300, we also add the luminosity function computed with the AGN variability correction and $\epsilon_{\rm r}=0.2$, shown as dotted lines.
The effects of AGN variability are more important for a width of 0.5 dex. In particular the luminosity function of EAGLE is in better agreement with the observational constraints for its bright end. TNG100 and TNG300 also include now brighter AGN, and better agree with the constraints. SIMBA also agrees better with the constraints at $z=0$, but overestimates the luminosity function at higher redshifts. This figure does not include a correction for AGN obscuration for the simulations.
}
\label{fig:agn_variability_03}
\end{figure*}

\begin{figure*}
\centering
\hspace*{-0.46cm}
\includegraphics[scale=0.54]{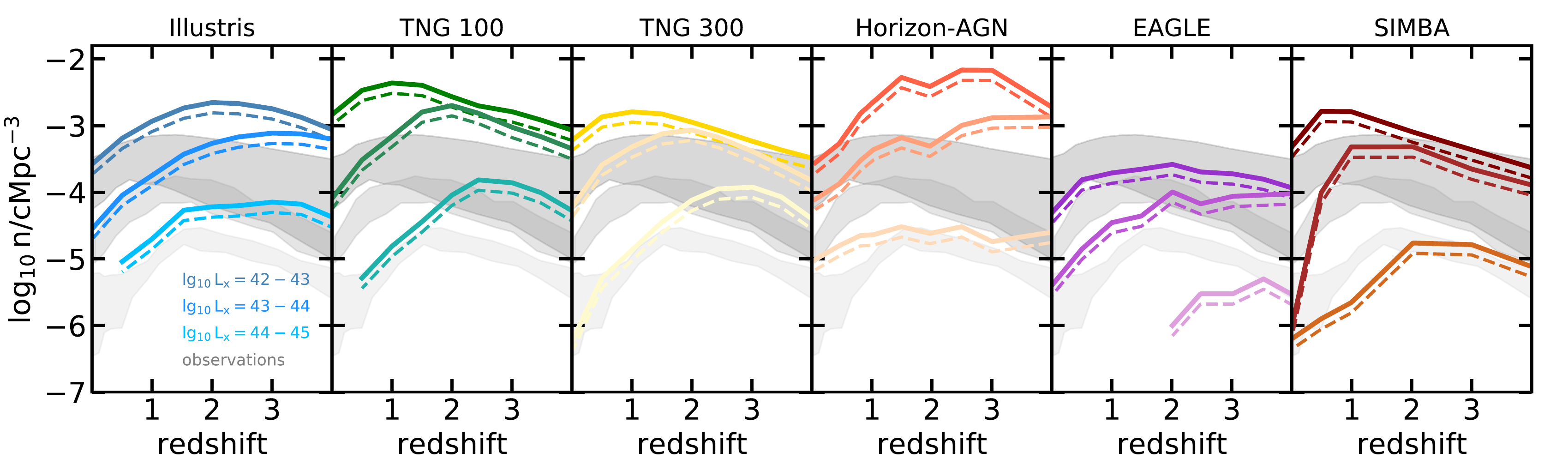}
\hspace*{-0.46cm}
\includegraphics[scale=0.54]{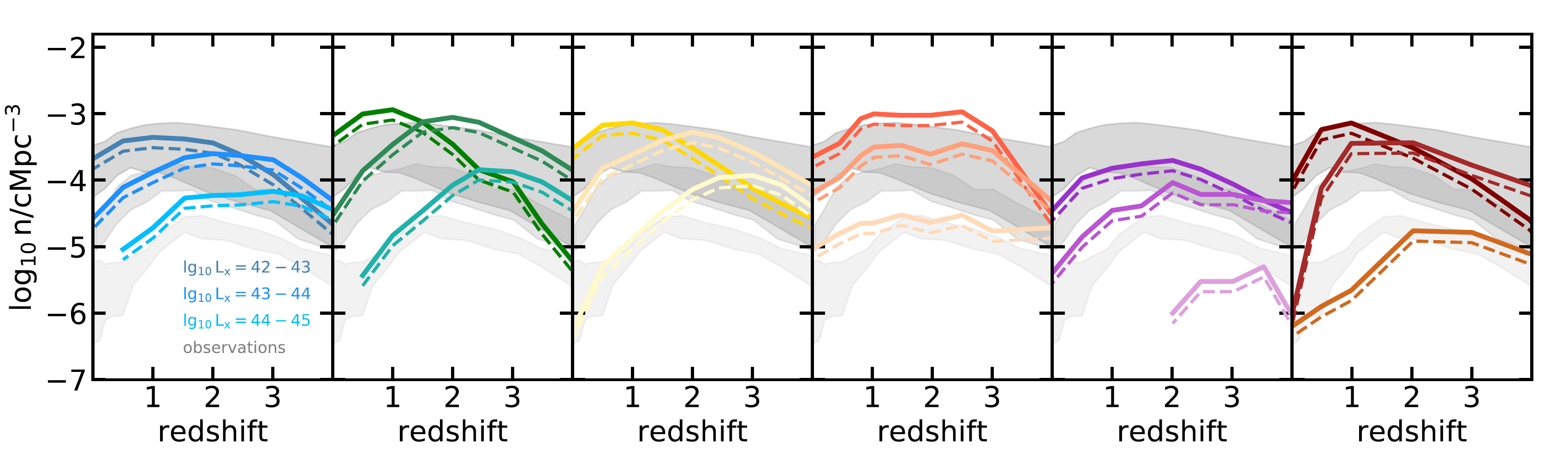}
\caption{{\it Top panels:} Evolution of the comoving number density of AGN binned in hard X-ray luminosity, for galaxies with $M_{\star}\geqslant 10^{9}\, \rm M_{\odot}$. {\it Bottom panels:} Same figure for galaxies with $M_{\star}\geqslant 10^{10}\, \rm M_{\odot}$.
We show the number density (solid lines) of AGN in the luminosity bins $\log_{10} L_{\rm x}/\rm (erg/s) = 42-43$ (top lines),  $\log_{10} L_{\rm x}/\rm (erg/s) = 43-44$, and  $\log_{10} L_{\rm x}/\rm (erg/s) = 44-45$ (bottom lines). We compare the simulation data with observational constraints of \citet{2014ApJ...786..104U,2015MNRAS.451.1892A,2015ApJ...802...89B}: the dark grey regions correspond to the constraints for faint AGN with $\log_{10} L_{\rm x}/\rm (erg/s) = 42-43$, the middle grey regions to AGN of $\log_{10} L_{\rm x}/\rm (erg/s) = 43-44$, and the light grey regions to brighter AGN with $\log_{10} L_{\rm x}/\rm (erg/s) = 44-45$. In addition to the solid lines, we also show the number densities for the simulations after applying a correction for Compton-thick AGN (the most obscured AGN); dashed lines assume that $30\%$ of the simulated AGN are Compton-thick AGN.
}
\label{fig:number_density}
\end{figure*}

\subsubsection{Impact of the simulation resolution in TNG}
The TNG100 and TNG300 simulations allow us to study the effect of volume and resolution on the luminosity function \citep[see also][]{2018MNRAS.479.4056W}. The luminosity function of TNG300 has a lower normalization for $\log_{10} L_{\rm x}/{\rm (erg/s)}\leqslant 44$ and $\log_{10} L_{\rm bol}/{\rm (erg/s)}\leqslant 45-46$. The gas density around BHs is less accurately resolved in TNG300, which can explain the fewer AGN at fixed AGN luminosity. The fewer number of $\log_{10} L_{\rm x}/{\rm (erg/s)}\leqslant 44$ AGN can be seen in Fig. 3 of \citet{2020arXiv200610094H}, with both fainter AGN powered by TNG300 BHs of $M_{\rm BH}\sim 10^{6}\, \rm M_{\odot}$ in galaxies of total stellar mass $M_{\star}\leqslant 10^{10}\, \rm M_{\odot}$, and by BHs of $M_{\rm BH}= 10^{6-7}\, \rm M_{\odot}$ in galaxies of $M_{\star}=10^{10-11}\, \rm M_{\odot}$, compared to the brighter TNG100 BHs.

While TNG100 and TNG300 have similar number densities of AGN with $L_{\rm bol}\gtrsim 10^{44}\, \rm erg/s$ and $L_{\rm bol}\gtrsim 10^{46}\, \rm erg/s$, the larger volume of the TNG300 (27 times larger volume than TNG100, and about 10 times larger than Horizon-AGN and SIMBA) produces even brighter AGN (which are not present in TNG100). The number density of these brightest AGN is in good agreement with observations at $z\leqslant 2$.

\subsubsection{Impact of AGN variability}
Cosmological simulations offer good statistics on the AGN population, but remain limited by their resolution. In particular, the region around the BHs is not sufficiently resolved (in space and time) to capture short timescale variability. 
Both the simulations that resolve the region near BHs at sub-pc scales \citep{2011ApJ...737...26N,2020arXiv200812303A} and the observations have shown that the accretion rate onto BHs can change by orders of magnitude over short timescales that are not resolved in large-scale cosmological simulations \citep[e.g.,][and references therein]{2017MNRAS.466.1462D,2014MNRAS.441.1615G}. 
In order to account for the impact of AGN variability on the AGN luminosity function, we modify the luminosity of each simulated AGN: we randomly draw a new AGN luminosity from a log-normal distribution centered on the initial AGN luminosity and with a width of 0.3 (Fig.~\ref{fig:agn_variability_03}, left panels) or 0.5 dex (right panels). Fig.~\ref{fig:agn_variability_03} shows only one realization of the AGN luminosity functions when we apply the variability models.
The impact of AGN variability is only noticeable for $L_{\rm x}\geqslant 10^{43.5}\, \rm erg/s$ for $z\geqslant 1$ (and for most simulations only for brighter AGN with $L_{\rm x}\geqslant 10^{44.5}\, \rm erg/s$ ), and $L_{\rm x}\geqslant 10^{43}\, \rm erg/s$ at $z=0$, and we find that this bright end of the luminosity function can be shallower for several simulations. The effect is limited for a log-normal distribution with a width of 0.3 dex, and more important for the distributions with 0.5 dex width. 
The largest effect is found in SIMBA, whose bright end is considerably extended to brighter AGN for the 0.5 dex width distribution model, leading to better agreement with the constraints of \citet{2015ApJ...802...89B} at $z=0$. EAGLE is the simulation producing the fewest bright AGN, and we also note a shallower bright end of the AGN luminosity function when accounting for AGN variability \citep[see also][]{2016MNRAS.462..190R}, and therefore better agreement with the measurements at all redshifts.

\subsubsection{Impact of other parameters}
In Fig.~\ref{fig:agn_variability_03}, we show the impact of the radiative efficiency $\epsilon_{\rm r}$: for Illustris, TNG100, and TNG300 (i.e., the first three rows) we show both the luminosity function with $\epsilon_{\rm r}=0.2$ (the parameter employed in the simulations) and $\epsilon_{\rm r}=0.1$ (parameter used for all the other simulations). A higher radiative efficiency increases the normalization of the luminosity functions, and the number of brighter AGN \citep[see also Appendix of][]{2019MNRAS.484.4413H}. 

We now discuss the impact of our method to compute the AGN luminosity. In this paper, we consider that AGN are either radiatively efficient, or inefficient (see Section 2.2). The main effect of considering AGN with $f_{Edd}\leqslant 0.1$ as inefficient is a decrease in the amount of AGN with $L_{\rm x}\leqslant 10^{44}\, \rm erg/s$ , especially for $z\leqslant 2$ \citep[see also Appendix of][]{2019MNRAS.484.4413H}. 
To understand the role of the $f_{\rm Edd}$ transition to define efficient and inefficient AGN, we compute the luminosity function with a transition at $f_{\rm Edd}=0.01$ (not shown here). The main effect is also an increase of the number of AGN with $L_{\rm x}\leqslant 10^{44}\, \rm erg/s$, with a lower amplitude than considering all AGN as efficient. While most simulations are above the observational constraints (TNGs, Horizon-AGN, Illustris) at $z=0$, we find a better agreement for EAGLE and SIMBA with this $f_{\rm Edd}$ transition.

Finally, we note here that the effect of the AGN variability on the X-ray luminosity function could be similar to allowing for dispersion in the bolometric correction used to compute the hard X-ray luminosity of the AGN, and this needs to be investigated in detail. Dispersion in the conversion from X-ray to bolometric luminosities was used in \citet{2020ApJ...903...85A} \citep[following][]{2010A&A...509A..38G,2014ApJ...786..104U} to compute the total radiation of AGN, and thus investigate the contribution of AGN to reionization.

\subsection{Comoving number density of AGN as a function of redshift}
\label{sec:number_density}
We now turn to quantify the time evolution of the number density of AGN with different luminosities. 
In Fig.~\ref{fig:number_density}, we show the redshift evolution of the comoving number density of AGN binned in hard X-ray luminosity\footnote{The spikes in the AGN number density of Horizon-AGN in Fig.~\ref{fig:number_density} (and in other figures of the paper) are triggered by higher levels of accretion at some given redshifts for which a new level of mesh refinement is added in the simulation. The new refinement level can also trigger spikes in the SFR history, and the effect has been discussed recently in \citet{2018MNRAS.477..983S}.}, considering galaxies with $M_{\star}\geqslant 10^{9}\,\rm M_{\odot}$ (top panels). We show the number density (solid lines) for AGN in the luminosity bins $\log_{10} L_{\rm x}/\rm (erg/s) = 42-43$ (top lines in each panel),  $\log_{10} L_{\rm x}/\rm (erg/s) = 43-44$, and  $\log_{10} L_{\rm x}/\rm (erg/s) = 44-45$ (bottom lines in each panel). 
A similar figure can be found in \citet{2016MNRAS.462..190R} for the EAGLE simulation.
Observational constraints on the number density for these AGN X-ray luminosities are shown in grey.
More precisely, we show the regions enclosed by the minimum and the maximum of the three observational constraints (all together) derived by \citet{2014ApJ...786..104U,2015MNRAS.451.1892A,2015ApJ...802...89B}. One can see that the faint AGN regime is the one suffering from the largest uncertainties in observations, especially for $z\geqslant 1$.
The observational constraints include a correction for moderately obscured AGN, and therefore we do not need to correct for Compton-thin AGN ($10^{22}<N_{\rm H}< 10^{24}\, \rm cm^{-2}$). 
However, the observations do not correct for heavily obscured objects with column densities of $N_{\rm H}\geqslant 10^{24}\, \rm cm^{-2}$, the Compton-thick AGN.
We therefore test the impact of applying an additional correction for Compton-thick AGN. The dashed lines assume that $30\%$ of the simulated AGN are heavily obscured and we remove them from our samples\footnote{Instead of completely removing these $30\%$ heavily obscured AGN from our samples, we could also have decreased their luminosities by, for example, one order of magnitude. These AGN would have moved from a given $L_{\rm x}$ bin to the fainter bin in Fig.~\ref{fig:number_density}.}. The fraction of Compton-thick AGN is hard to constrain in observations, and could be more than $30\%$ \citep[e.g.][]{2007A&A...463...79G,2014MNRAS.437.3550M}. Moreover the fraction could also depend on the AGN luminosity, and redshift. The uncertainties induced by the $30\%$ heavily obscured AGN that we use here are lower than the differences among the observational constraints. \\

Almost all the simulations produce too many AGN in the range $\log_{10} L_{\rm x}/\rm (erg/s) = 42-44$ (top solid lines and top grey shaded constraints), with Horizon-AGN producing the highest number of those.
However, the EAGLE simulation is in very good agreement with the observational constraints for the faint AGN of $\log_{10} L_{\rm x}/\rm (erg/s) = 42-43$ at all redshifts, but the agreement is on average poorer for more luminous AGN with $\log_{10} L_{\rm x}/\rm (erg/s) = 43-44$ (except for $z=2-3$).
Yet the EAGLE simulation produces too few bright AGN of $\log_{10} L_{\rm x}/\rm (erg/s) = 44-45$, at any redshift, while the other simulations produce more of these bright AGN and obtain a better agreement with observations, at least for $z\leqslant 2$. 
Most of the simulations, except SIMBA and Horizon-AGN for which a good agreement is found, overproduce the number of these bright $\log_{10} L_{\rm x}/\rm (erg/s) = 44-45$ AGN at high redshift $z\geqslant 2$.
In general, we find that many of the simulations form too many AGN of any luminosity at high redshift. This suggests that BH growth is too efficient at high redshift. Having a higher fraction of heavily obscured AGN at high redshift would decrease the discrepancy with observations. 

\begin{figure*}
\centering
\hspace*{-0.6cm}
\includegraphics[scale=0.486]{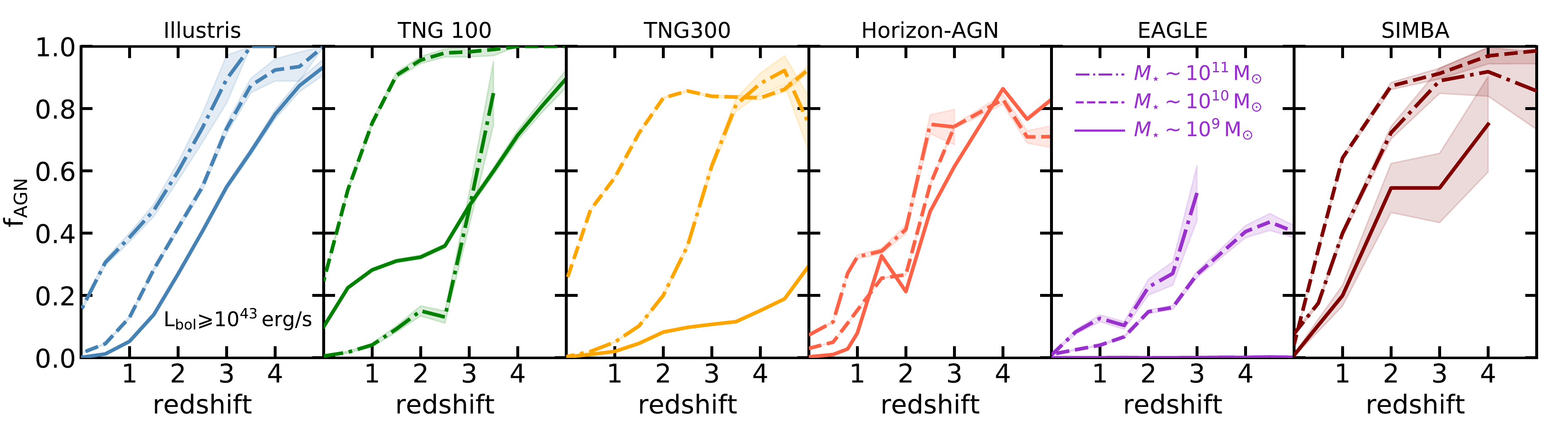}
\hspace*{-0.6cm}
\includegraphics[scale=0.486]{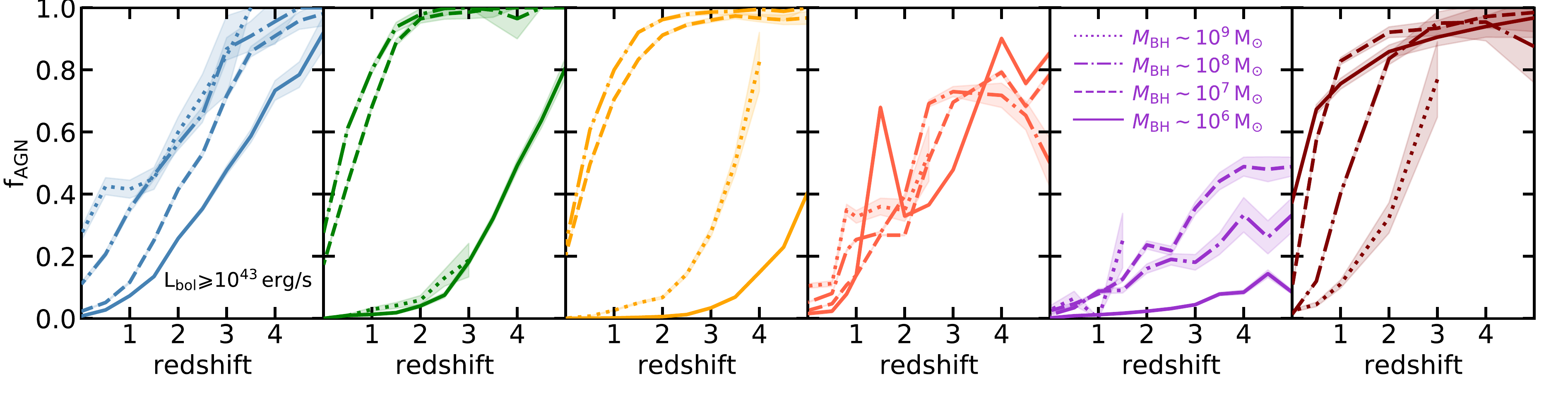}
\caption{Fraction of AGN with $L_{\rm bol}\geqslant 10^{43}\, \rm erg/s$ in galaxies of different masses $M_{\star}\sim 10^{9},\, 10^{10}, \,10^{11}\, \rm M_{\odot}$ (top panels), and similarly different BH masses of $M_{\rm BH}\sim 10^{6},\, 10^{7}, \,10^{8},\, 10^{9}\, \rm M_{\odot}$ (bottom panels). Only redshift bins with more than 5 galaxies are included here. Shaded areas represent Poisson error bars. The fractions of AGN are higher at higher redshifts for all the simulations, but there is no consensus on the time evolution of the AGN fractions for different ranges of BH or galaxy masses.}
\label{fig:agn_occ_all}
\end{figure*}

We can conclude here that it is hard for a given simulation to produce a number density of AGN in agreement with these observational constraints at {\it both} high (e.g., $z>2$) and low redshifts ($z<2$), but also for {\it both} fainter AGN (e.g., $\log_{10} L_{\rm x}/\rm (erg/s) = 42-43$) and brighter AGN (e.g., $\log_{10} L_{\rm x}/\rm (erg/s) = 44-45$). Simulations generally reproduce one of these aspects, but fail
in other regimes.

\subsubsection{Peak of the AGN number density}
In addition to the relative number of AGN that we have discussed above, the trend with redshift is also very informative. 
The shape of the number density function of all the simulations is similar to the overall shape in observations: all the simulations have increasing number densities of AGN (of any luminosity) at high redshift, peak at some redshift, and then have decreasing number densities when moving towards lower redshifts. 
However, the redshift at which the turn-over takes place is not in precise agreement with the observations for all the simulations.
In observations, we see what we call the {\it downsizing} effect: brighter AGN peak (in number density) at earlier times \citep{2014ApJ...786..104U,2015MNRAS.451.1892A}, and fainter AGN at later times. We also find this trend in the simulations, in a clear way for TNG and SIMBA, and in a less obvious way for the Illustris, Horizon-AGN, and EAGLE simulations.
The TNG number density of AGN with $\log_{10} L_{\rm x}/\rm (erg/s) = 42-44$ peaks at roughly the same redshift as in the observations. The brightest AGN population with $\log_{10} L_{\rm x}/\rm (erg/s) = 44-45$ peaks at much earlier times ($z\sim2.5-3$) in the simulations than in observations ($z\sim1.5-2$).

\subsubsection{Uncertainties: the impact of galaxy stellar mass limits}
The galaxy stellar mass limit considered above is the first aspect that could affect our comparison with observations.
In the top panels of Fig.~\ref{fig:number_density}, we have only included AGN in galaxies of $M_{\star}\geqslant 10^{9}\, \rm M_{\odot}$ to homogenize the resolution limit over all the simulations. The differences between simulations and observations could thus arise if the observational samples include lower-mass galaxies. This is unlikely since these galaxies are difficult to detect in optical wavelengths, which is needed to estimate their redshift. 
Indeed, in observational samples most X-ray detected AGN are found to reside in more massive galaxies than $M_{\star}\geqslant 10^{9}\, \rm M_{\odot}$, e.g., in galaxies with $M_{\star}\geqslant 10^{10}\, \rm M_{\odot}$ \citep{2009ApJ...693....8B,2010ApJ...720..368X,Aird2012,2013ApJ...775...41A,2013ApJ...770...40M,2018MNRAS.474.1225A}. We apply this latter stellar mass cut to compute the number density of the AGN in Fig.~\ref{fig:number_density} (bottom panels). Considering only galaxies with $M_{\star}\geqslant 10^{10}\, \rm M_{\odot}$ significantly affects the results: i) the number density of the faint AGN with $\log_{10}\,L_{\rm x}/(\rm erg/s)=42-43$ is reduced, particularly at high redshift, leading to a better agreement with observational constraints for all the simulations, ii) with a smaller amplitude the number density of AGN with $\log_{10}\,L_{\rm x}/(\rm erg/s)=43-44$ is also reduced, but they are still overproduced in simulations with respect to the observations, iii) the number of the brightest AGN is not affected.
Many faint to intermediate AGN in the simulations are located in galaxies with stellar mass in the range $M_{\star}=10^{9}-10^{10}\, \rm M_{\odot}$, causing the changes described above. In other words, some simulations produce too many faint AGN (especially at high redshift) in low-mass galaxies of $M_{\star}=10^{9}-10^{10}\, \rm M_{\odot}$. Overall, these changes do not affect our main conclusion that all the simulations generally do not agree with observational constraints in all the regimes (faint/bright AGN, low/high redshift).
We investigate the correlations between AGN, host galaxies and redshift, in the next paper of our series.

\subsubsection{Uncertainties: obscuration effects}
AGN obscuration could also trigger differences between the observations and simulations.
If obscuration of Compton-thick AGN mainly arises from large amounts of gas and/or dust in the AGN host galaxies rather than small regions close to the AGN \citep[but see][]{2017MNRAS.465.4348B}, a fixed fraction of Compton-thick AGN (as we use here) would also be an overly simplistic approach, and would affect the shape and normalization of the number density. Similarly, observations could also under-estimate the number of Compton-thick AGN due to small-scale obscuration, particularly at low luminosity.

 \begin{figure*}
\centering
\includegraphics[scale=0.49]{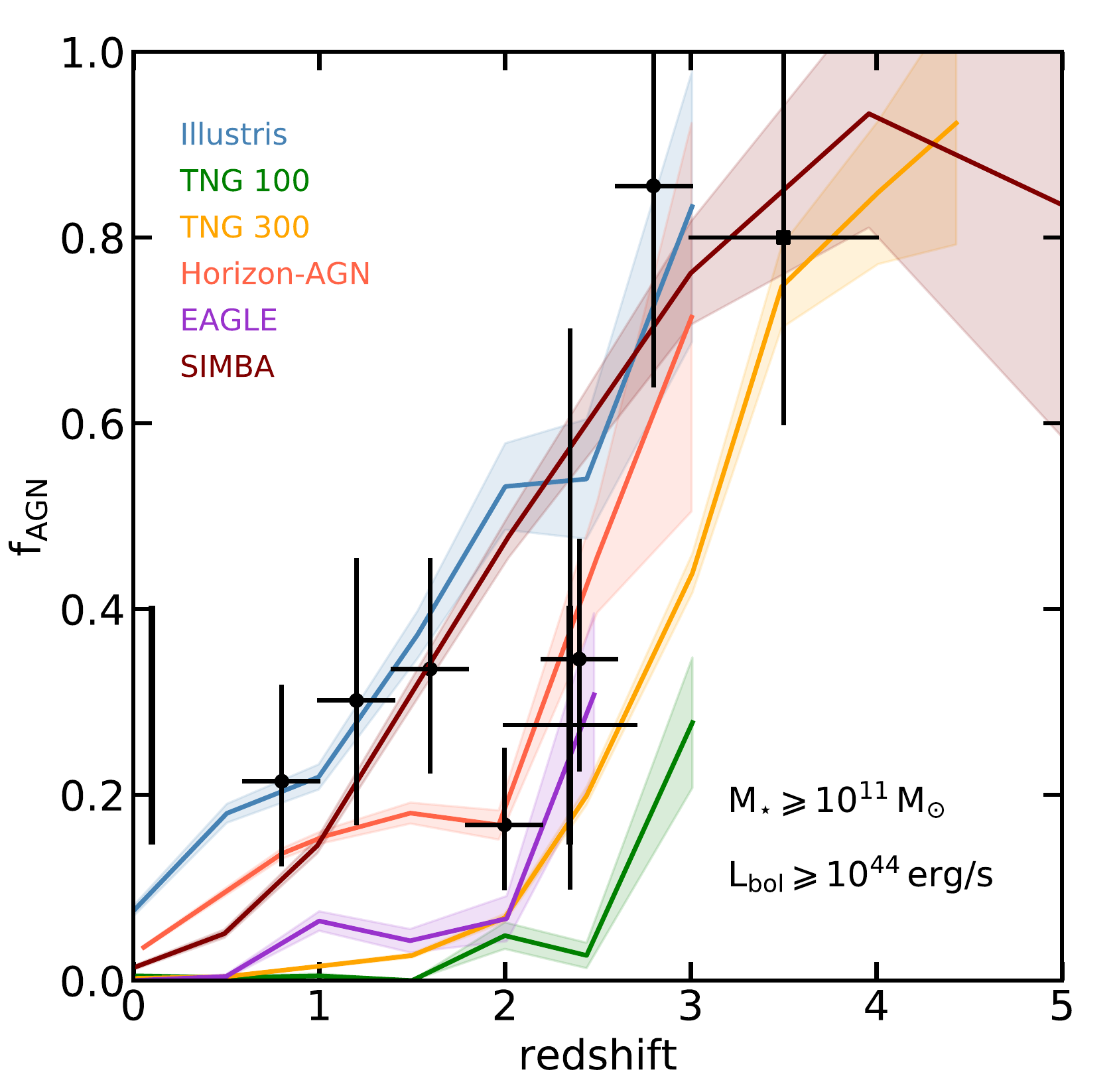}
\includegraphics[scale=0.49]{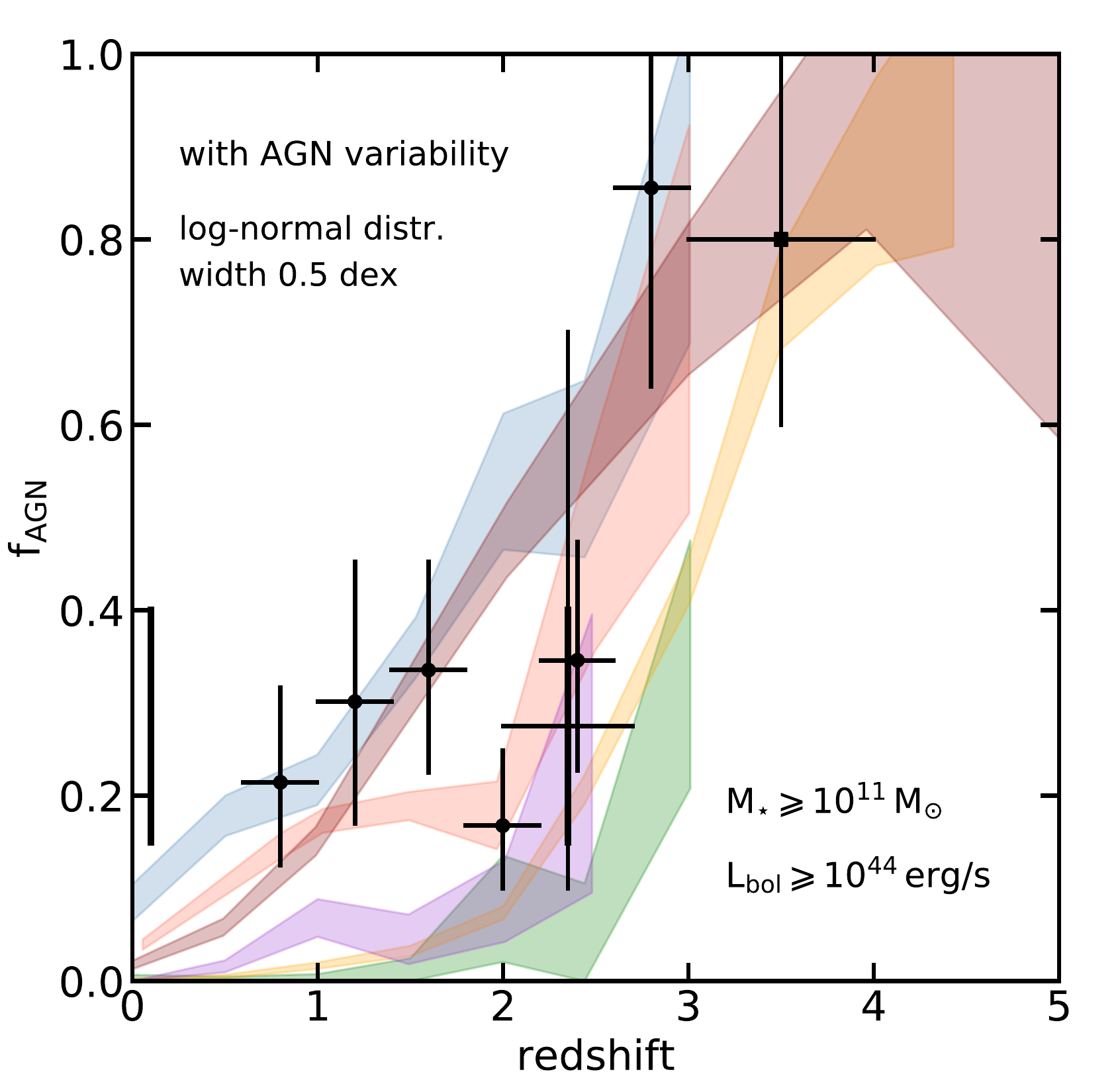}
\caption{{\it Left panel:} Fraction of AGN with $L_{\rm bol}\geqslant 10^{44}\, \rm erg/s$ in massive galaxies of $M_{\star}\geqslant 10^{11}\, \rm M_{\odot}$ (right panels). Only redshift bins with more than 10 galaxies are included.
The regime of massive galaxies is important as we expect that a large fraction of these galaxies should be quiescent. We show in black existing observational constraints for this regime of massive galaxies \citep{2017ApJ...842...21M,2016MNRAS.457..629C,2007ApJ...669..776K,Kauffmann2003}. At high redshift, all the simulations have a large fraction of AGN in massive galaxies, in agreement with observations. However, at lower redshift ($z\leqslant 2$) there is a plateau in observations at $20-40\%$ of the galaxies hosting an AGN, which is not well reproduced in some of the simulations. {\it Right panel:} Effect of AGN variability with a log normal distribution of width 0.5 dex. Shaded regions show the minimum and maximum values of the 15th-85th percentiles of the AGN fraction distributions over several realizations of our AGN variability model.} 
\label{fig:agn_occ_massivegal}
\end{figure*}

\section{Results: AGN fraction in galaxies}
\label{sec:agn_fraction}
\subsection{Time evolution of the galaxy AGN fraction}
In Fig.~\ref{fig:agn_occ_all} (top panels), we show the fraction of AGN with $L_{\rm bol}\geqslant 10^{43}\, \rm erg/s$ in galaxies of different masses as a function of redshift. This limit represents more or less the peak of the AGN bolometric luminosity function (it depends on the simulations and redshift), and allows us to include the AGN that could be detectable by the current available instruments (e.g., Chandra, XMM-Newton). Here, we define the AGN fraction as the number of galaxies hosting an active BH divided by the number of galaxes hosting a BH (active or not); we do not include galaxies which do not host a BH.
To understand how the AGN occupation depends on the host galaxies, we divide the simulated galaxies into three samples with different stellar masses: $M_{\star}\sim 10^{9}\, \rm M_{\odot}$ (solid lines), $M_{\star}\sim 10^{10}\, \rm M_{\odot}$ (dashed lines), and $M_{\star}=10^{11}\, \rm M_{\odot}$ (dotted-dashed lines).
In the bottom panels of Fig.~\ref{fig:agn_occ_all} we instead split the data in BH mass bins: $M_{\rm BH}\sim 10^{6}, \rm M_{\odot}$ (solid lines), $M_{\rm BH}=10^{7}\, \rm M_{\odot}$ (dashed lines), $M_{\rm BH}=10^{8}\, \rm M_{\odot}$ (dashed-dotted lines), and $M_{\rm BH}=10^{9}\, \rm M_{\odot}$ (solid lines).

The fraction of AGN is always higher at high redshifts, for all the galaxy stellar mass bins. The increase with redshift up to $z\sim 3$ was also found in observations for galaxies with $M_{\star}\geqslant 10^{9.5}\, \rm M_{\odot}$ \citep{2018MNRAS.474.1225A}.
The fraction of AGN varies strongly from one simulation to another. In Illustris, TNG100, Horizon-AGN, and SIMBA\footnote{In Fig.~\ref{fig:agn_occ_all}, we show the AGN fractions in galaxies of $M_{\star}\sim 10^{9}\, \rm M_{\odot}$ for SIMBA, but we do not discuss this in the text since the BH seeding generally takes place in galaxies of $M_{\star}\sim 10^{9.5}\, \rm M_{\odot}$ in this simulation.}, all galaxies with $M_{\star}\sim 10^{9}, 10^{10}\, \rm M_{\odot}$ have a probability of $>80\%$ to host an efficient accretor at $z\sim 4$. In EAGLE, the fraction of galaxies hosting an AGN is always lower than the other simulations, as discussed in the following.
With time, the fraction of galaxies hosting efficient accretors decreases. This decrease can be linear with redshift: in Illustris the AGN fractions decrease with the same slope from $z=4$ to $z=0$. We can also identify some different trends in the other simulations.
As discussed with Fig.~\ref{fig:lum_bhmass}, the evolution with time of the median $L_{\rm bol}$ in TNG100, TNG300, and SIMBA, for BHs of $M_{\rm BH}\leqslant 10^{8}\, \rm M_{\odot}$ is mild for $z\geqslant 2$ compared to the evolution in Illustris and Horizon-AGN.
As a consequence, TNG100, TNG300 and SIMBA present a relatively small decrease of the AGN fractions in the redshift range $z=4-2$ for $M_{\star}\geqslant 10^{10}\, \rm M_{\odot}$. After this ($z<2$), the decrease of the AGN fractions is more pronounced in these simulations.

For more massive galaxies of $M_{\star}\sim 10^{11}\, \rm M_{\odot}$ (dashed-dotted lines), the fraction of AGN is even lower in TNG100/TNG300 and SIMBA.  
We find that the strong AGN feedback operating in the massive TNG100/TNG300/SIMBA galaxies self-regulate the BHs and significantly decreases the number of rapid accretors at redshift $z\leqslant 3$. From the bottom panels of Fig.~\ref{fig:agn_occ_all}, we see that these BHs are among the most massive with masses of $M_{\rm BH}\sim 10^{9}\, \rm M_{\odot}$ for TNG100/TNG300 or $M_{\rm BH}\sim 10^{8}\, \rm M_{\odot}$ for SIMBA.
Interestingly, we find higher or similar fractions of AGN with $L_{\rm bol}\geqslant 10^{43}\, \rm erg/s$ in galaxies of $M_{\star}\sim 10^{11}\, \rm M_{\odot}$ (dashed-dotted lines) than of $M_{\star}\sim 10^{10}\, \rm M_{\odot}$ (dashed lines) in the Illustris, Horizon-AGN, and EAGLE simulations. This shows that at least in some simulations, massive galaxies of $M_{\star}\sim 10^{11}\, \rm M_{\odot}$ are statistically capable of feeding AGN, as the $M_{\star}\sim 10^{10}\, \rm M_{\odot}$ galaxies. In these simulations, we do not find strong differences between the fraction of AGN powered by $M_{\rm BH}\sim 10^{8}\, \rm M_{\odot}$ and $M_{\rm BH}\sim 10^{9}\, \rm M_{\odot}$ BHs.
By $z=0$, the Horizon-AGN, EAGLE and SIMBA simulations have AGN fractions of $\leqslant 10\%$ for all the galaxy mass bins presented here. This is the case for the least massive ($M_{\star}\geqslant 10^{9}\, \rm M_{\odot}$) and most massive galaxies ($M_{\star}\geqslant 10^{11}\, \rm M_{\odot}$) of TNG100. However, the TNG100 simulation still has an AGN fraction of $\sim 20\%$ in galaxies of $M_{\star}\geqslant 10^{10}\, \rm M_{\odot}$, which corresponds to an efficient growth phase between low gas content phases due to SN feedback and AGN feedback. We also note that the massive galaxies ($M_{\star}\geqslant 10^{11}\, \rm M_{\odot}$) in Illustris still have a high fraction of AGN ($\sim 20\%$) due to a less efficient AGN feedback.

In EAGLE, the number of AGN is lower than in the other simulations, as shown in Fig.~\ref{fig:LF} and Fig.~\ref{fig:number_density}. The AGN fraction is $<0.4$ at $z=4$, and decreases towards lower redshifts. 
We note that the fraction in low-mass galaxies of $M_{\star}\sim 10^{9}\, \rm M_{\odot}$ in EAGLE is close to zero for all redshifts. These low-mass galaxies host BHs of $M_{\rm BH}\sim 10^{5}-10^{6}\, \rm M_{\odot}$ whose accretion is strongly stunted by SN feedback \citep{2018MNRAS.481.3118M,2017MNRAS.468.3935H,2017MNRAS.472L.109A}.
SN feedback affects the growth of BHs in galaxies with $M_{\star}<10^{10}\, \rm M_{\odot}$ from high redshift to low redshift \citep{2018MNRAS.481.3118M}. 
After the phase of SN regulation, BHs starts growing in mass efficiently. At $z\geqslant 2$, this phase starts in galaxies of $M_{\star}\geqslant 10^{9.5}\, \rm M_{\odot}$ with BHs of $M_{\rm BH}\sim 10^{6}\, \rm M_{\odot}$; these BHs power the $\sim 10\%$ AGN fraction (bottom panel).

We have demonstrated that the fraction of efficient accretors with $L_{\rm bol}\geqslant 10^{43}\, \rm erg/s$ varies in time (differently in all simulations), and depends on the host galaxy stellar mass. At $z\leqslant 2$, these differences in the fractions of galaxies hosting an AGN could help us to constrain the sub-grid physics of the simulations, and particularly the efficiency of AGN feedback. We develop this point in the following subsection.

In observations, there are no clear trends in the duty cycle with galaxy or BH mass. At $z\sim 0$, while \citet{2010A&A...516A..87S} identify a decrease with BH mass, a mild evolution was reported in \citet{2010MNRAS.406..597G}.
More precisely, \citet{2015MNRAS.447.2085S} find almost no evolution with BH mass for type 1 AGN with $M_{\rm BH}=10^{7}-10^{9}\, \rm M_{\odot}$ and $z=1.2-2$. We define AGN with a cut in the bolometric luminosity, while the definition of \citet{2010A&A...516A..87S} is based on Eddington ratios. Nevertheless, we find that most simulations show an evolution of the duty cycle with BH mass in the same mass and redshift range, with the exception of EAGLE and Horizon-AGN which shows very little evolution. \citet{2015MNRAS.447.2085S} identify a strong decrease with BH mass for lower redshift $z<1.2$; this trend is found in most simulations as well.

The increase of the AGN fraction at $z\sim 0$ that we find in all simulations between galaxies with $M_{\star}\sim 10^{9}\, \rm M_{\odot}$ and those with $M_{\star}\sim 10^{10}\, \rm M_{\odot}$ was also found in observations \citep{2019MNRAS.488...89M}. The increase with galaxy stellar mass in the low-redshift Universe was also found in \citet{Aird2012}.

\subsection{AGN fraction in massive galaxies}
The number of AGN in massive galaxies through cosmic time is crucial for assessing the role of AGN feedback in these galaxies.
Observational constraints are based on relatively poor number statistics \citep{2017ApJ...842...21M,2016MNRAS.457..629C,2007ApJ...669..776K}, with samples including 10 or fewer galaxies of $M_{\star}\geqslant 10^{11}\, \rm M_{\odot}$, but they provide us with a first insight into the fraction of AGN in massive galaxies up to $z\sim 3.5$. The sample of \citet{2017ApJ...842...21M} finds an AGN fraction of $\geqslant 80\%$ with $L_{\rm bol}\sim 10^{44}-10^{46}\, \rm erg/s$ in 6 galaxies ($M_{\star}\sim1.5-4\times 10^{11}\, \rm M_{\odot}$,$3<z<4$).
\citet{2016MNRAS.457..629C} also find that $\geqslant 80\%$ of their galaxy sample from zFOURGE host an AGN in the range $2.6<z<3.2$.
At lower redshift, \citet{2007ApJ...669..776K} study a sample of 11 galaxies ($M_{\star}\sim 3\times 10^{11}\, \rm M_{\odot}$, $2<z<2.7$) with some of them hosting an AGN of $L_{\rm bol}\sim 10^{44}\, \rm erg/s$ or evidence for narrow-line emission, and therefore, find an AGN fraction of $\sim 20\%$. 
We also report the estimates of $\sim 20-40\%$ of \citet{2016MNRAS.457..629C} at lower redshifts ($0.5\leqslant 2.5$).
We reproduced all these different constraints as black crosses in Fig.~\ref{fig:agn_occ_massivegal}. About $80\% $ of the massive galaxies host an AGN at high redshift ($z\geqslant 2.5$). At $z<2.5$, the observational constraints on the fraction of AGN varies in the range $20-40\%$ \citep[SDSS data,][]{Kauffmann2003}.

We show in Fig.~\ref{fig:agn_occ_massivegal} the fraction of simulated AGN with $L_{\rm bol}\geqslant 10^{44}\, \rm erg/s$ in massive galaxies of $M_{\star}\geqslant 10^{11}\, \rm M_{\odot}$. 
Simulations of $\sim 100 \, \rm cMpc$ side length start forming galaxies of $\sim 10^{11}\, \rm M_{\odot}$ only at $z\sim 3-4$. The simulation TNG300 (and SIMBA) with its larger volume allows us to investigate the fraction of AGN in massive galaxies at much earlier times \citep{2019MNRAS.484.4413H}.  
We find the same overall trend in all the simulations: a high fraction of massive galaxies host an AGN at high redshift, and the fraction decreases toward lower redshift. While this trend is in good agreement with the observations, the AGN fractions found in simulations can vary substantially. The difference in the fractions for the Illustris and TNG simulations reaches up to $50\%$ in the redshift range $z=3-2$, for example.
In general, we see here that in the simulations there is no consensus on the possible sharp decrease of the AGN fraction at $z\sim 2.5$ found in the observations.
There is also no consensus on the fraction of AGN found in these massive galaxies, even at relatively low redshifts $z\leqslant 2$. Some of the simulations produce very low fractions of AGN for these redshifts, which could indicate a too efficient self-regulation of these BHs by their AGN feedback.\\

The AGN fractions strongly depend on the model that we use to compute the AGN bolometric luminosity; i.e. whether we assume that all AGN are radiatively efficient or not, as well as on the radiative efficiency $\epsilon_{\rm r}$. To illustrate this we show in Fig.~\ref{fig:agn_occ_massivegal_bis} the same figure Fig.~\ref{fig:agn_occ_massivegal} but assuming the same radiative efficiency of $\epsilon_{\rm r}=0.1$ for all the simulations, instead of $\epsilon_{\rm r}=0.2$ and that all AGN are radiatively efficient \citep[as assumed in many analyses of simulations, but however disfavored by observations in massive galaxies whose BHs often lie in the range $f_{\rm Edd}=0.01-0.1$,][]{2013MNRAS.432..530R}. The latter results in an enhancement of the fraction of AGN in all the simulations, leading to a better agreement with observations for $z<2$.

We show in Fig.~\ref{fig:agn_occ_massivegal} (right panel) the impact of AGN variability on the fraction of AGN in massive galaxies. AGN variability broaden the range of possible values of the AGN fraction, including at $z<1$ when several simulations have very low fractions, but does not affect the conclusions of this section.

\section{Predictions for upcoming or planned X-ray missions}
\label{sec:predictions}
In this section, we predict the number of AGN from the six large-scale cosmological simulations that would be detectable by the upcoming Athena mission \citep{2013arXiv1306.2307N}, and the AXIS \citep[][and references therein]{2020arXiv200809133M} and LynX \citep{2018arXiv180909642T} planned missions.
In the previous sections, we showed that the different simulations all predict different populations of AGN, not always in perfect agreement with current observational constraints. 
However, the predictions that we derive below are important as they cover the broad range of subgrid modeling (BH seeding, accretion, feedback, galaxy physics) employed in simulations.

\begin{figure}
\centering
\hspace*{-0.2cm}
\includegraphics[scale=0.55]{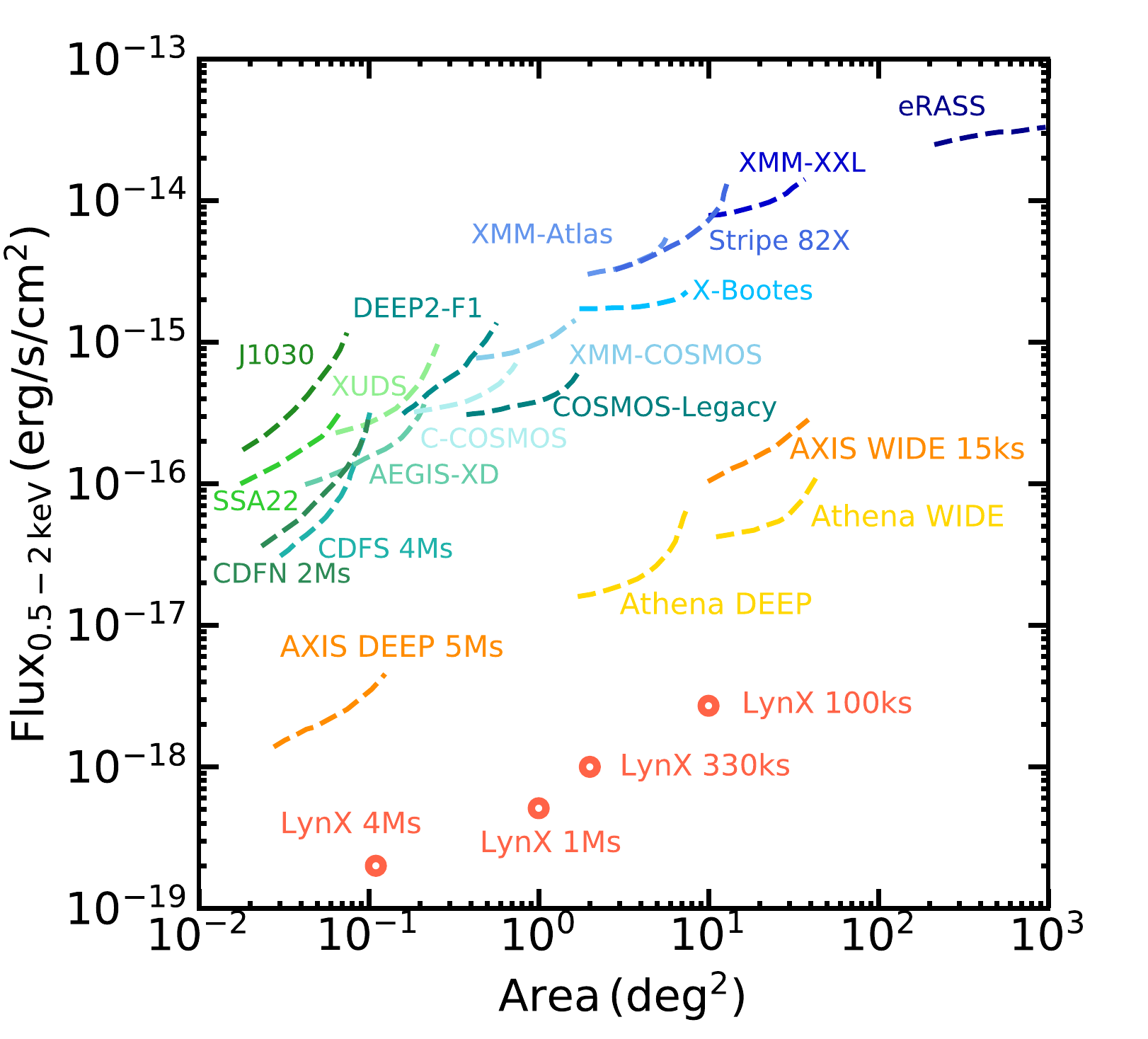}
\caption{Sensitivity curves of current X-ray missions (green to blue colors). For all surveys, the curves represent the sensitivity within the $20-80\%$ total survey area. 
We show the curve of the new Athena mission in yellow, and of the AXIS and LynX concept missions in orange and red, respectively.}
\label{fig:sensitivity}
\end{figure}

\subsection{The landscape of X-ray surveys, and AGN detection in new X-ray missions}
Athena, AXIS and LynX all have different luminosity thresholds to detect AGN.
Their sensitivity curves express the flux that can be reached by the observations of a given instrument as a function of the sky area covered by the survey. We show the 0.5-2 keV sensitivity curves of Athena, AXIS \citep{2020arXiv200809133M} and LynX (private communication with Alexey Vikhlinin and Niel Brandt), in Fig.~\ref{fig:sensitivity}. 
Athena, AXIS, and LynX will/would increase by one order of magnitude the sensitivity in the X-ray band at fixed area compared to previous surveys\footnote{The sensitivity curves for previous X-ray missions are taken from \citet{2016ApJ...819...62C,2020arXiv200809133M}} (in Fig.~\ref{fig:sensitivity}, green to blue colors).
The sensitivity of a given survey depends on its exposure time and size.
At fixed parameters, the sensitivity of LynX should be more than 3 times better than the sensitivity of AXIS. Indeed, while having the same telescope design, the AXIS effective area is 3.3 times smaller than LynX.

To predict the population of AGN that could be detected by Athena, AXIS, and LynX, we define several possible surveys. All have been already discussed by the different mission teams and/or in the literature. The parameters of these surveys are all different (i.e., in terms of survey size, exposure time, sensitivity), and thus, our predictions for the missions below {\it cannot} be compared with one another. Predictions for any other survey parameters are available upon request.
We use two different surveys for Athena, as described in \citet{2013arXiv1306.2307N}, two possible surveys for LynX (private communication with LynX researchers), and we follow the papers of \citet{2019BAAS...51g.107M} and \citet{2020arXiv200809133M} to define two surveys for AXIS. We report the parameters of these surveys in Table~\ref{table:missions}. We convert the 0.5-2 keV sensitivity curves into 2-10 keV luminosity detection limits, as explained in Appendix~\ref{sec:lum_limit}. 
In general, LynX should detect AGN fainter by one order of magnitude than AXIS, at all redshifts. The WIDE field of AXIS could reach a similar sensitivity of the WIDE survey of Athena, but for reduced exposure time (e.g., $15\, \rm ks$ instead of $90\, \rm ks$).

\begin{table}
\caption{Characteristics of several surveys for Athena, AXIS, and LynX: covered area, flux sensitivity (0.5-2 keV band) that could be reached, and exposure time.}
\begin{center}
\begin{tabular}{llll}
\hline
Surveys & Area ($\rm deg^{2}$) & $F$ (erg/s/cm$^{2}$) & Exposure \\
\hline
Athena DEEP & 5.28 & $1.6-6.5 \times 10^{-17}$ & 1 Ms \\
Athena WIDE & 47.52 & $4.2-11. \times 10^{-17}$ & 90 ks\\
\hline
LynX DEEP & 0.11 & $2.0\times10^{-19}$ & 4 Ms\\
LynX WIDE & 10 & $2.0\times10^{-18}$  & 100 ks \\
\hline
AXIS DEEP & 0.16 & $1.4-4.5 \times 10^{-18}$ & 5 Ms\\
AXIS WIDE & 50 & $1.0-2.8 \times 10^{-16}$ & 15 ks\\
\hline
\end {tabular}
\end{center}
\label{table:missions}
\end{table}

\begin{figure*}
\centering
\includegraphics[scale=0.58]{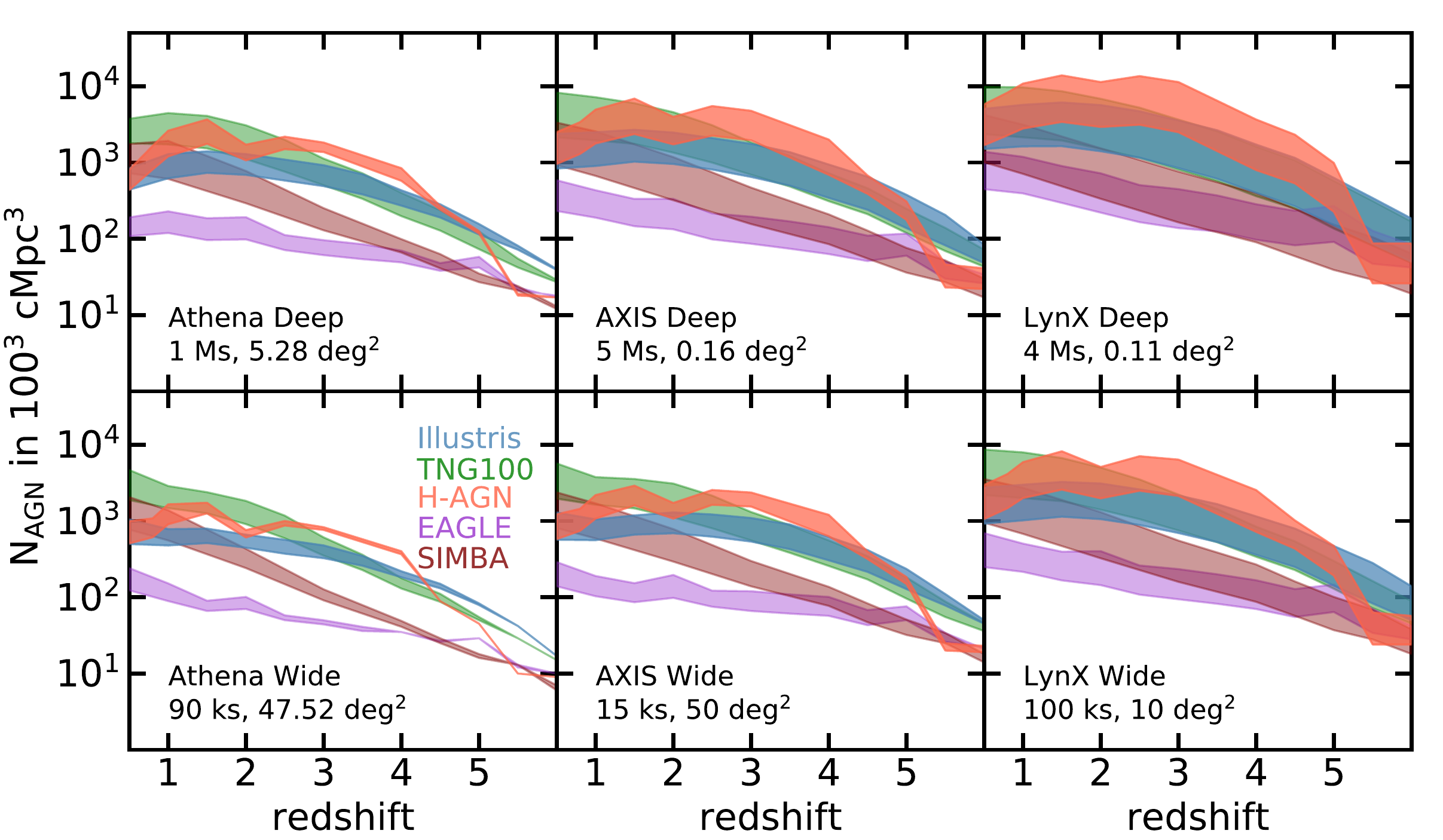}
\includegraphics[scale=0.58]{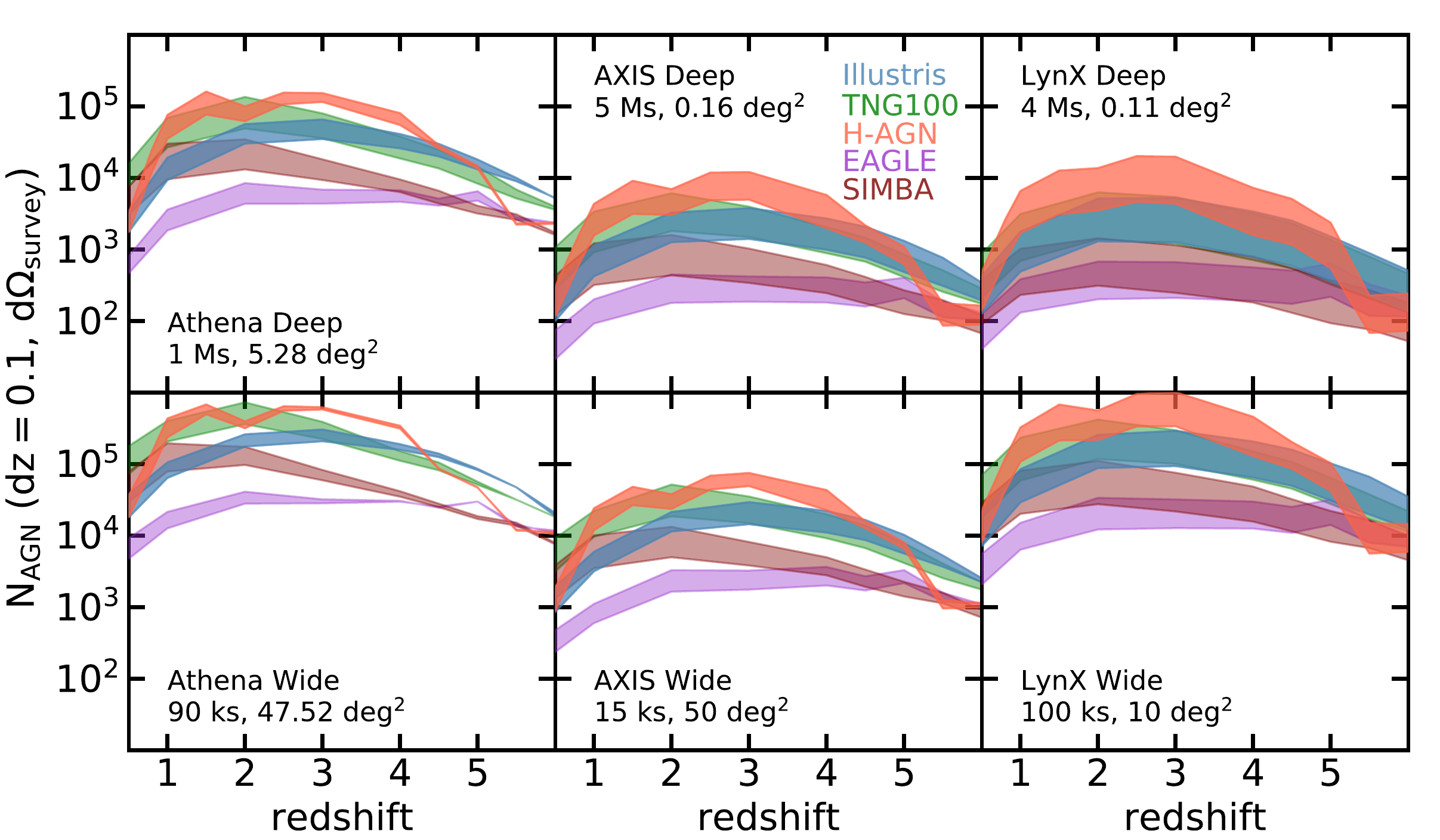}
\caption{{\it Top panel:} Number of AGN detections per $\rm (100 cMpc)^{3}$ in galaxies of $M_{\star}\geqslant 10^{9}\, \rm M_{\odot}$ for the sensitivity of a couple of possible Athena, AXIS, and LynX surveys. The number of AGN are not comparable between the different surveys, as we show different exposure times and sizes of surveys. 
The shaded areas bracket the number of detections per $\rm (100 cMpc)^{3}$ when corrected for obscured AGN (our second AGN luminosity- and redshift-dependent model) when we remove the AGN 
(lower edge of the shaded region) and when we decrease their luminosity (upper edge).
The simulations predict different number of detectable AGN. At low redshift ($z<2$), EAGLE predicts a number of AGN one order of magnitude lower than the other simulations. At high redshift ($z>4$), Illustris and TNG100 predict more AGN than the other simulations. 
{\it Bottom panel:} Number of detectable AGN in the field of view of the given Athena, AXIS, and LynX surveys, and for redshift slices of ${\rm d}z=0.1$. }
\label{fig:number_AGN_100cMpc_redshift}
\end{figure*}

\subsection{Predictions for the number of detectable AGN for the different simulations}
We show the number of detectable AGN $\rm N_{\rm AGN}$ in a volume of $\rm (100 cMpc)^{3}$ in Fig.~\ref{fig:number_AGN_100cMpc_redshift} (top panel), for all the surveys described in Table~\ref{table:missions}. We also provide similar predictions but presented as ${\rm N_{AGN}}/{\rm d}z/{\rm d}\Omega_{\rm survey}$ in Fig.~\ref{fig:number_AGN_100cMpc_redshift} (bottom panel); i.e. the number of detectable AGN per slice of redshift and for the field of view of the different surveys.
The shaded regions bracket, for each simulation, the number of detections when corrected for obscured AGN with our second AGN luminosity- and redshift-dependent obscuration model (see Fig.~\ref{fig:obscuration_model}). With this model, we either remove the AGN from the samples (lower edges of the shaded regions in Fig.~\ref{fig:number_AGN_100cMpc_redshift}) or we decrease their hard X-ray luminosity by one order of magnitude (upper limits). Other models are tested in Appendix~\ref{sec:obscuration}.
We precise here that the impact of obscuration in the observed 0.5-2 keV band is redshift dependent, and also that the missions will have 2-10 keV sensitivity, which is less affected by obscuration.

We provide in Table~\ref{tab:best_fit} the best-fit for the number density of AGN per $\rm (100 cMpc)^{3}$ shown in Fig.~\ref{fig:number_AGN_100cMpc_redshift} for the second AGN luminosity- and redshift-dependent obscuration model. 
These best-fits can be used to prepare the future X-ray missions; i.e., to bracket how many AGN could be detected, and investigate the optimal size of the mission surveys.

\begin{figure*}
\centering
\includegraphics[scale=1.133]{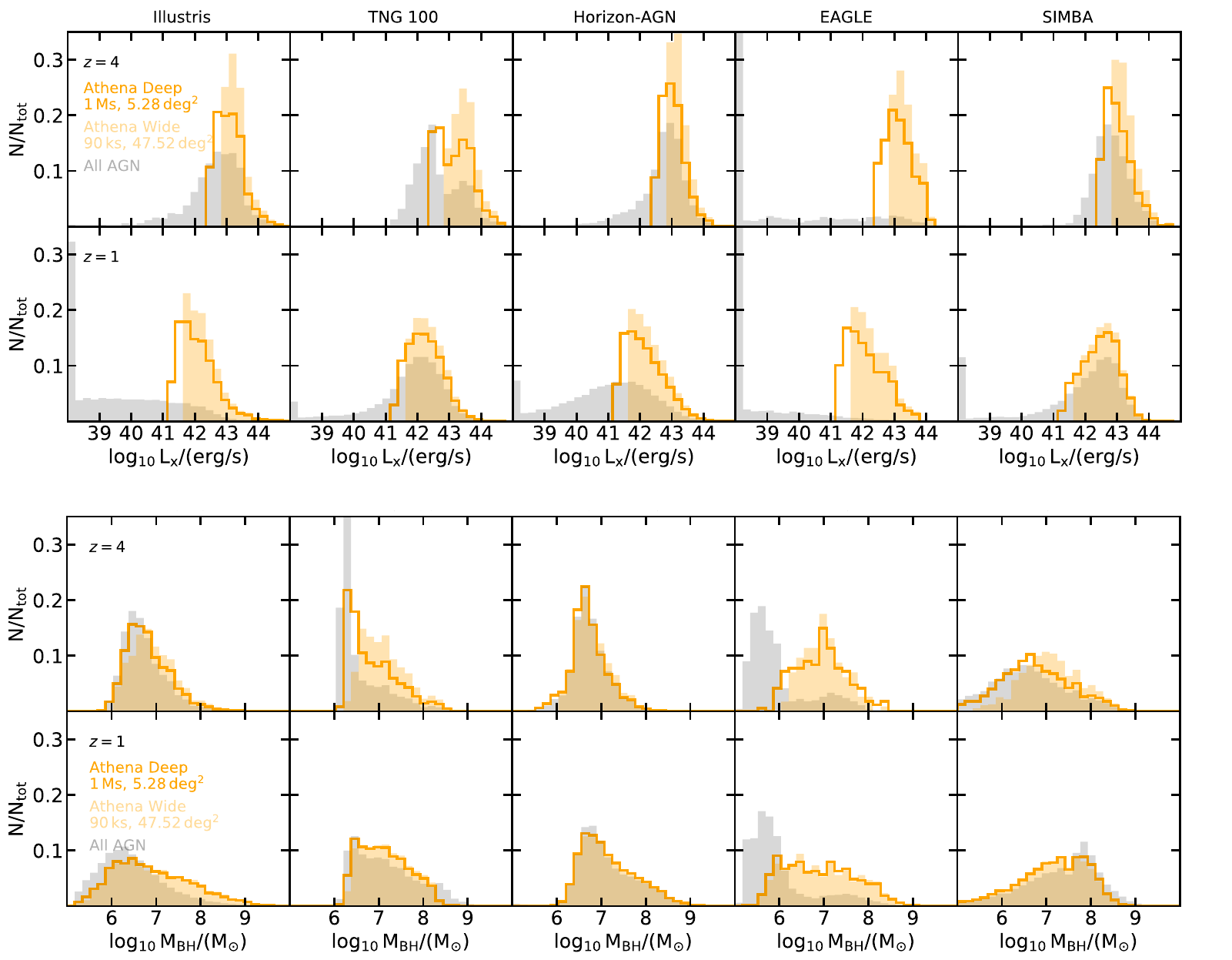}
\caption{Athena deep (1 Ms of exposure time, 5.28 $\rm deg^{2}$) and wide (90 ks, 47.52 $\rm deg^{2}$) surveys. {\it Top panels:} 
Normalized distributions of the hard (2-10 keV) X-ray luminosity of the AGN that are detectable. For reference, we show in grey the distribution of all the AGN with $L_{\rm x}\geqslant 10^{38}\, \rm erg/s$ in the simulations.
{\it Bottom panels}: Normalized distributions of BH mass corresponding to the detectable AGN.
We only include galaxies with $M_{\star}\geqslant 10^{9}\, \rm M_{\odot}$ in all the panels.}
\label{fig:summary_athena}
\end{figure*}

The number density of detections varies strongly from one simulation to another. At $z\geqslant 5$, the fewest AGN would be detected in Horizon-AGN and SIMBA, with between 20 to 150 detections per $\rm (100 \, cMpc)^{3}$ at the sensitivity of the LynX Deep and Wide surveys, about 20--80 for the AXIS Deep and Wide surveys, and from less than 10 to 40 for Athena surveys. 
Illustris and TNG100 predict more AGN to be detected at the same redshifts, with e.g., 140--180 detections per $\rm (100 \, cMpc)^{3}$ with LynX Deep and Wide surveys, 50-90 with AXIS surveys, and 20-40 with the Athena surveys. The number of AGN that could be detected increases with decreasing redshift until $z\sim 2$, at which point the number of AGN stabilizes or decreases (e.g., for Horizon-AGN, Illustris). At $z=1$, we find more similar predictions for the total number of AGN detectable with the missions for Horizon-AGN, Illustris, and SIMBA. TNG100 is the simulation predicting the highest number of AGN to be uncovered.

When considering the entire AGN population, independently of their BH mass or galaxy stellar mass, the Athena, AXIS, and LynX missions will be able to constrain the AGN population produced by cosmological simulations.
More precisely, there is more than an order magnitude of difference in the number of detectable AGN in the different simulations. Thus, being able to detect fainter AGN in new surveys will be crucial to discriminate between simulation sub-grid models. The shape of the number of detections with redshift also varies from one simulation to another, and it could also be used to constrain the modeling.
Depending on the simulated population of AGN, the impact of obscuration can sometimes be significant (large shaded areas in Fig.~\ref{fig:number_AGN_100cMpc_redshift}). This is a large source of uncertainty when comparing simulations to observations, and unfortunately obscuration and the intrinsic number of AGN produced by the simulations are degenerate.

\begin{figure*}
\centering
\includegraphics[scale=1.133]{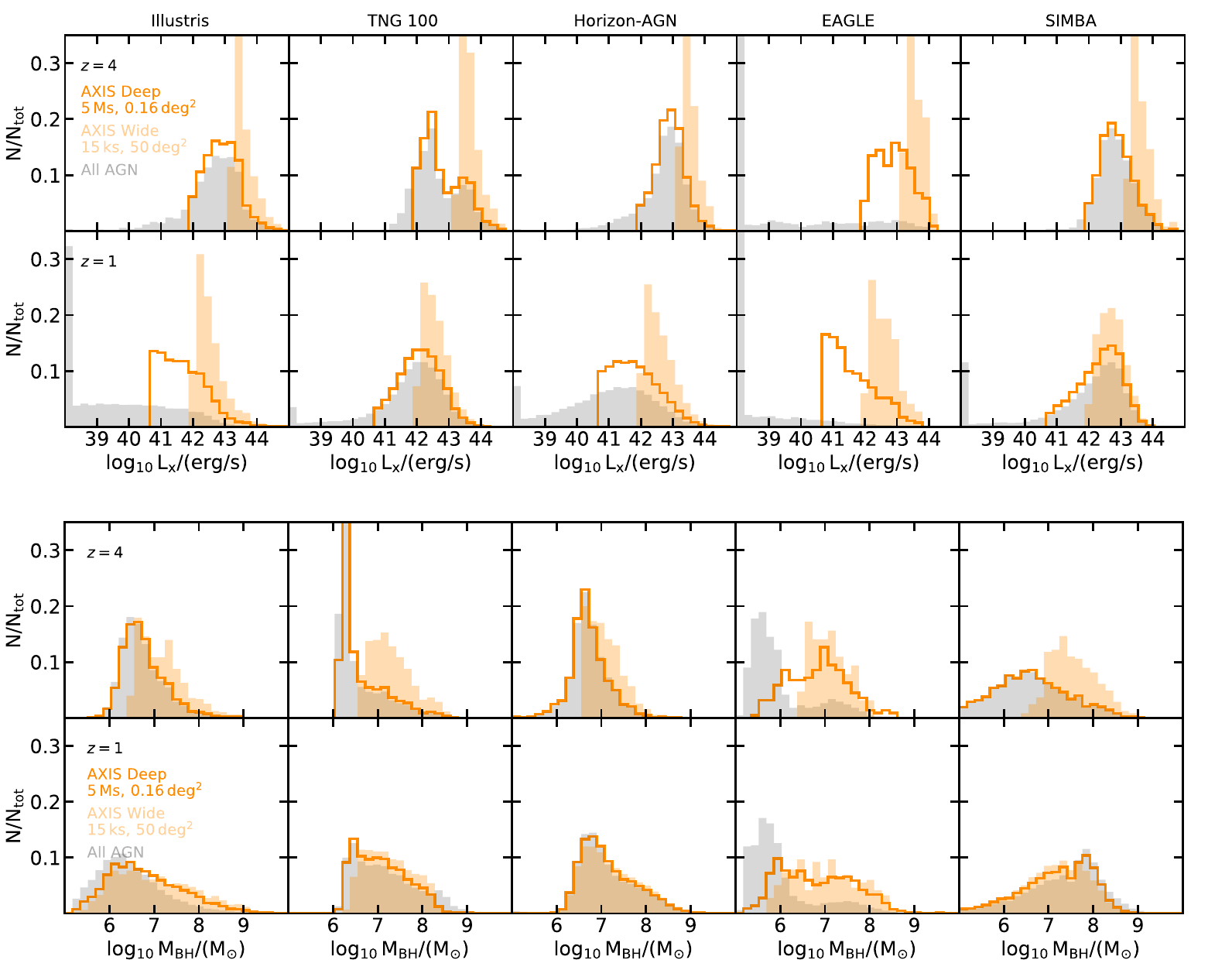}
\caption{AXIS deep (5 Ms of exposure time, 0.16 $\rm deg^{2}$) and wide (15 ks, 50 $\rm deg^{2}$) surveys. {\it Top panels:} 
Normalized distributions of the hard (2-10 keV) X-ray luminosity of the AGN that are detectable. 
Distributions of all the AGN with $L_{\rm x}\geqslant 10^{38}\, \rm erg/s$ is shown in grey.
{\it Bottom panels}: Normalized distributions of BH mass corresponding to the detectable AGN.
Only $\geqslant 10^{9}\, \rm M_{\odot}$ galaxies are included.}
\label{fig:summary_axis}
\end{figure*}

\begin{figure*}
\centering
\includegraphics[scale=1.133]{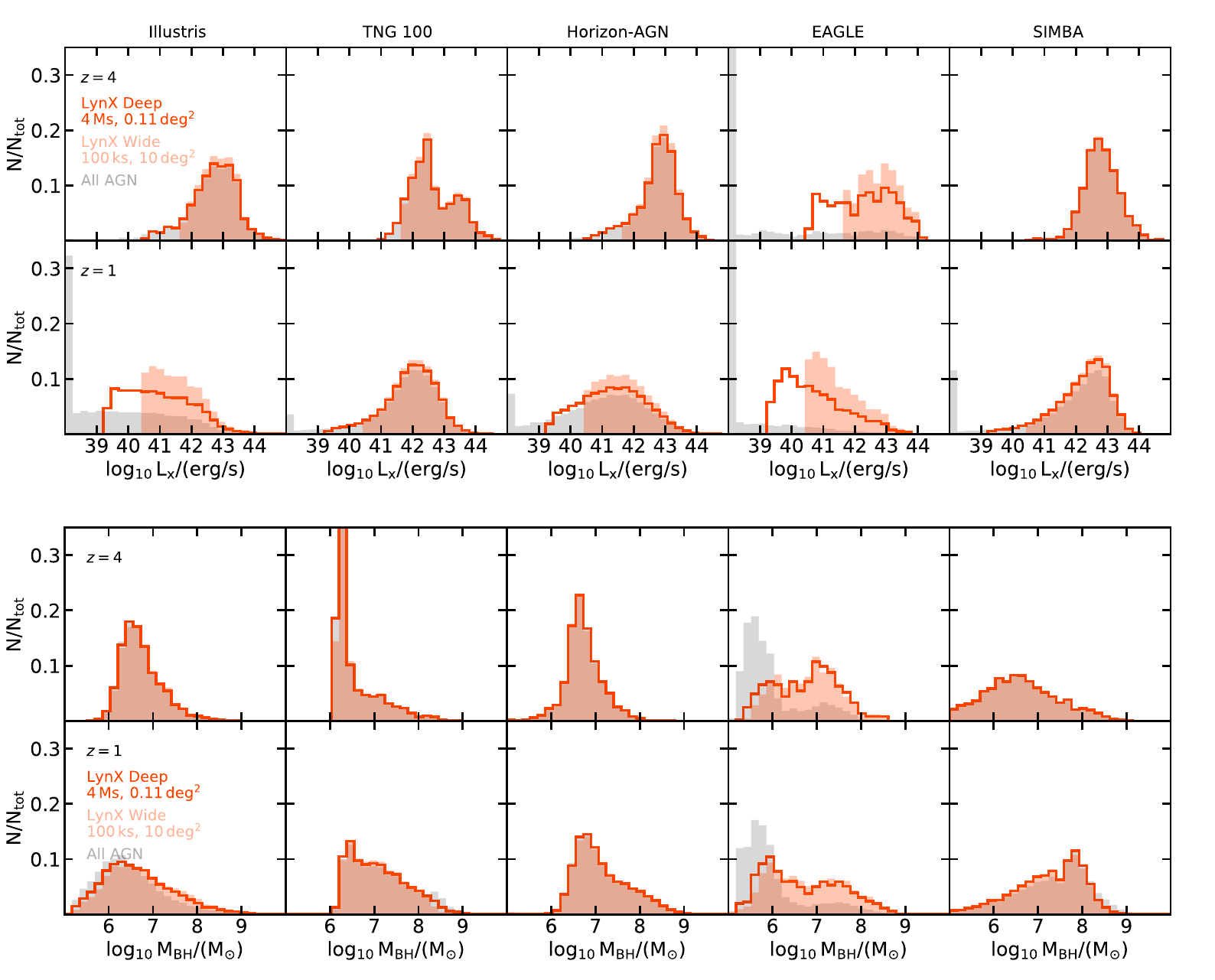}
\caption{LynX deep (4 Ms of exposure time, 0.11 $\rm deg^{2}$) and wide (100 ks, 10 $\rm deg^{2}$) surveys. {\it Top panels:} 
Normalized distributions of the hard (2-10 keV) X-ray luminosity of the AGN that are detectable. Distributions of all the AGN with $L_{\rm x}\geqslant 10^{38}\, \rm erg/s$ is shown in grey.
{\it Bottom panels}: Normalized distributions of BH mass corresponding to the detectable AGN.
Only $\geqslant 10^{9}\, \rm M_{\odot}$ galaxies are included. The high sensitivity of LynX could make accessible almost the full $L_{\rm x}$ and $M_{\rm BH}$ distributions predicted by the Illustris, TNG100, Horizon-AGN, and SIMBA simulations. At $z\leqslant 1$, there are more and more quiescent BHs with $L_{\rm x}<L_{\rm x,\, LynX}$ that will be missed. In EAGLE, the $L_{\rm x}$ and $M_{\rm BH}$ distributions that could be obtained with LynX (even with a very large survey) would still not have the same shape as the EAGLE BH population. This is because of the significant number of faint and quiescent AGN at all times in EAGLE. }
\label{fig:summary_lynx}
\end{figure*}

\subsection{Populations of AGN and BHs to be uncovered by the new X-ray observatories}
The sensitivity of Athena, AXIS, and LynX being different, these missions will have access to different populations of AGN and BHs, as shown in Fig.~\ref{fig:summary_athena}, Fig.~\ref{fig:summary_axis}, and Fig.~\ref{fig:summary_lynx}. 
These figures show the distributions of AGN hard X-ray luminosity of the detectable AGN, and the corresponding BH mass distributions (although not measurable by the X-ray missions) for $z=4$ and $z=1$. 
Deep surveys (higher sensitivity) are shown with the darkest color, and the wide surveys (lower sensitivity) with the lightest color and shaded histograms.
For reference, we show in grey in all the panels the distribution of the intrinsic AGN population produced by the simulations.  
The sensitivity of the Athena wide survey of $50\, \rm deg^{2}$ will capture AGN with $L_{\rm x}\geqslant 10^{43}\, \rm erg/s$ at high redshift, and AGN with $L_{\rm x}\geqslant 10^{42}\, \rm erg/s$ at low redshift.
In the TNG100 simulation at $z=4$, Athena would see the peak of the $L_{\rm x}$ luminosity distribution corresponding to the efficient accretors, that are mostly powered by $\leqslant 10^{8}\, \rm M_{\odot}$ BHs in TNG100. However, the Athena wide survey would not detect any AGN of the fainter peak of the luminosity distribution. In TNG100, these AGN are powered by massive BHs with $M_{\rm BH}\geqslant \rm a \, few \, 10^{8}\, \rm M_{\odot}$ entering in the kinetic mode of AGN feedback, and responsible for regulating themselves (see bimodality in the Eddington ratio distribution of TNG100, Fig.~\ref{fig:fedd_ratio}) and quenching their host galaxies. 
In EAGLE, a significant population of the BHs are not efficient accretors, and have low luminosities. While Athena will go deeper than the current X-ray missions, it will be insufficient to detect most of the AGN population in EAGLE (see the mismatch of the grey and yellow $L_{\rm x}$ and $M_{\rm BH}$ distributions in Fig.~\ref{fig:summary_athena}). The mission would see the AGN powered by BHs of $M_{\rm BH}\sim 10^{6-8.5}\, \rm M_{\odot}$ at $z=4$, but not the lower-mass BHs which constitute most of the BH population in EAGLE.
In some simulations such as TNG100, the Athena deep survey will start uncovering the faint regime of the AGN, and provide us with a distribution of BH masses more similar to the intrinsic distribution produced by the simulations. However, it will not be sufficient to access the full spectrum of the AGN population.

The AXIS mission will have a higher sensitivity than Athena. We find that the AXIS wide survey and the Athena deep survey (as we defined them) provide similar results.
Now, looking at the deep AXIS survey in Fig.~\ref{fig:summary_axis}, we see that in theory it would provide us with distributions of $L_{\rm x}$ and $M_{\rm BH}$ very consistent with the simulation intrinsic distributions. If the Universe hosts an AGN population similar to the EAGLE simulation, i.e. with globally fainter AGN than the other simulations, we would still miss a significant fraction of the AGN population with such a deep AXIS survey.

LynX will have the highest sensitivity, orders of magnitude better than the current X-ray facilities, and about one order of magnitude higher than Athena. In Fig.~\ref{fig:summary_lynx}, we find that the LynX mission will indeed probe much fainter AGN, with e.g., $L_{\rm x}\geqslant 10^{41.5}\, \rm erg/s$ at $z=4$, and $L_{\rm x}\geqslant 10^{40.5}\, \rm erg/s$ at $z=1$, for the wide survey that we defined. The deep survey should access even fainter AGN, with e.g., $L_{\rm x}\geqslant 10^{40.5}\, \rm erg/s$ at $z=4$, and $L_{\rm x}\geqslant 10^{39}\, \rm erg/s$ at $z=1$. We find that with LynX the distributions of AGN luminosities (and corresponding BH masses) would be representative of the intrinsic simulated population of BHs, except for EAGLE. 
The observed distribution at high redshift would highlight a relatively more significant population of massive BHs while in reality the population of lower-mass BHs would be larger in EAGLE.

\begin{figure*}
\centering
\includegraphics[scale=0.45]{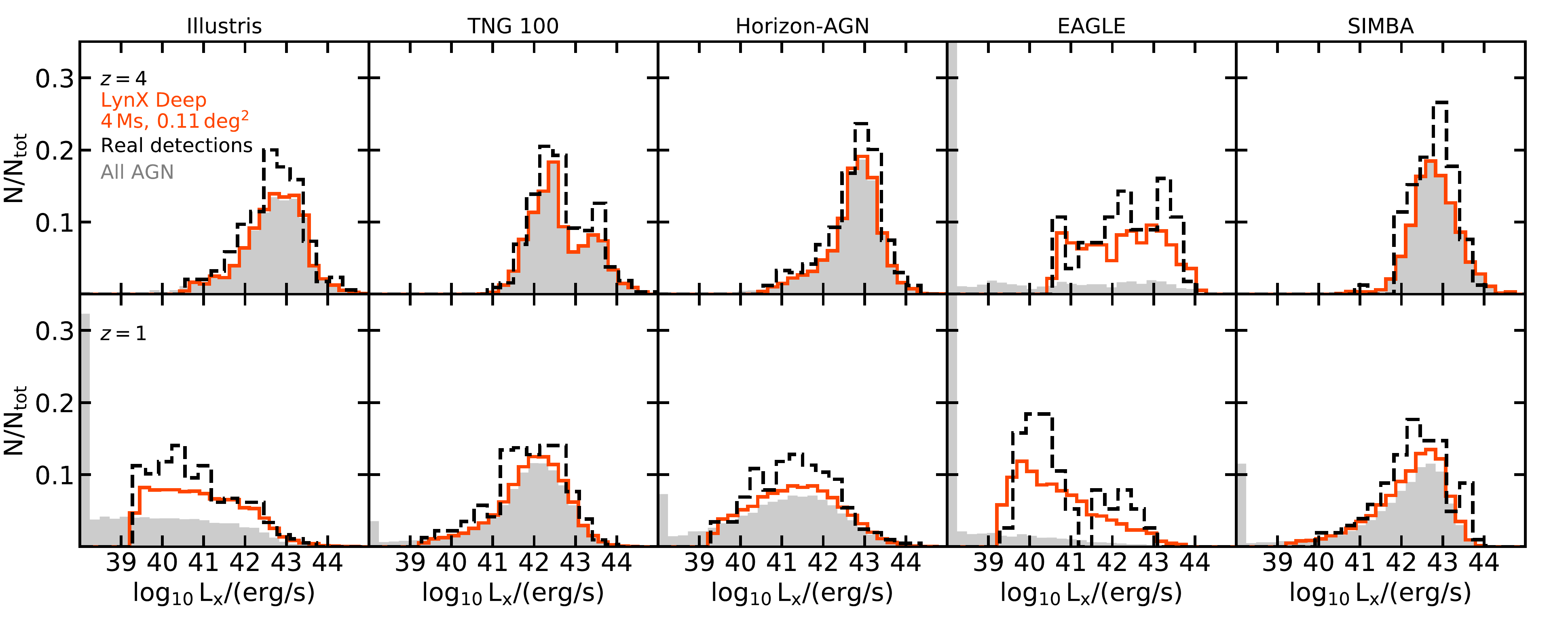}
\hspace*{-0.28cm}
\includegraphics[scale=0.45]{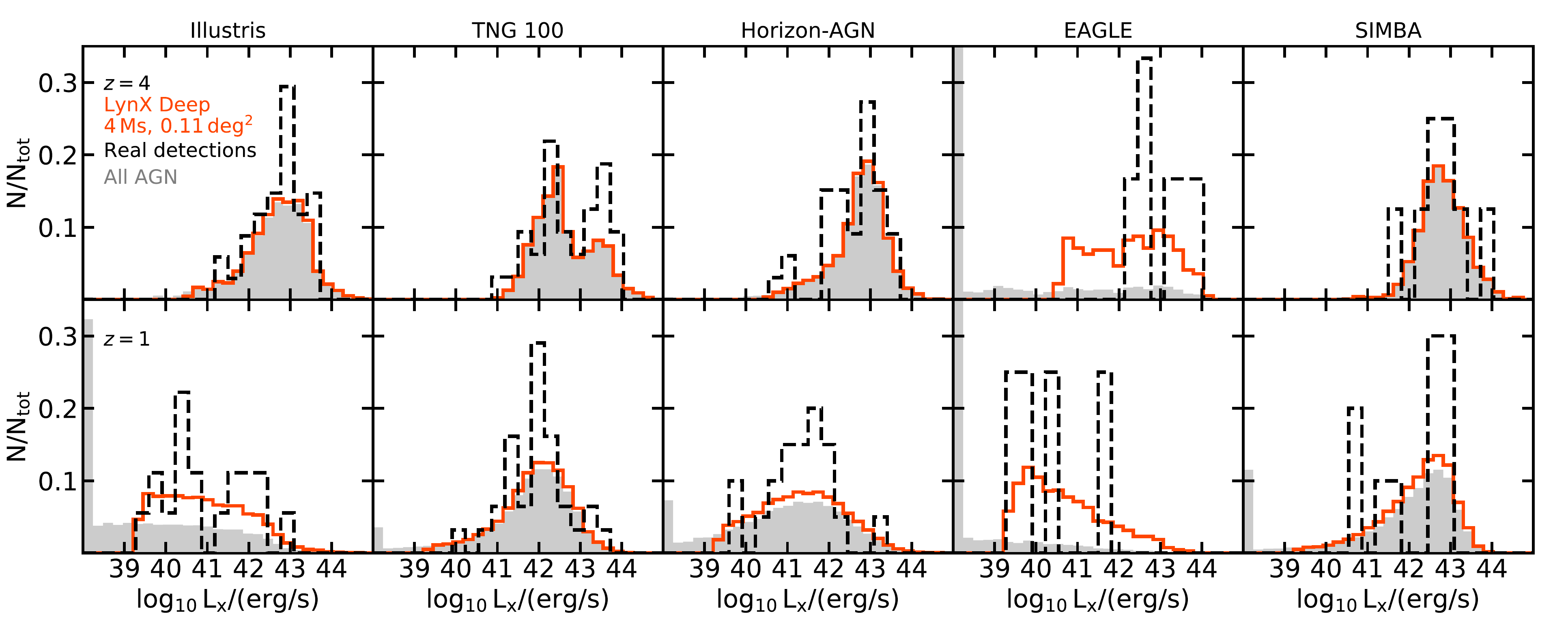}
\caption{Illustration of how small fields of view would affect our understanding of the AGN $L_{\rm x}$ distribution. We show in grey the intrinsic AGN population produced by simulations, in red the population that could be detected by the deep survey of LynX (simply assuming $L_{\rm AGN}\geqslant L_{\rm limit, survey}$) in each simulation. In black we show how the distributions would look like with the real number of AGN detections in a deep field of view of the LynX mission, assuming a redshift accuracy of either ${\rm d}z=0.01$ (top panels) or ${\rm d}z=0.001$ (bottom panels).}
\label{fig:summary_lynx_smallfield}
\end{figure*}

\subsection{Will small fields of view allow for enough detections ?}
In Fig.~\ref{fig:summary_athena}, Fig.~\ref{fig:summary_axis}, and Fig.~\ref{fig:summary_lynx}, we showed the $L_{\rm x}$ and $M_{\rm BH}$ distributions that could be accessible by the different surveys (simply selecting all AGN with $L_{\rm x}\geqslant L_{\rm limit \, survey}$). The number of AGN detections depends on the size of the surveys, and for a very small field of view if for example only 10 AGN are detected, the obtained distributions would not be representative of the intrinsic distributions. To investigate this, we randomly select from our samples the number of AGN that would be observable for the different surveys (as shown in Fig.~\ref{fig:number_AGN_100cMpc_redshift}), i.e., for their fields of view $\Omega_{\rm survey}$, given redshift slices, and including the effect of obscuration. 
We test three different redshift slices of ${\rm d}z=0.1, 0.01, 0.001$. ${\rm d}z=0.1$ represents the uncertainty of redshift estimate of the host galaxies at high redshift, and the two other ${\rm d}z$ are purposely smaller to be conservative. Measuring redshifts with an uncertainty of ${\rm d}z=0.01$ is difficult but possible. Redshift accuracy of ${\rm d}z=0.001$ will be extremely difficult for a large number of sources, especially for faint objects.

%\subsubsection{Wide surveys} 
For all the wide Athena, AXIS and LynX surveys used here, and for ${\rm d}z=0.01$ and $z=1-4$, there would be enough AGN detections in the survey's $\Omega_{\rm survey}$ to recover the $L_{\rm x}$ and $M_{\rm BH}$ distributions, for all the simulations. This is with the exception of EAGLE at low redshift, e.g. $z=1$, for which the low number of detections may not allow us to completely recover the shape of the intrinsic $L_{\rm x}$ distribution for the small slices of redshifts.

%\subsubsection{Deep surveys}
The intrinsic $L_{\rm x}$ distributions of the simulations can also be recovered for the deep field surveys, which have small fields of view, in large redshift slices of ${\rm d}z=0.1$ and ${\rm d}z=0.01$.
More precisely, we find that the deep field of Athena ($\rm 5.28\, deg^{2}$) provides a sufficient number of detections to recover the intrinsic distributions even for ${\rm d}z=0.001$. 
For the AXIS deep survey, the $L_{\rm x}$ and $M_{\rm BH}$ distributions can be recovered for ${\rm d}z\geqslant 0.01$ for all the simulations except EAGLE.
We find similar results for the deep field of LynX. We show the impact of a deep survey of $0.11 \,\rm deg^{2}$ LynX in Fig.~\ref{fig:summary_lynx_smallfield} with in black the distributions that we would get from the actual detections of the survey.
In the case of ${\rm d}z=0.001$, i.e., if very high precision is obtained for the host galaxy redshift, it will be difficult to recover the intrinsic distributions predicted by any of the simulations at a given redshift with the few detections (see Fig.~\ref{fig:summary_lynx_smallfield}, bottom panels).

In general, these new upcoming and concept missions will have the power to provide us with a unique view of the distribution of BHs, because they will be able to observe fainter AGN than current X-ray facilities. If the surveys cover a sufficiently large field of view, they will probe BH populations that we could compare with results from cosmological simulations. Some of the simulations have sufficiently different distributions of AGN luminosity that we could be able to constrain sub-grid models of BH physics.

\section{Discussion}

\subsection{Interpretation of the differences between the number of AGN in simulations and in observations}
Throughout this paper we have demonstrated that some simulations overproduce, while some underestimate, the number of AGN with respect to observational constraints, particularly for $z\geqslant 1$. Combining all the diagnostics from this paper, i.e. the AGN luminosity function (Fig.~\ref{fig:LF}), the Eddington ratio distribution for different BH mass bins (Fig.~\ref{fig:fedd_ratio}), the fraction of AGN in massive galaxies (Fig.~\ref{fig:agn_occ_massivegal}), to the diagnostics studied in \citet{2020arXiv200610094H}, i.e. the $M_{\rm BH}-M_{\star}$ diagram and the BH mass function, we can now explain the causes of such differences with observations. We caution here that our interpretation relies on the accuracy of the observational constraints, and that we would need to review it as new constraints become available.

\subsubsection{Simulation with a lower luminosity function than current observational constraints}
The AGN population produced in EAGLE is generally in better agreement with current observational constraints than the other simulations.
However, we find a lower luminosity function than the constraints in the bright regime, i.e. for $L_{\rm x}\geqslant 10^{44}\, \rm erg/s$ (Fig.~\ref{fig:LF}). 
The lower luminosity function likely comes from a lower fraction of AGN powered by relatively massive BHs of $M_{\rm BH}\sim 10^{8}-10^{9}\rm \, M_{\odot}$. This is shown by the lower BH mass function for this regime with respect to observational constraints \citep[see Fig.~11 of][]{2020arXiv200610094H}, but also by the absence of BHs with Eddington ratios $f_{\rm Edd}>-2$ for BHs of $M_{\rm BH}\sim 10^{8}-10^{9}\rm \, M_{\odot}$ (Fig.~\ref{fig:fedd_ratio}). Compared to the other simulations, there are fewer BHs in this mass range, and they are not accreting at high accretion rates. The smaller accretion rates result from the combination of the strong SN feedback, AGN feedback, and the modified Bondi accretion model of EAGLE which takes into account angular momentum \citep{2015MNRAS.454.1038R}.

\subsubsection{Simulations with a higher luminosity function than current observational constraints}
In simulations such as Horizon-AGN and Illustris, weak SN feedback could be responsible for the larger number of AGN in relatively low-mass galaxies with $M_{\star}\leqslant 10^{10} \, \rm M_{\odot}$ compared to observations.
Excluding the least massive galaxies from the AGN luminosity function, for example by applying a cut of $M_{\star}\geqslant 10^{9.5}\, \rm M_{\odot}$ \citep[see also][]{2016MNRAS.460.2979V}, leads to a lower normalization of the AGN luminosity function and better agreement with current observational constraints. This could indicate that a large fraction of the AGN located in the low-mass galaxies are not sufficiently regulated by SN feedback. This can be connected to the galaxy mass functions produced by simulations, and their agreement with observational constraints. The SN feedback model of Illustris \citep{2014Natur.509..177V} has been modified in the TNG simulations to correct for the larger stellar mass function in the low-mass regime \citep[see Fig.~4 in][for a comparison between the Illustris and TNG models]{2017arXiv170302970P}. 
The TNG simulations also overproduce the AGN luminosity function. The high number of AGN could be explained by the fast growing BHs with $M_{\rm BH}\leqslant 10^8 \, \rm M_{\odot}$ in galaxies of $M_{\star}=10^{10}-10^{10.5}\, \rm M_{\odot}$. These fast-growing BHs are found just after the regime where BHs and galaxies are regulated by SN feedback and before the regime where AGN are regulated by AGN feedback. Indeed, we find that the Eddington ratio distributions of the TNG BHs with $M_{\rm BH}\leqslant 10^{8}\, \rm M_{\odot}$ peak at higher $f_{Edd}$ than in the observational constraints available at $z=0$ (Fig.~\ref{fig:fedd_ratio}). Since the ratios are proportional to $\dot{M}_{\rm BH}/M_{\rm BH}$, the higher ratios indicate too high accretion rates for the given mass of the BHs. 
We can also connect this to the higher normalization of the TNG100 BH mass function for $M_{\rm BH}\leqslant 10^{8}\, \rm M_{\odot}$ \citep[Fig.~11 of][]{2020arXiv200610094H}. Interestingly, we note better agreement of the BH mass function for TNG300 with observations, and also better agreement of the AGN luminosity function of TNG300. 
SIMBA is an interesting case: this simulation produces a BH mass function not too different from the EAGLE mass function for $M_{\rm BH}\leqslant 10^{8.5}\, \rm M_{\odot}$ at $z>1$ \citep[see Fig.~11 of][]{2020arXiv200610094H}, and in agreement with observations, but produces an AGN luminosity function with a higher normalization than EAGLE.
From the Eddington ratio distributions shown in Fig.~\ref{fig:fedd_ratio}, the main difference between SIMBA and EAGLE is the ability (and number) of BHs with $M_{\rm BH}\geqslant 10^{7}\rm \, M_{\odot}$ to accrete efficiently and to power a large number of bright AGN with $L_{\rm bol}\geqslant 10^{45}\, \rm erg/s$ \citep[see Fig.~2 of][]{2020arXiv200610094H}.  
Our results highlight that different populations of BHs can be responsible for the higher normalization of the AGN luminosity functions produced in simulations compared to the current observational constraints.

\subsection{Moving forward: how to discriminate between simulation models with observations}
In this paper we have investigated the properties of the AGN populations produced by large-scale cosmological simulations. 
With various diagnostics (e.g., Eddington ratio distribution, AGN luminosity function, fraction of galaxies hosting an AGN), we showed that the simulations studied here do not reach a consensus on the population of AGN that they form.  
Given that none of the simulations was calibrated on AGN properties, their agreement with current observations is reasonable in some regimes, but still need to be worked on in some other regimes, as explained below. Some observational constraints also carry large uncertainties, e.g., at high redshift (e.g., AGN fraction in massive high-redshift galaxies, number density of faint AGN at $z\geqslant 4$), which makes the improvement of cosmological simulations a difficult task.

One major difference among the simulations is the population of relatively faint AGN with $L_{\rm x}\sim 10^{42-44}\, \rm erg/s$ at $z\geqslant 1$. Indeed, the AGN luminosity function of different simulations can vary by more than one order of magnitude for, e.g., $L_{\rm x}\sim 10^{43}\, \rm erg/s$ (Fig.~\ref{fig:LF}, Fig.~\ref{fig:number_density}).
Moreover, the population of faint AGN is powered by different BHs (and hosted in different galaxies) in the simulations.
In TNG100 and SIMBA, the population of faint AGN is significantly driven by massive BHs in massive quiescent galaxies, while in other simulations such as Illustris and EAGLE, the faint AGN are powered by lower-mass BHs in relatively low-mass galaxies \citep{2020arXiv201102501S}. The distinction of these populations in Horizon-AGN is not as strong, and faint AGN are powered by the two populations.
We find (Fig.~\ref{fig:number_density}) that the large discrepancies between simulations and observations in the number of faint AGN could arise from simulations producing too many of these faint AGN in low-mass galaxies of $M_{\star}=10^{9}-10^{10}\, \rm M_{\odot}$ at $z\geqslant 1$, compared to observations where most AGN are found in galaxies of $M_{\star}\geqslant10^{10}\, \rm M_{\odot}$. In observations there is a good agreement among studies for the space density of bright AGN with $L_{\rm x}\geqslant 10^{43}\, \rm erg/s$ for $z\geqslant 1$ \citep[e.g.,][]{2009ApJ...693....8B,2011ApJ...741...91C}, but the regime of faint AGN with $L_{\rm x}\sim 10^{42}-10^{43}\, \rm erg/s$ has larger uncertainties, and even more for higher redshift  \citep[e.g.,][and our Fig.~\ref{fig:number_density}]{2014ApJ...786..104U,2015MNRAS.451.1892A,2015ApJ...802...89B,2015A&A...578A..83G}. A source of uncertainty could be the fraction of heavily obscured Compton-thick AGN. We note that recent studies have investigated the fraction of obscured and unobscured AGN up to high redshift of $z\sim 4-5$ \citep[e.g.,][and references therein]{2005ApJ...635..864L,2015MNRAS.451.1892A,2016ApJ...827..150M,2017MNRAS.465.1915R,2020ApJ...903...85A}.
This regime needs to be exploited to constrain further the sub-grid models of the simulations, here likely the models of accretion and of AGN feedback. While there is a diversity of AGN feedback implementations, most simulations (except SIMBA) employ variations of the Bondi model. Because the uncertainties on accretion processes are large, and accretion modeling can strongly impact the AGN populations, our work motivates the needs to explore more models.
Today, exploring the faint regime is very challenging given the current X-ray facilities, but should become achievable with the Athena mission to be launched in $\sim 2030$, and the AXIS and LynX concept missions. Measuring the redshift of the AGN host galaxies could be possible, e.g., with 30m telescopes, allowing us to understand the co-evolution of the faint AGN and their galaxies.
Both observations and simulations/theory will benefit from new high sensitivity X-ray surveys.

Observationally, moving to the faint regime will be challenging as the contribution from X-ray binaries (XRBs) to the galaxy total X-ray emission is still unknown.
XRBs could contribute significantly as shown in both observations  \citep[e.g.,][]{2018ApJ...865...43F,2019ApJS..243....3L}, and simulations \citep{2020arXiv201102501S}. In the latter, the XRB contribution is estimated as a function of galaxy mass and SFR. These quantitites both differ from one simulation to another, leading to strong discrepancies in the XRB contribution among simulations \citep{2020arXiv201102501S}.

In this paper, we also find that the fraction of galaxies hosting an AGN varies strongly from one simulation to another, and this could open a new path to constrain simulation sub-grid physics. This is interesting both at low and high redshift, and both in low- and high-mass galaxies, for different reasons. 
The fraction of massive quenched galaxies produced at $z\leqslant 3$ differs in all the simulations, but is in good agreement with observational constraints at first order  \citep[e.g.,][]{2015MNRAS.446..521S,2017MNRAS.tmp..224K,2019MNRAS.485.4817D,2020arXiv200800004D}. AGN are responsible for quenching massive galaxies in these simulations, and as we have demonstrated in this paper the AGN do not have the same properties in all the simulations. The population of AGN needs to be investigated in more detail, and compared to observations.
To understand the cosmic evolution of the Universe, galaxy formation, and the co-evolution between BHs and galaxies, large-scale cosmological simulations need to reproduce {\it both} the population of galaxies and the population of BHs and AGN. From Fig.~\ref{fig:agn_occ_massivegal}, several simulations appear to lack AGN in massive galaxies at $z\leqslant 2$ (when using the radiative efficiency $\epsilon_{\rm r}$ used in the simulations) when compared to observational constraints. Therefore, AGN feedback in simulations could be efficient enough to quench as many massive galaxies as observed, but somehow too efficient in regulating the AGN themselves. This paradox needs to be addressed.\\

In observations, the fraction of massive galaxies with $M_{\star}\geqslant 10^{11}\, \rm M_{\odot}$ hosting an AGN is high ($80\%$) at $z\geqslant 3$, and lower ($\leqslant 50\%$) at $z<3$ (Fig.~\ref{fig:agn_occ_massivegal}).
While the time evolution is quite uncertain in observations due to the low number statistics at high redshift, this could be tested further in lower-mass galaxies.
Indeed, we find that the time evolution of the AGN fraction in the simulations
is somewhat similar in lower-mass galaxies with e.g., $M_{\star}\sim 10^{10}\, \rm M_{\odot}$ (Fig.~\ref{fig:agn_occ_all}). This work motivates the need to test this in observations, with multi-wavelength surveys to measure the galaxies' properties and X-ray detections of AGN.\\

The fraction of low-mass galaxies of $M_{\star}\sim 10^{9}\, \rm M_{\odot}$ hosting an AGN at low redshift can also vary from one simulation to another. This has important implications for BH formation and the ability of BHs to growth at early times. Comparing in detail the AGN fraction of these galaxies to the current systematic search of AGN in local dwarf galaxies \citep{2019arXiv191109678G,2013ApJ...775..116R,2015ApJ...809L..14B,2018ApJ...863....1C,2018MNRAS.478.2576M,2020MNRAS.492.2268B} will bring us a new way of constraining the BH physics in cosmological simulations, particularly the seeding, but probably also AGN feedback. Recently, there has been evidence for AGN outflows in dwarf galaxies \citep{2019ApJ...884...54M,2020arXiv201009008L}, and in simulations \citep{2020arXiv200710342K}.
We will investigate the AGN/low-mass galaxy regime connection in detail in our future work.

\subsection{AGN obscuration and AGN variability}
 The obscuration of the AGN is a key aspect when comparing simulated and observed populations of AGN. It may be the key to explain some of the discrepancies that we found in this paper. Gas and dust content along the line-of-sight can limit or extinguish the radiation coming from the AGN \citep[][for a review]{2015A&ARv..23....1B}. Obscuration and particularly the fraction of AGN that could be obscured has been an active field of research, but the picture is still unclear \citep[e.g.,][and references therein]{Ueda2003,2014ApJ...786..104U,2015ApJ...802...89B,2014MNRAS.437.3550M,2017MNRAS.469.3232G}. There is also evidence for obscuration at high redshifts \citep[e.g., $z>4$][]{2014A&A...562A..67G,2020A&A...642A.149V}, and even a possible large fraction of $\sim 50\%$ obscured AGN in the range $z\sim 4-6$ \citep{2018MNRAS.473.2378V}.

In this paper, we have used several obscuration models to make predictions for future observational surveys.  
Unsurprisingly, we found that obscuration can play a crucial role, and can lead to large uncertainties in the number of AGN that will be detectable with the upcoming Athena mission, and AXIS and LynX concept missions. Constraining the obscured fraction of AGN as a function of redshift, and BH/galaxy properties (e.g., BH and galaxy mass, AGN luminosity, galaxy compactness, sizes, SFR) appears crucial for deriving more accurate predictions for these new missions. Given the discrepancies that we found through this paper with current AGN observations, addressing AGN obscuration in the near future is essential.
On the simulation side, we could employ simulations with higher resolution to link galaxies and their BHs to their column density (possibly with machine learning techniques). Such a model would give us the likely column density and obscuration level for a given BH in a given galaxy; a model easily applicable to post-processing analyses of large-scale cosmological simulations.

Finally, the uncertainty on AGN obscuration in observations can only result in a possible increase of the number of observed AGN. Therefore, regimes in the simulations for which fewer AGN are produced than in observations are likely robust and can already be used to improve next-generation simulations. This is mostly the case for bright AGN at low redshift in half of the simulations studied here, and more globally for the EAGLE simulation. This could be due to too efficient AGN feedback.

More analyses of the bright AGN regime are needed, especially as the high luminosity end of the luminosity function is sensitive to time variability, as shown in this paper. The impact of AGN variability that we find e.g. on the AGN luminosity function could also be found for variations of the radiative efficiency, which is a constant parameter in all cosmological simulations.

\section{Conclusions}
In this paper, the second in our series, we have analyzed the six Illustris, TNG100, TNG300, Horizon-AGN, EAGLE, and SIMBA large-scale cosmological simulations. We focused on the populations of AGN produced by the simulations and how their properties evolve with both redshift and their BH/galaxies properties. Our goals were to understand how the simulations differ and to identify features in the AGN population that could help us to rule out and/or improve the sub-grid physics modeling in simulations. Through this paper we have addressed several comparisons between the simulated AGN populations and observational constraints. 
We have also predicted what population of BHs the next-generation X-ray missions Athena, AXIS, and LynX, could detect, and how accessing the fainter AGN regime will help us to constrain further the physics of cosmological simulations.
We summarize our main findings below.

\begin{itemize}
    \item The AGN populations produced by Illustris, TNG100, Horizon-AGN, EAGLE, and SIMBA, are different on many aspects.
    These differences are caused by the various sub-grid models of BH and galaxy formation physics.
    
    \item 
    Some simulations show a strong decrease of the median BH bolometric luminosity in massive galaxies (Fig.~\ref{fig:lum_bhmass}), due to AGN feedback. The trend is not as strong in current observational constraints.
    
    \item The distribution of Eddington ratios moves towards lower $f_{\rm Edd}$ ratios with time, for all the simulations (Fig.~\ref{fig:fedd_ratio}). However, the simulations do not have the same shapes and peak at different ratios for different BH masses. 
    At $z=0$, the simulations are in agreement with the observations of \citet{2004ApJ...613..109H} for BHs of $M_{\rm BH}\leqslant 10^{8}\, \rm M_{\odot}$, but not for more massive BHs: simulations either produce AGN with higher $f_{\rm Edd}$ (Illustris) or lower $f_{\rm Edd}$ distributions (TNG, SIMBA). A good agreement is found for Horizon-AGN.

    \item Considering all the AGN ($L_{\rm bol}>10^{43}\, \rm erg/s$) and independently of their BH masses, we find good agreement for the mean $f_{\rm Edd}$ with observations \citep{2013ApJ...764...45K} at high redshift ($z=4-3$) for all the simulations. However, at $z=0$ we note that the simulations predict higher mean $f_{\rm Edd}$.     

    \item The simulations produce different AGN luminosity functions (bolometric and hard X-ray, Fig.~\ref{fig:LF}); the discrepancies can be $\rm \geqslant 1\, dex$ at fixed luminosity. Compared to observations, most of the simulations produce too many AGN of any luminosity at high redshift, and the agreement improves towards lower redshift. We find the opposite behavior for EAGLE, which agrees better with constraints at high redshift.
    
    \item Several simulations produce very few of the brightest AGN (Fig.~\ref{fig:LF}). However, we found that the bright end of the luminosity function is sensitive to AGN short timescale variability (Fig.~\ref{fig:agn_variability_03}), which is not resolved in the simulations.

    \item The simulations have AGN number densities peaking at different redshifts (Fig.~\ref{fig:number_density}). In observations, brighter AGN peak at a higher redshift than fainter AGN \citep{2014ApJ...786..104U,2015MNRAS.451.1892A,2015ApJ...802...89B}. In simulations, we find that only the TNG and SIMBA simulations clearly present the same trend, while peaking at different redshifts.
    
    \item Several simulations produce too many faint AGN with $\log_{10}\, L_{\rm x}/(\rm erg/s)=42-43$ (and $\log_{10}\, L_{\rm x}/(\rm erg/s)=43-44$ for some of them) in low-mass galaxies of $M_{\star}=10^{9}-10^{10}\, \rm M_{\odot}$ with respect to observations (Fig.~\ref{fig:number_density}), especially at $z\geqslant 1$.
    
    \item All the simulations have a hard time producing a population of AGN in good agreement with observational constraints at {\it both} high and low redshift, but also for {\it both} faint and brighter AGN (Fig.~\ref{fig:number_density}).

    \item The fraction of galaxies, of a given mass,  hosting an AGN varies from simulation to simulation, with the largest differences for galaxies of $M_{\star}\sim 10^{9}\, \rm M_{\odot}$ and $M_{\star}\sim 10^{11}\, \rm M_{\odot}$ (Fig.~\ref{fig:agn_occ_all}). The differences in the low-mass regime have important implications for the search of AGN in local dwarf galaxies (see the discussion in section 7.1). For the massive galaxies, we find that the AGN fraction of all the simulations has the same trend as observations (high fraction at high redshifts, decreasing with time), but some simulations have a lower fraction at $z<2$ due to strong AGN feedback.
    
    \item The lower fraction of AGN found in simulated massive galaxies with respect to observations at low redshift, indicates that simulations all need AGN feedback that must be efficient enough to quench star formation, but somehow should not completely limit the AGN activity itself. 
    
\end{itemize}

  New constraints on the AGN population will come from the upcoming Athena X-ray mission, and the concept missions AXIS and LynX. We showed that the population and properties of AGN vary from one simulation to another, and that all individual simulations have a hard time producing AGN in agreement with current observations for all the luminosity and redshift regimes. However, our analysis of 6 large-scale cosmological simulations covers many different modelings, and thus is powerful to better estimate the uncertainties on the predicted AGN population to be detected by the new missions.

\begin{itemize}
    \item 
    All the simulations studied here predict different total numbers of AGN detections as a function of redshift, sometimes varying by more than one order of magnitude (Fig.~\ref{fig:number_AGN_100cMpc_redshift}). 
    
    \item The sensitivity of these missions will be more than one order of magnitude better than current X-ray facilities, and will allow us to observe fainter AGN. We find that Athena, AXIS, and LynX will make it possible to observe and recover the intrinsic distribution of AGN luminosity (and the corresponding BH mass distribution) to different levels (Fig.~\ref{fig:summary_athena}, Fig.~\ref{fig:summary_axis}, Fig.~\ref{fig:summary_lynx}). 
    The shape of the AGN luminosity (and BH mass) distributions being different for all the simulations, the new missions will provide us with a crucial pathway to constrain simulation sub-grid models.

\end{itemize}

In the next paper of our series we will link the BHs to their host galaxies. Particularly, we will investigate in detail the connections between the simulated AGN populations and the star-forming properties of their host galaxies in large-scale cosmological simulations.

\section*{Acknowledgment}
We thank Alexey Vikhlinin and Niel Brandt for very fruitful discussions on the next-generation X-ray missions. We thank the anonymous reviewer for valuable suggestions that helped improve the paper.
JA acknowledges support from a UKRI Future Leaders Fellowship (grant code: MR/T020989/1). DAA acknowledges support by NSF grant AST-2009687 and by the Flatiron Institute, which is supported by the Simons Foundation. 
YRG acknowledges the support of ``Juan de la Cierva Incorporation'' fellowship (IJC2019-041131-I). MV acknowledges support through NASA ATP grants 16-ATP16-0167, 19-ATP19-0019, 19-ATP19-0020, 19-ATP19-0167, and NSF grants AST-1814053, AST-1814259, AST-1909831 and AST-2007355.

\section*{Data Availability Statement}
The data from the Illustris and the TNG100 simulations can be found on their respective websites: https://www.illustris-project.org, https://www.tng-project.org. The data from the EAGLE simulation can be obtained upon request to the EAGLE team at their website: http://icc.dur.ac.uk/Eagle/. 
The data from the SIMBA simulation can be found on the website: http://simba.roe.ac.uk/.
Some catalogs of the Horizon-AGN simulation are available at: https://www.horizon-simulation.org/data.html, and some others are available on request.

\appendix

\section{Impact of parameters/models on the AGN luminosity function}
In this paper, we discussed the AGN luminosity function for the Illustris, TNG100, TNG300, Horizon-AGN, EAGLE, and SIMBA simulations. In the main text of the paper, we choose a given set of parameters/models:
\begin{itemize}
    \item We choose the radiative efficiency that was used in the different simulations to compute AGN luminosity, i.e., $\epsilon_{\rm r}=0.2$ for Illustris, TNG100, TNG300, and $\epsilon_{\rm r}=0.1$ for Horizon-AGN, EAGLE, and SIMBA. In practice, varying $\epsilon_{\rm r}=0.2$ in our post-processing analysis simply shifts the AGN luminosity to fainter or brighter luminosity. 
    \item To compute AGN luminosity, we considered AGN radiatively efficient if $f_{\rm Edd}>0.1$ and radiatively inefficient otherwise. A fraction of the AGN are fainter than when assuming that all AGN are radiatively efficient. This mostly affects the faint end of the AGN distribution. 
    \item To compute the X-ray luminosity of the AGN we use the bolometric correction of \citet{Hop_bol_2007}.
    \item Finally, we did not correct for AGN variability, that can not be captured in the simulations, in most of the paper. We showed the possible impact of AGN variability in Fig.~\ref{fig:agn_variability_03}.
\end{itemize}
Here, we investigate the impact of these parameters.

\subsection{Impact of the model to compute AGN luminosity on the AGN luminosity function}
\begin{figure*}
\centering
\hspace*{-0.3cm}
\includegraphics[scale=0.499]{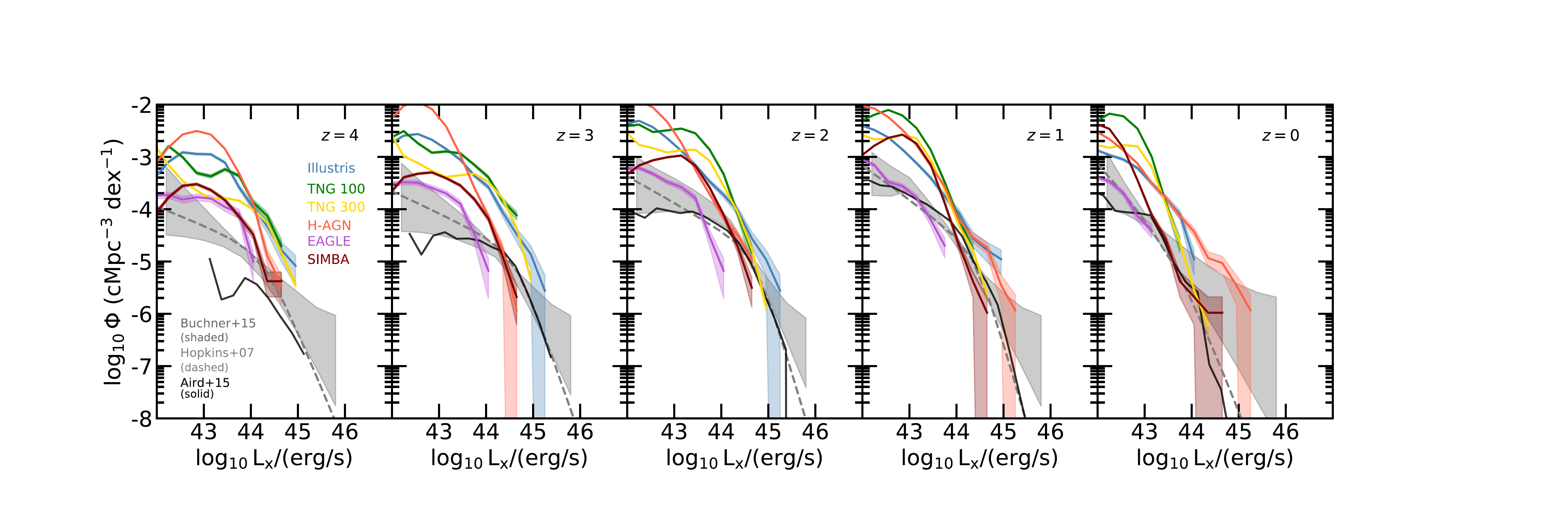}
\hspace*{-0.3cm}
\includegraphics[scale=0.499]{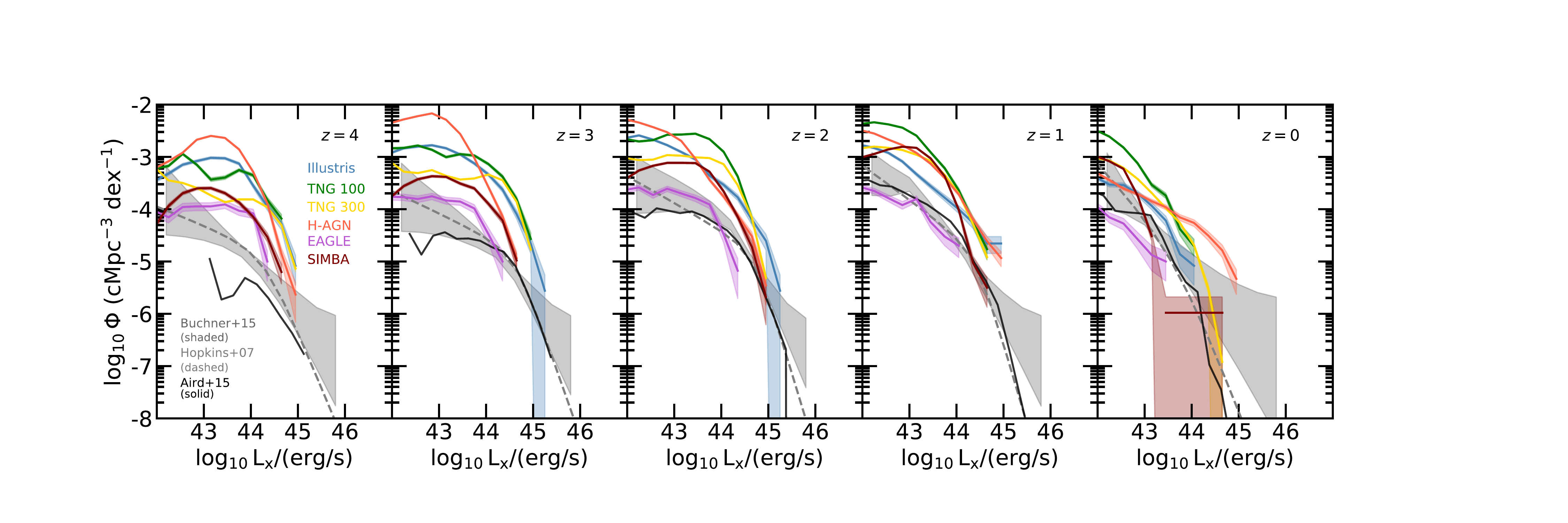}
\caption{{\it Top panels:} Hard X-ray (2-10 keV) AGN luminosity function for Illustris, TNG100, TNG300, Horizon-AGN, EAGLE, and SIMBA. Observational constraints from \citet{Hop_bol_2007,2015ApJ...802...89B,2015MNRAS.451.1892A} and are shown in grey. Here, we assume that all BHs are radiatively efficient AGN. Compared to the results presented in Fig.~\ref{fig:LF}, this assumption increases the number of AGN, particularly in the faint end of the luminosity function, i.e. for $\log_{10} L_{\rm x}\rm /(erg/s)\leqslant 44$. {\it Bottom panels:} Same as Fig.~\ref{fig:LF} but with the bolometric correction of \citet{2020A&A...636A..73D} instead of \citet{Hop_bol_2007}. The luminosity functions slightly shift towards more luminous AGN.}
\label{fig:LF2}
\end{figure*}

Often used in the analysis of cosmological simulations, the assumption that all AGN are radiatively efficient can increase significantly the number of AGN in certain luminosity bins. In Fig.~\ref{fig:LF2} (top panels) we show the hard X-ray (2-10 keV) AGN luminosity functions when making this assumption.
Compared to our previous model to compute the AGN luminosities, the main consequence is an increase of the number of AGN 
for luminosities of $L_{\rm x}\leqslant 10^{44}\, \rm erg/s$ and of $L_{\rm bol}\leqslant 10^{45}\, \rm erg/s$ at all redshifts.

In Fig.~\ref{fig:LF2} (bottom panels), we show the impact of the bolometric correction on the hard X-ray luminosity function, and use the correction of \citet{2020A&A...636A..73D} instead of \citet{Hop_bol_2007}. The main effect is a shift in the luminosity function of all simulations towards more luminous AGN. The choice of the bolometric correction does not affect the conclusions of the paper. 

\subsection{Impact of the model to compute AGN luminosity on the AGN fraction in massive galaxies}
\begin{figure}
\centering
\includegraphics[scale=0.49]{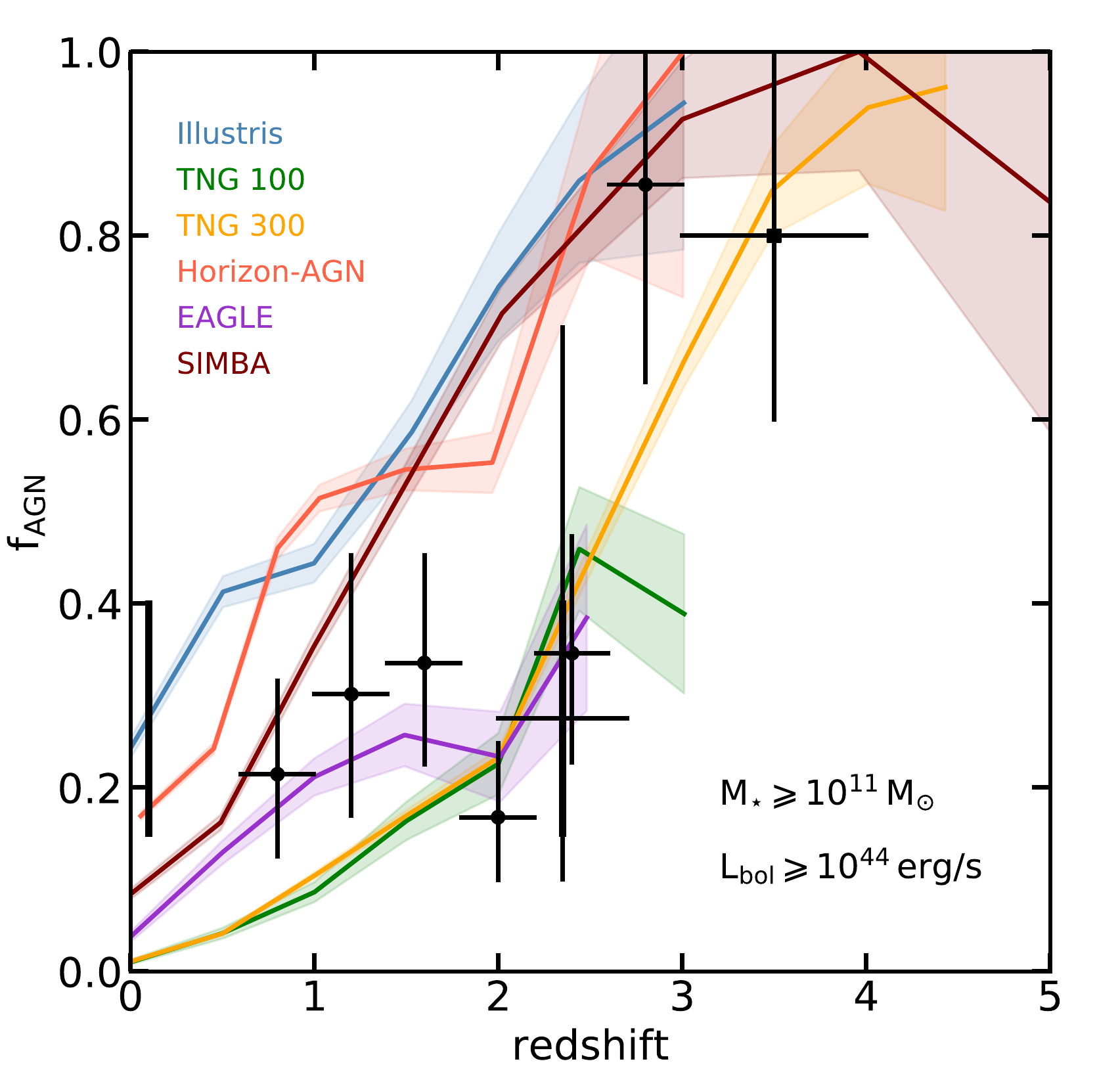}
\caption{Fraction of AGN with $L_{\rm bol}\geqslant 10^{44}\, \rm erg/s$ in massive galaxies of $M_{\star}\geqslant 10^{11}\, \rm M_{\odot}$ (right panels). Same as Fig.~\ref{fig:agn_occ_massivegal}, but assuming $\epsilon_{\rm r}=0.1$ for all the simulations and that all AGN are radiatively efficient when computing their luminosities. 
The number of AGN in Illustris and TNG is reduced slightly when using $\epsilon_{\rm r}=0.1$, but the dominant effect is an enhancement of the fractions of AGN in all the simulations when we assume that all AGN are radiatively efficient.
}
\label{fig:agn_occ_massivegal_bis}
\end{figure}

We show in Fig.~\ref{fig:agn_occ_massivegal_bis} the same figure as Fig.~\ref{fig:agn_occ_massivegal} but using the same radiative efficiency of $\epsilon_{\rm r}=0.1$ for all the simulations (instead of $\epsilon_{\rm r}=0.2$ for the Illustris and TNG simulations) and we assume that all the simulated AGN are radiatively efficient. This means that AGN with $f_{\rm Edd}\leqslant 0.1$ are not considered as radiatively inefficient, and therefore now have higher luminosity. Consequently, we find higher fractions of AGN for in all the simulations.
This particularly affects the AGN fraction at $z\leqslant 2.5$ in these massive galaxies \citep[see also][]{2019MNRAS.484.4413H}.

\section{Predictions for new X-ray missions}
\subsection{Detection limits}
\label{sec:lum_limit}
\begin{figure}
\centering
\includegraphics[scale=0.55]{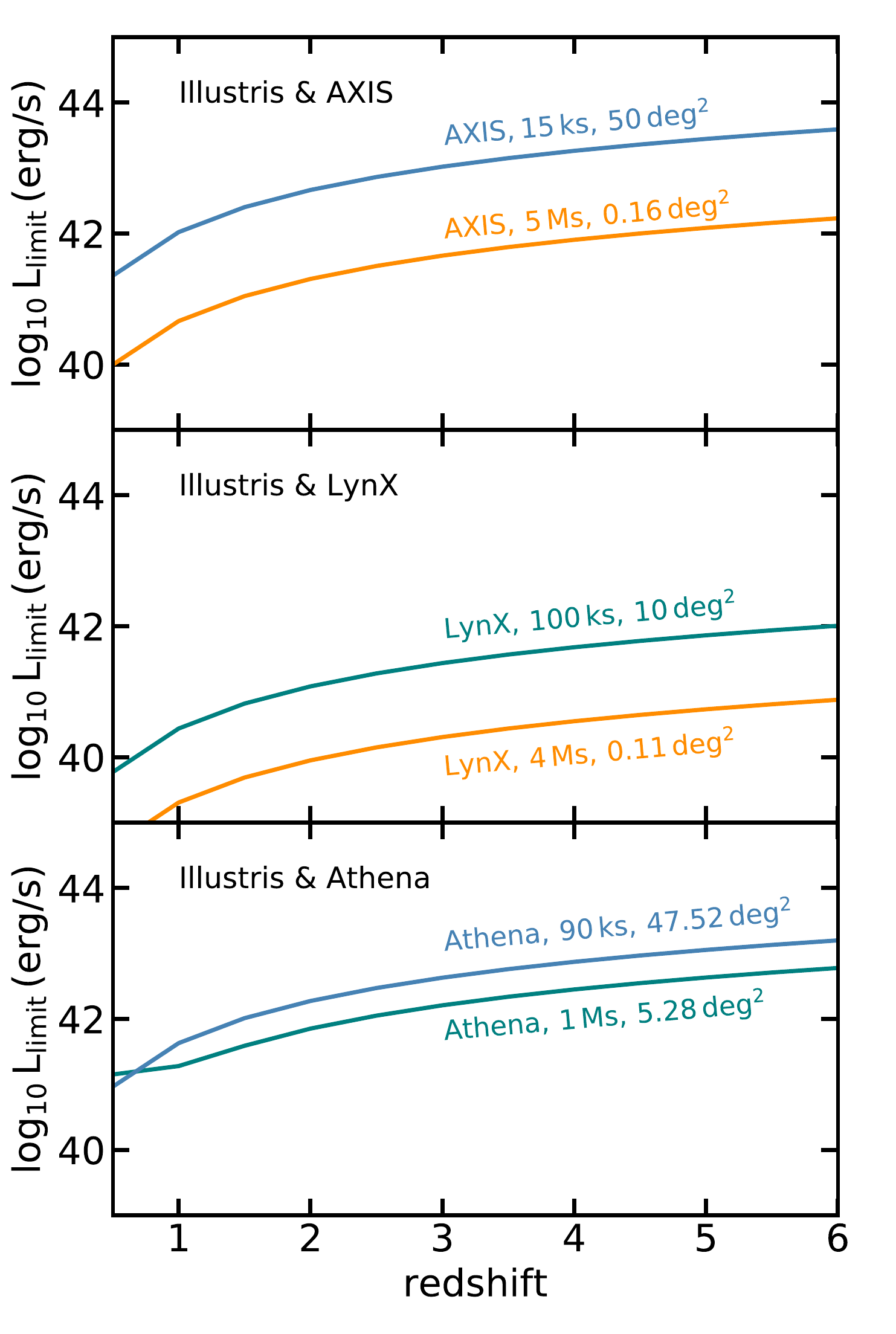}
\caption{Hard (2-10 keV) X-ray luminosity limits to detect an AGN with possible Athena, AXIS, and LynX surveys. We derive these limits from the sensitivity curve, exposure time and size of the mission surveys.}
\label{fig:lum_limit}
\end{figure}

We convert the 0.5-2 keV sensitivity curves shown in Fig.~\ref{fig:sensitivity} to the 2-10 keV band for each of the surveys by applying the following K-correction, and assuming $\gamma=1.4$:
\begin{eqnarray}
F_{\rm 2-10\, keV}=F_{\rm 0.5-2 \,keV} \, \frac{10^{2-\gamma}-2^{2-\gamma}}{2^{2-\gamma}-0.5^{2-\gamma}} \, (1+z)^{\gamma-2}.
\end{eqnarray}
The sensitivity limit to detect a source depends on the area of sky covered by the surveys and exposure time. 
We derive a fixed flux sensitivity for each survey, i.e. that we do not vary the sensitivity across the survey field of view. In practise, the sensitivity is higher at the edges of the single pointing's field of view, which can be improved by overlapping the pointings. Here, we do not enter in such detail to compute the detection limit since the Athena, Axis and LynX surveys are not yet finalized.
We express the flux limits as 2-10 keV X-ray luminosity limits with the expression $L_{\rm detection}=4\pi D_{\rm L}^{2} \times F_{\rm detection}$, with $D_{\rm L}^{2}$ the luminosity distance. The luminosity distance depends on redshift and cosmology; we employ the cosmology of each of the simulations, respectively.
As an example, we show the 2-10 keV luminosity limits to detect an AGN in Illustris (similar limits are found for the other simulations), in Fig.~\ref{fig:lum_limit}.

\subsection{Impact of obscuration models}
\label{sec:obscuration}
\begin{figure}
\centering
\includegraphics[scale=0.55]{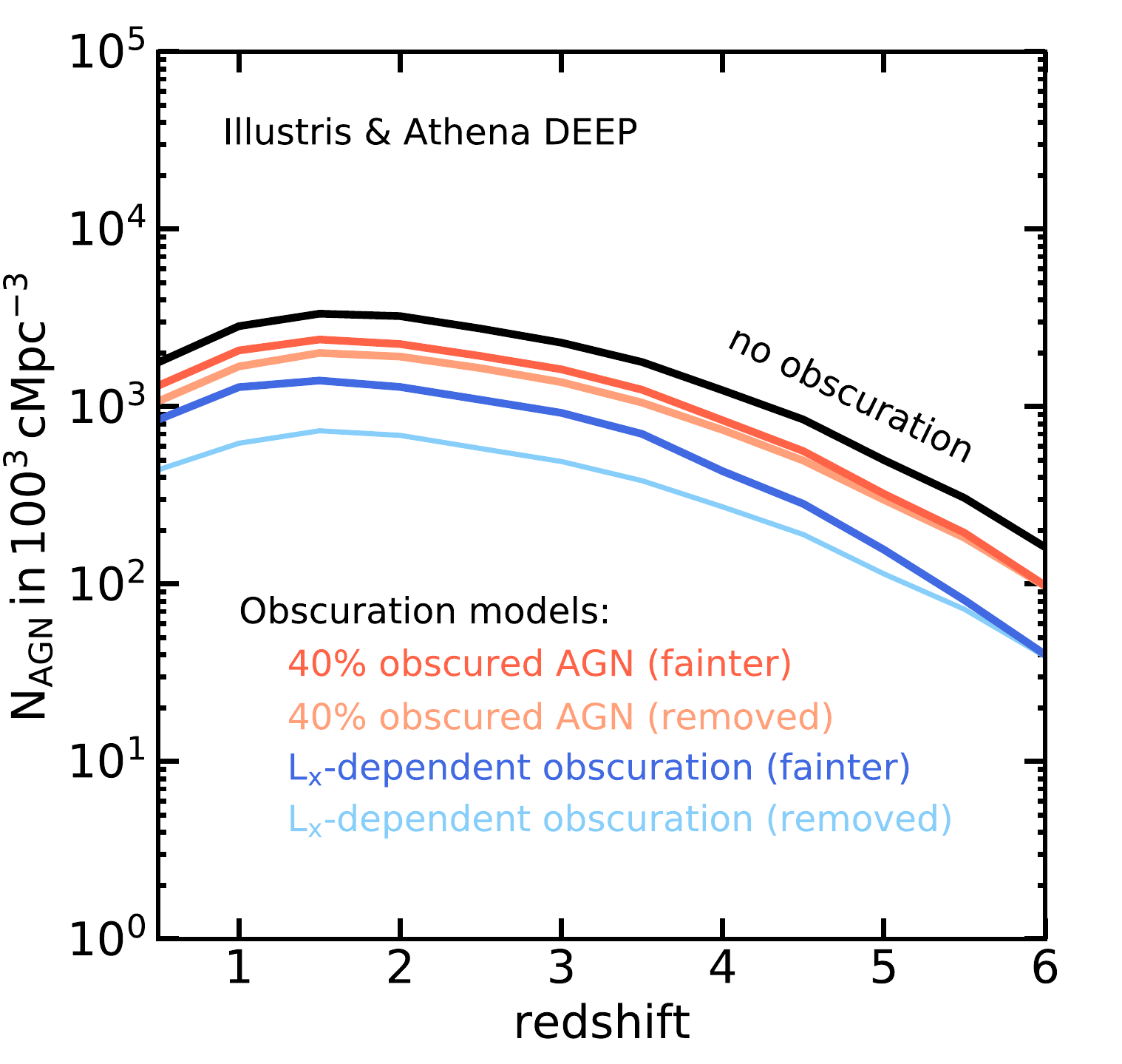}
\caption{Effect of our AGN obscuration models on the total number of detectable AGN in $\rm 100^{3}\, cMpc^{3}$, for the Illustris simulation and the Athena DEEP survey. A similar effect is found for the other simulations and surveys. The first models assume that $40\%$ of the AGN are heavily obscured, and the two other models assume an anti-correlation between the fraction of obscured AGN and their luminosity. We either remove the obscured AGN from the samples ({\it removed}), or decrease their luminosity ({\it fainter}).}
\label{fig:effect_obscuration}
\end{figure}

Our predictions on the number of AGN to be detected by X-ray missions are impacted by the number of AGN that could be obscured. As the fraction of obscured AGN is still uncertain, we test in Fig.~\ref{fig:effect_obscuration} several obscuration models.
We show our AGN luminosity- and redshift-dependent model, when we either remove the obscured AGN (light blue line), or decrease their luminosity (dark blue line).
We also test an obscuration model with $40\%$ of obscured AGN  (independently of redshift, and AGN luminosity), shown as red and orange lines.  
The $40\%$ obscuration model leads to a higher number of detectable AGN compared to our second AGN luminosity- and redshift-dependent obscuration models shown in Fig.~\ref{fig:effect_obscuration}.

\subsection{Best fits for the number of detectable AGN detectable}
In Table~\ref{tab:best_fit}, we provide the best-fit equations for the number of AGN (per $(100\, \rm cMpc)^{3}$) observable with the upcoming Athena mission and NASA concept missions AXIS and LynX, for all the simulations (1st to 5th rows). These predictions are shown in Fig.~\ref{fig:number_AGN_100cMpc_redshift} (top panels).
In the last row, we also provide the mean number of AGN per $(100\, \rm cMpc)^{3}$ over all the simulations $\rm \bar{N}_{\rm AGN}(z)$. Using $\rm (1/8{-}2)\times \bar{N}_{\rm AGN}$ allows to enclose all the simulations. 

\begin{table*}
    \caption{Best-fit equations for the number of AGN (per $(100\, \rm cMpc)^{3}$) observable with the upcoming Athena mission and NASA concept missions AXIS and LynX. We use the possible surveys described in Table~\ref{table:missions}, and our second model for obscuration depending on both redshift and AGN X-ray (2-10 keV) luminosity: obscured AGN are either removed ({\it removed} model), or detected as fainter AGN ({\it faint} model). The two models represent the upper and lower edges of the shaded regions in Fig.~\ref{fig:number_AGN_100cMpc_redshift}.
    The equations follow: $N_{\rm AGN}(z)/(100\, \rm cMpc)^{3}=a\,x + b\, x^{2}+c\, x^{3}+d \, x^{4}+e\, x^{5}+f$. We provide the fits for all the simulations in the 1-5th rows. We also add the mean AGN detections over all the simulations $\rm \bar{N}_{\rm AGN}(z)$ in the 6th row, and a region enclosing all the simulations can be defined by $\rm (1/8{-}2)\times \bar{N}_{\rm AGN}$.}
    \centering
    \begin{tabular}{|l|l|ll|}
    \hline
    Simulations & Missions & Coefficients (a,b,c,d,e,f) & Coefficients (a,b,c,d,e,f)\\
    && Obsc. $z$-,$L_{\rm x}$-dependent model (removed) & Obsc. $z$-,$L_{\rm x}$-dependent model (faint) \\
    \hline
    Illustris & Athena DEEP   & (1062.6,-540.7,102.6,-8.5,0.3,25.9) & (2489.9,-1425.4,331.4,-37.5,1.7,-80.7) \\
     & Athena WIDE   & (98.0,-48.3,-6.7,3.0,-0.2,455.5) & (-1293.7,904.1,-321.5,50.8,-2.9,1493.3)\\
     & AXIS DEEP   & (451.6,-66.7,-74.3,20.4,-1.5,617.2) & (957.2,-258.1,-105.0,33.0,-2.4,1950.6)\\
     & AXIS WIDE   & (-438.4,578.7,-255.5,43.4,-2.5,662.1) & (-2526.6,2251.9,-822.4,126.8,-7.0,2070.9)\\
     & LynX DEEP   & (1344.0,-848.9,157.1,-9.7,0.0,1014.9) & (4352.2,-1788.0,-10.5,63.3,-5.7,3260.7)\\
     & LynX WIDE   & (656.0,-199.0,-43.1,17.2,-1.3,633.5) & (1390.1,-261.3,-195.9,55.8,-4.0,2093.5)\\
    \hline
    TNG100 & Athena DEEP   & (1679.4,-1675.9,522.9,-70.8,3.6,1298.1)& (9655.1,-7707.3,2302.1,-306.7,15.3,557.1) \\
     & Athena WIDE   & (-489.5,-162.6,53.7,-2.9,-0.2,2138.0)& (-7123.9,4064.8,-1317.0,207.8,-12.3,7279.0) \\
     & AXIS DEEP   & (818.4,-1013.9,295.7,-36.3,1.7,1930.7) & (1505.7,-3208.2,984.1,-118.2,5.0,8134.5)\\
     & AXIS WIDE   & (-182.6,-296.1,82.6,-7.0,0.1,2096.6)&(-8160.3,5413.5,-1942.2,320.6,-19.3,8504.9) \\
     & LynX DEEP   & (569.6,-811.9,216.4,-23.0,0.9,2234.6)& (6421.4,-6216.0,1766.8,-219.5,10.3,7954.3) \\
     & LynX WIDE   & (691.1,-888.8,248.4,-28.9,1.2,2019.2) &(4075.4,-5140.0,1589.7,-205.9,9.8,7677.8)\\
    \hline
    Horizon-AGN & Athena DEEP   & (4611.1,-2967.2,951.7,-159.0,10.5,-1274.2) &(14974.7,-10892.3,3568.2,-558.7,33.6,-4536.0) \\
     & Athena WIDE   & (3322.2,-2431.7,829.0,-138.9,9.0,-701.3) &(6250.7,-5155.7,1833.6,-303.8,19.0,-1140.4)\\
     & AXIS DEEP   & (2780.3,-979.6,121.2,-14.4,1.5,-224.2)&(15381.7,-9970.1,3146.2,-519.2,34.2,-3280.4) \\
     & AXIS WIDE   & (1622.1,-594.8,157.5,-39.1,3.7,-125.8) &(6266.8,-4489.1,1632.4,-300.5,21.0,-1074.5)\\
     & LynX DEEP   & (2929.7,-192.5,-467.6,118.4,-8.2,318.6)&(13390.5,-1884.4,-1193.1,297.5,-17.8,-293.6) \\
     & LynX WIDE   & (2902.1,-854.9,16.4,8.7,0.1,-190.4)&(16153.3,-9860.2,3065.4,-522.3,35.9,-3185.2) \\
    \hline
    EAGLE & Athena DEEP   & (129.2,-113.5,36.3,-5.2,0.3,68.1)& (403.4,-328.7,102.2,-14.2,0.7,60.1) \\
     & Athena WIDE   & (-154.2,83.3,-24.5,3.5,-0.2,181.)& (-438.5,240.8,-69.0,9.7,-0.5,407.5) \\
     & AXIS DEEP   & (-114.9,20.7,-1.7,0.1,-0.0,283.5)& (-567.5,263.1,-73.5,10.6,-0.6,807.6) \\
     & AXIS WIDE   & (-191.5,124.9,-41.0,6.3,-0.4,206.9)& (-530.1,350.5,-113.0,17.0,-1.0,475.2) \\
     & LynX DEEP   & (120.4,-273.0,110.5,-17.9,1.0,443.9)& (-85.4,-418.8,179.0,-28.1,1.5,1518.9) \\
     & LynX WIDE   & (-28.6,-54.5,24.4,-3.9,0.2,273.9)& (-799.9,438.4,-137.8,21.4,-1.3,994.1) \\
    \hline
    SIMBA & Athena DEEP   & (145.8,-429.9,171.9,-27.2,1.5,740.3) & (5441.9,-5390.3,1982.3,-320.4,19.2,157.1)\\
     & Athena WIDE   & (-502.4,14.9,35.0,-7.5,0.5,1013.1) & (-1411.7,-205.3,260.5,-56.3,3.8,2783.0)\\
     & AXIS DEEP   & (-714.9,130.3,3.9,-3.7,0.3,1249.3) & (-1107.6,-681.6,378.9,-64.9,3.8,4020.2)\\
     & AXIS WIDE   & (-594.5,105.8,1.7,-2.5,0.2,1077.9) & (-1672.3,187.5,51.7,-14.0,0.9,3152.5)\\
     & LynX DEEP   & (-899.6,225.4,-20.4,-0.7,0.2,1402.6) & (-2532.4,92.9,170.3,-38.8,2.6,5423.8)\\
     & LynX WIDE   & (-802.7,183.8,-11.5,-1.6,0.2,1307.5) & (-1298.1,-574.1,344.7,-59.9,3.5,4295.7)\\
     \hline
     Mean   & Athena DEEP & (312.8,-14.5,-68.8,16.3,-1.0,549.2) & (3336.0,-2040.7,459.4,-45.6,1.7,238.0) \\
     over all & Athena WIDE &  (-404.9,320.6,-143.8,25.7,-1.6,882.1) & (-2162.3,1337.9,-469.8,76.4,-4.5,2569.4)\\
     simulations & AXIS DEEP   &  (-700.9,852.2,-381.0,64.8,-3.8,1191.4) & (-1413.5,1598.9,-785.8,141.6,-8.6,3773.3)\\
     & AXIS WIDE   &  (-917.4,851.4,-328.6,51.6,-2.8,1089.3) & (-3462.0,2744.2,-999.5,157.3,-8.9,3297.2)\\
     & LynX DEEP   &  (-910.4,-1202.9,-564.3,99.4,-6.0,1616.2) & (-3344.1,4889.9,-2290.4,397.5,-23.4,5967.8)\\
     & LynX WIDE   &  (-762.6,961.2,-432.6,73.9,-4.3,1259.5) & (-1372.4,1857.0,-924.9,164.5,-9.8,4017.5)\\
     \hline
    \end{tabular}
    \label{tab:best_fit}
\end{table*}

\bibliographystyle{mn2e}
\bibliography{biblio_complete,biblio_complete-AGN}

\end{document}